\documentclass[aps,prx,twocolumn,superscriptaddress,showpacs]{revtex4-1}
\usepackage{amsmath}
\usepackage{color}
\usepackage{graphicx}
\usepackage{epstopdf}
\usepackage{epsfig}
\usepackage{amsfonts}
\usepackage[naturalnames]{hyperref}
\usepackage{hypcap}
\usepackage{verbatim}
\usepackage{braket}
\usepackage{mathrsfs}
\usepackage{setspace}

\hypersetup{colorlinks = true, urlcolor = blue, linkcolor = blue, citecolor = blue}

\begin{document}


\title{Electron temperature and tunnel coupling dependence of zero-bias and almost-zero-bias conductance peaks in Majorana nanowires}

\author{F. Setiawan}\email{setiawan@umd.edu}
\affiliation{Condensed Matter Theory Center, Joint Quantum Institute, Station Q Maryland, and 
Department of Physics, University of Maryland, College Park, Maryland 20742-4111, USA}
\author{Chun-Xiao Liu}
\affiliation{Condensed Matter Theory Center, Joint Quantum Institute, Station Q Maryland, and 
Department of Physics, University of Maryland, College Park, Maryland 20742-4111, USA}
\author{Jay D. Sau}
\affiliation{Condensed Matter Theory Center, Joint Quantum Institute, Station Q Maryland, and 
Department of Physics, University of Maryland, College Park, Maryland 20742-4111, USA}
\author{S. Das Sarma}
\affiliation{Condensed Matter Theory Center, Joint Quantum Institute, Station Q Maryland, and 
Department of Physics, University of Maryland, College Park, Maryland 20742-4111, USA}

\date{\today}

\begin{abstract}
A one-dimensional semiconductor nanowire proximitized by a nearby superconductor may become a topological superconductor hosting localized Majorana zero modes at the two wire ends in the presence of spin-orbit coupling and Zeeman spin splitting (arising from an external magnetic field).  The hallmark of the presence of such Majorana zero modes is the appearance of a zero-temperature quantized zero-bias conductance peak in the tunneling spectroscopy of the Majorana nanowire.  We theoretically study the temperature and the tunnel coupling dependence of the tunneling conductance in such nanowires to understand possible intrinsic deviations from the predicted conductance quantization.  We find that the full temperature and the tunneling transmission dependence of the tunnel conductance does not obey any simple scaling relation, and estimating the zero-temperature conductance from finite-temperature and finite-tunnel-broadening tunneling data is difficult in general. A scaling relation, however, does hold at the extreme weak-tunneling low-temperature limit where the conductance depends only on the dimensionless ratio of the temperature and tunnel broadening. We also consider the tunneling contributions from nontopological Andreev bound states which may produce almost-zero-bias conductance peaks, which are not easy to distinguish from the Majorana-induced zero-bias peaks, finding that the nontopological almost-zero-modes associated with Andreev bound states manifest similar temperature and transmission dependence as the topological Majorana modes. We comment on the Zeeman splitting dependence of the zero-bias conductance peak for finite temperature and tunnel coupling.

\end{abstract}

\maketitle


\section{Introduction}\label{sec:introduction}

In 2001 essentially simultaneous theoretical works predicted (1) the generic existence of localized Majorana zero modes (MZMs) at the ends of a one-dimensional (1D) topological superconducting wire~\cite{Kitaev2001Unpaired}, and (2) that such MZMs generically manifest  at zero temperature ($T = 0$) a quantized zero-bias differential conductance peak of 2$e^2/h$ in the standard normal-superconductor (NS) tunneling spectroscopy through the 1D wire~\cite{Sengupta2001Midgap}.  It was pointed out~\cite{Kitaev2001Unpaired} that such paired MZMs at the wire ends, if sufficiently separated from each other leading to non-overlapping isolated Majorana modes~\cite{Read2000Paired}, would manifest non-Abelian braiding statistics which could be utilized to carry out intrinsically fault-tolerant topological quantum computation without the need for quantum error correction~\cite{Nayak2008Non-Abelian,Alicea2012New,DasSarma2015Majorana}. The subject took on particular significance when a practical route to creating topologically superconducting 1D semiconductor nanowires was proposed in 2010 combining proximity-induced superconductivity, spin-orbit coupling, and spin splitting~\cite{Sau2010Generic,Lutchyn2010Majorana,Oreg2010Helical,Sau2010Non}. In such proximitized nanowires, topological superconductivity with localized MZMs is achieved when the  spin-splitting energy is above a critical strength [the so-called topological quantum phase transition (TQPT) point] determined by the induced superconducting gap and chemical potential. The interest in the subject grew a great deal, and the subject remains among the most active research areas in condensed matter physics, when experimentalists observed~\cite{Mourik2012Signatures, Deng2012Anomalous, Das2012Zero, Churchill2013Superconductor, Finck2013Anomalous, Zhang2016Ballistic, Deng2016Majorana, Chen2016Experimental} the ``predicted" zero-bias peak (ZBP) in the differential tunneling conductance of semiconductor nanowires precisely following the theoretical proposal.  The actual reported conductance values of the experimental ZBPs, however, vary wildly (from $\sim 0.02 e^2/h$ to above $\sim 2e^2/h$) from experiment to experiment with the early experiments in rather disordered nanowires consistently reporting conductance values substantially (by more than one order of magnitude) below $2e^2/h$, whereas very recent experimental data from Copenhagen~\cite{Marcus2017PrivateCom} and Delft~\cite{Kouwenhoven2017PrivateCom} on clean and ballistic nanowires reported ZBPs with 2$e^2/h$ values (and often values $> 2e^2/h$).  What is, however, remarkable is that the ZBP develops in all these experiments only in the presence of finite Zeeman spin splitting (induced by an externally applied magnetic field) exactly as theory predicts, and the zero-field subgap conductance is relatively unremarkable in the clean system~\cite{liu2017phenomenology}. The subject has been reviewed in multiple recent review articles to which we refer for more details and extensive references~\cite{Nayak2008Non-Abelian,Alicea2012New,DasSarma2015Majorana, Beenakker2013Search, Leijnse2012Introduction, Stanescu2013Majorana, Elliott2015Colloquium, stanescu2016introduction, sato2016majorana, lutchyn2017realizing}.
 
The important question is, of course, whether the observed ZBP in nanowire experiments is indeed the `predicted' MZM signature, and this question has been (and continues to be) actively discussed in the theoretical literature~\cite{Lee2012Zero,Liu2012Zero,Dmitry2012Class,Pikulin2012Zero,Kells2012Near,Liu2017Andreev,Liu2017Role,Chiu2017Conductance,Prada2017Measuring,Clarke2017Experimentally,Lin2012Zero}.  Although observing a ZBP is only a necessary condition for the existence of MZMs, consistent reproducible observations of robust ZBPs in semiconductor nanowires with the measured conductance values close to 2$e^2/h$ at the lowest experimental temperatures would provide considerable support to the underlying cause being Majorana modes.  The nanowire conductance depends on many parameters, most of which are not well known experimentally.  Some of these parameters are the induced superconducting gap (which in turn depends on the parent superconducting gap and all the complex details of the proximity effect including the coupling between the parent superconductor and the semiconductor nanowire~\cite{Cole2015Effects,Stanescu2017Proximity,Cole2016Proximity}), spin-orbit coupling, Zeeman spin splitting, detailed properties of the nanowire (e.g., effective mass, $g$-factor, wire length and cross-section, number of occupied subbands), disorder (in the nanowire and in the environment including the parent superconductor), chemical potential in the wire, tunnel coupling, and temperature.  Even in the simplest possible idealized situation (assuming clean long wires with no disorder), the tunneling conductance depends on (at least) three independent variables: temperature, tunnel coupling, and Zeeman splitting.  (Chemical potential is also a relevant variable, but we assume that it remains fixed in the experiment.)  In the current work we investigate theoretically the parametric dependence of the conductance on these variables in a systematic manner to provide insight into recent experimental data reporting ZBPs with conductance values $\sim 2e^2/h$~\cite{nichele2017scaling,Kouwenhoven2017PrivateCom,HaoZhang2017PrivateCom,Marcus2017PrivateCom}. These reported observations of ZBPs with 2$e^2/h$ conductance values have made a critical analysis of the theoretical functional dependence of the subgap tunnel conductance on the nanowire system parameters a matter of considerable importance. We mention that some of the experimental nanowires~\cite{nichele2017scaling,Kouwenhoven2017PrivateCom,HaoZhang2017PrivateCom,Marcus2017PrivateCom} are actually based on 2D electron gas-based lithographic 1D systems as were realized and described in Refs.~\cite{suominen2017scalable,Shabani-2016-Two}.  In our work, we do not distinguish between pure nanowires and 2D-based 1D structures, calling them both ``nanowires" in this context,  since the applicable minimal model for the underlying MZM and ABS physics is the same in both situations.

One important point to emphasize right away is that at $T = 0$, the MZM zero-bias conductance must necessarily be quantized at 2$e^2/h$, independent of tunnel coupling and Zeeman splitting values, although it is possible for the zero-bias conductance not to have a peak or maximum~\cite{Setiawan2015Conductance,Wimmer2011Quantum}. This, of course, assumes the long wire limit (so that MZMs are isolated at the wire ends) and no disorder (and that the spin splitting is in the topological regime above the TQPT point so that the system is indeed a topological superconductor). Tunnel coupling at the NS tunneling contact can affect the width (or the broadening) of the MZM ZBP, but the zero-bias conductance must be 2$e^2/h$ at $T =0$ independent of the tunnel barrier strength.  With increasing temperature, however, the ZBP strength must decrease with concomitant increase in the ZBP width as thermal broadening starts contributing to the ZBP width.  This has an immediate implication:  If the experimentally observed ZBP height is already at or above the $2e^2/h$ conductance value at a finite temperature, then the corresponding zero-temperature ZBP strength is necessarily higher than the quantized value, casting doubt on the MZM as the underlying mechanism unless multichannel effects in the nanowire are playing a role~\cite{Wimmer2011Quantum}. However, such multichannel effects would invalidate the utility of ZBP as a probe for MZMs and therefore must be eliminated by an appropriate design of the point contact. Additionally, it is possible that the fridge temperature is different from the electron temperature, but this makes the issue even more problematic since the electron temperature is typically higher than the fridge temperature.  A ZBP with strength $>2e^2/h$ suggests the possibility of a subgap conductance arising from a nontopological (or trivial) Andreev bound state (ABS) in the nanowire as has been recently discussed in Ref.~\cite{Liu2017Andreev}.  The subgap conductance contributed by ABSs behaves similar to that by MZMs in the nanowire systems under consideration except that the ABS-induced ZBPs happen to be manifestly nontopological.  In view of the experimental observation of ZBPs with conductance value exceeding $2e^2/h$ already at finite temperature, we also investigate the dependence of ABS-induced ZBP on electron temperature, tunnel coupling, and spin splitting in order to discern any possible difference between MZM- and ABS-induced ZBP.  We note in this context the important point emphasized in Ref.~\cite{Liu2017Andreev}:  Given the present experimental imperfections of the $2e^2/h$ quantization, there is no way to know for sure whether a particular ZBP is topological or trivial simply on the basis of tunneling measurements since the TQPT point (i.e., the critical value of Zeeman splitting separating the trivial superconductivity at lower magnetic field from the topological superconductivity at higher field) is unknown whereas in the theory, it is manifestly known by construction.
 
Our work is partially motivated by the appearance of a recent experimental paper~\cite{nichele2017scaling} reporting an interesting scaling behavior in the experimental ZBP data as a function of temperature and tunnel coupling, as was predicted a long time ago by Sengupta \textit{et al.}~\cite{Sengupta2001Midgap} in the very weak tunneling (i.e., high-barrier-strength limit) --- see Eq.~\eqref{eq:fit} below, which is equivalent to Eq.~(12) in Ref.~\cite{Sengupta2001Midgap} and Eq.~(1) in Ref.~\cite{nichele2017scaling}.  This experimental paper~\cite{nichele2017scaling} reports the interesting finding that the measured ZBP tunneling conductance in semiconductor nanowires scales with respect to a single dimensionless scaling parameter given by the electron temperature expressed in units of the intrinsic tunnel broadening.  This scaling is construed to be consistent with the MZM behavior.  Our work based on the theoretically computed ZBP conductance as a function of electron temperature and tunnel coupling finds the following results. (1) The scaling behavior is only approximate and breaks down systematically for higher temperatures and higher transmission strengths (in fact, scaling applies only asymptotically in the regime where both temperature and tunnel broadening are much smaller than the induced topological gap). (2) any scaling applies equally well to ZBPs arising from both MZMs and ABSs so that a clear distinction between topological and trivial physics is not feasible based only on such scaling plots although a distinction is possible, as a matter of principle, based on the absolute magnitude of the zero-temperature ZBP conductance value itself. We point out that the possibility that ABS might reflect similar scaling behavior for the subgap conductance as MZM is mentioned in Ref.~\cite{nichele2017scaling} without any details.
 
We provide detailed numerical results for the calculated differential conductance properties as a function of electron temperature, intrinsic tunnel broadening, and Zeeman splitting for both MZMs and ABSs, and carefully investigate the scaling behavior of conductance with respect to temperature and tunnel broadening (i.e., whether the conductance can be expressed purely in terms of a single dimensionless parameter combining the two). In the current work, we critically investigate the extent to which the subgap conductance may be expressed in terms of a dimensionless scaling variable and the associated constraints on such a scaling behavior.  We also numerically connect the intrinsic subgap tunnel broadening with the above-gap tunnel conductance since they both arise from the tunnel coupling properties.  Experimentally, the tunnel barrier properties (i.e., the tunneling transmission coefficient or the tunnel barrier strength) are not known, and often the above-gap normal conductance is used (somewhat erroneously in our view~\cite{liu2017phenomenology}) to estimate the intrinsic subgap tunneling properties.  We also study in some depth how the ZBP strength varies as a function of the Zeeman spin splitting in various regimes in order to make direct comments on the experimental findings. 

We mention that dissipative broadening, arising from extrinsic (and unknown) sources (e.g., vortices in the environment, coupling to fermionic baths in the environment, stray nonthermal quasiparticles, etc.) also provide ZBP broadening, which has nothing whatsoever to do with the thermal and tunnel broadening being considered in the current work.  Earlier works by us have considered the possible role of dissipative broadening in Majorana nanowires~\cite{Liu2017Role,DasSarma2016How}, and therefore, we do not consider dissipation at all in the current work.  The goal of the current work is to address the intrinsic effects of finite temperature and tunnel coupling in Majorana physics, and any finite dissipation will have strong effects on the results presented in the current work, particularly at low temperature and small tunnel broadening, completely eliminating any `scaling' behavior in the conductance as a function of temperature and tunnel broadening.  Since the details of dissipation in actual nanowires is unknown, we do not see much point in discussing dissipation in the current work although it is straightforward to include dissipation in the theory following the procedure of Refs.~\cite{Liu2017Role,DasSarma2016How}.

The rest of the paper is organized as follows. In Sec.~\ref{sec:model}, we begin by discussing the NS junction model and the numerical method used to calculate the differential conductance. We proceed to give the results of the differential conductance for an MZM in a semiinfinite nanowire in Sec.~\ref{sec:majorana} for two cases, namely Zeeman-independent and Zeeman-dependent bulk superconducting gap. (The reason for choosing a magnetic-field dependent bulk superconducting gap here is that experiments often find that the bulk gap is suppressed by the applied magnetic field leading to its eventual collapse at a large enough field~\cite{Deng2016Majorana,nichele2017scaling}.) In particular, we show that the zero-bias differential conductance peak is a function of a dimensionless parameter (the ratio of temperature to tunnel broadening) only in the regime where the tunnel broadening and temperature are both much smaller than the proximity-induced superconducting gap. In Sec.~\ref{sec:ABS}, we show that the same scaling function also holds for zero-energy ABSs in the low-temperature weak-tunneling regime, and the generic dependence of the subgap conductance arising from MZMs and ABSs is pretty similar qualitatively. We end with a discussion of the experimental situation and open questions in Sec.~\ref{sec:conclusion}. The goal of the current work  is mainly the presentation of a large number of numerical simulations of semiconductor Majorana nanowire properties as a function of electron temperature, tunnel coupling, and Zeeman splitting with the emphasis on understanding the recent experimental data~\cite{nichele2017scaling,Kouwenhoven2017PrivateCom,HaoZhang2017PrivateCom,Marcus2017PrivateCom} in disorder-free ballistic nanowires manifesting hard induced gap at zero magnetic field and large conductance ($\sim 2e^2/h$) ZBPs at finite magnetic field. 

\section{Model}\label{sec:model}
\begin{figure}
\begin{center}
\includegraphics[width=\linewidth]{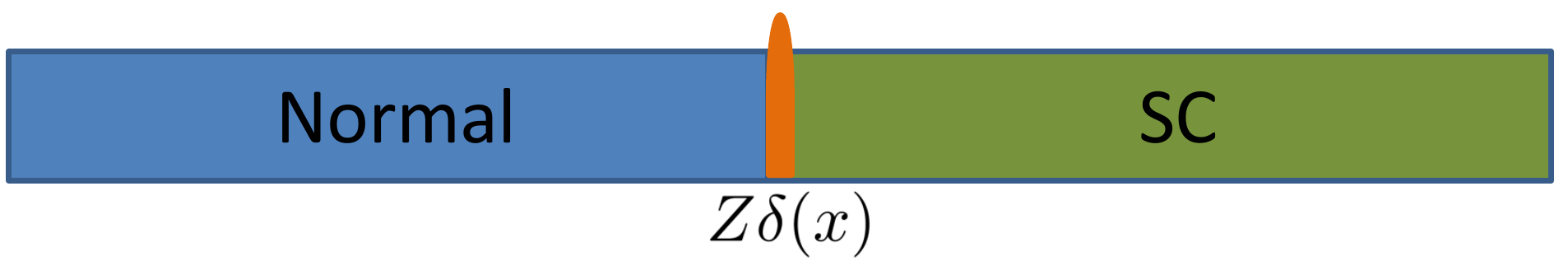}
\end{center}
\caption{(color online) Schematic diagram of a 1D NS junction where the barrier at the interface is modeled as a delta function with strength $Z$. Note that the interface delta-function barrier implicitly has a length scale inherent in its definition so that $Z$ itself is dimensionless with $Z$ = 0 ($\infty$) defining zero (infinite) barrier.}\label{fig:NS} 
\end{figure}

We consider a 1D NS junction where the normal region is a semiinfinite spin-orbit-coupled normal lead and the superconductor is a spin-orbit-coupled nanowire proximitized by an $s$-wave superconductor in the presence of an externally applied magnetic field creating a Zeeman spin splitting~\cite{Lutchyn2010Majorana,Oreg2010Helical}. As in the well-known Blonder-Tinkham-Klapwijk (BTK) work~\cite{Blonder1982Transition}, the tunnel barrier at the junction is modeled as a delta-function barrier with a strength $Z$ at the NS interface (see Fig.~\ref{fig:NS}). This tunnel barrier parameter $Z$ is inversely related to the tunnel coupling between the normal lead and the superconductor: high (low) value of $Z$ corresponds to small (large) tunnel coupling which in turn corresponds to small (large) junction transparency. The value of $Z$ cannot be determined experimentally since the microscopic details of the junction are generally unknown. In the experiment, the junction transparency (which is inversely related to $Z$) is usually taken to be the normal-state conductance, i.e., the above-gap conductance at high voltages. This extraction of tunnel coupling (or $Z$) using the above-gap normal-state conductance is not quite accurate since the normal-state conductance (1) depends on many  variables (e.g., density of states) in addition to tunnel coupling, and (2) is not uniquely defined as it depends on the bias voltage~\cite{liu2017phenomenology}. In addition, the normal-state conductance could contain the contribution of several channels with different transparencies, which makes it difficult to extract the average transparency~\cite{Beenakker1992Transport}. In the case where there is a ZBP in the NS conductance spectrum, the ZBP width or level broadening (at zero temperature) is inversely proportional to the barrier strength $Z$. In principle, therefore, the tunnel coupling determines three different quantities: the barrier strength $Z$ in the BTK theory, the level broadening of subgap conductance peak, and normal-state conductance.  We consider $Z$ to be the fundamental variable since it directly determines the interface scattering properties in the quantum mechanical transport theory.  

The Hamiltonian of the junction in the particle-hole space is
\begin{align}
H_j = \frac{1}{2}\int dx\Psi^{\dagger}_j(x)\mathcal{H}_j(x)\Psi_j(x),
\end{align}
where the Nambu spinor is $\Psi_j(x) = (\psi_{j\uparrow}(x),\psi_{j\downarrow}(x),\psi^\dagger_{j\downarrow}(x),-\psi^\dagger_{j\uparrow}(x))^{\mathrm{T}}$ with $\psi^\dagger_{j\sigma}(x)$ and $\psi_{j\sigma}(x)$ being respectively the operators that create and annihilate an electron with spin $\sigma$ in
region $j= \mathrm{lead}$ and $j = \mathrm{NW}$ (nanowire). The Hamiltonian of the normal lead and nanowire in the Bogoliubov-de Gennes (BdG) form are given by
\begin{subequations}\label{eq:hamiltonian}
\begin{align}
\mathcal{H}_{\mathrm{lead}} &=  \left(-\frac{\hbar^2\partial_x^2}{2m^*} - \mu_{\mathrm{lead}}\right)\tau_z -i \alpha \partial_x \tau_z\sigma_y + V_Z \sigma_x , \\
\mathcal{H}_{\mathrm{NW}} &= \left(-\frac{\hbar^2\partial_x^2}{2m^*} - \mu\right)\tau_z -i \alpha \partial_x \tau_z\sigma_y + V_Z \sigma_x + \Delta \tau_x,\label{eq:H_NW}
\end{align}
\end{subequations}
where $\mu_{\mathrm{lead}}$ ($\mu$) is the chemical potential of the lead (nanowire), $\alpha$ is the strength of spin-orbit coupling, $V_Z$ is the spin-splitting Zeeman field, $\Delta$ is the proximity-induced $s$-wave pairing potential from a parent superconductor, and $\sigma_{x,y,z}$ ($\tau_{x,y,z}$) are the Pauli matrices in the spin (particle-hole) subspace.  In the following, we will work in the unit where the reduced Planck constant $\hbar$, Boltzmann constant $k_{\mathrm{B}}$, and the spin-orbit length $\ell_{\mathrm{SO}}= \hbar^2/(m^*\alpha) = 0.1$ $\mu$m, which corresponds to the electron effective mass $m^*=0.015m$ (the effective electron mass in InSb nanowires~\cite{Mourik2012Signatures} with $m$ being the bare electron mass) and $\alpha$ = 0.5 eV{\AA}, . We set $\mu_{\mathrm{lead}} = 25\Delta_0$ [where $\Delta_0 \equiv \Delta (V = 0)$] throughout the numerical simulations performed in this paper. (Results do not qualitatively depend on the choice of $\mu_{\mathrm{lead}}$ as long as it is large.)

We calculate the differential conductance $G \equiv dI/dV$ of the NS junction for two different cases: (1) an MZM in a semiinfinite nanowire, and (2) a zero-energy ABS arising from a smoothly varying chemical potential profile in a quantum dot at the NS interface of a finite-length nanowire~\cite{Liu2017Andreev}. To this end, we numerically calculate the scattering matrix~\cite{Setiawan2015Conductance} from the numerical transport package Kwant~\cite{Kwant} with the BdG Hamiltonian~[Eq.~\eqref{eq:hamiltonian}] discretized into a 1D tight-binding chain. Using the scattering matrix, we can calculate the zero-temperature differential conductance (in unit of $e^2/h$) as
\begin{equation}\label{eq:conductance}
G_0(V) = 2 + \sum_{\sigma,\sigma' = \uparrow \downarrow} \left(|r^{\sigma \sigma'}_{eh}(V)|^2 -  |r^{\sigma \sigma'}_{ee}(V)|^2 \right), 
\end{equation}
where $r_{eh}$ and $r_{ee}$ are the Andreev and normal reflection amplitudes, respectively. The constant $2$ in Eq.~\eqref{eq:conductance} is due to the fact that we consider a one-subband lead with \textit{two} spin channels. We note that in this paper we also consider the nanowire to have one subband with two spin channels. This is the extensively used minimal model for studying Majorana physics in nanowires.

The finite-temperature conductance is calculated by convoluting the zero-temperature conductance with the derivative of the Fermi function $df/dE$, i.e.,
\begin{align}\label{eq:GT}
G(V) &= - \int_{-\infty}^{\infty} dE G_0(E)\frac{df(E-V)}{dE}, \nonumber\\
&= -\int_{-\infty}^{\infty} dE G_0(E)\left[\frac{1}{4T \cosh^2((E-V)/(2T))}\right].
\end{align}
Note that this finite-temperature convolution formula applies independent of the origin of the $T=0$ subgap conductance $G_0(V)$ in Eq.~\eqref{eq:conductance} --- in particular, both MZM and ABS subgap conductances obey exactly the same convolutions and will thus manifest similar finite temperature behavior if $G_0(V)$ is similar. In the following sections, we use Eq.~\eqref{eq:GT} to calculate the conductance for the nanowire in the presence of an MZM (Sec.~\ref{sec:majorana}) and an ABS (Sec.~\ref{sec:ABS}).

\section{Majorana in a semiinfinite nanowire}\label{sec:majorana}
In this section, we compute the differential conductance of a semiinfinite Majorana nanowire in the topological regime for two different cases, namely $V_Z$-independent and $V_Z$-dependent bulk superconducting gap. Our reason for considering a semiinfinite wire (in contrast to finite-length wires) is to avoid Majorana oscillations inherent in finite systems~\cite{Cheng2009Splitting,DasSarma2012Splitting}, which would complicate the interpretation of our theoretical results. A semiinfinite wire, by definition, has an isolated MZM at the wire end (near the NS tunnel junction) and has no Majorana oscillations.

\subsection{Zeeman-independent bulk gap}
\begin{figure}[h!]
\begin{center}
\includegraphics[width=\linewidth]{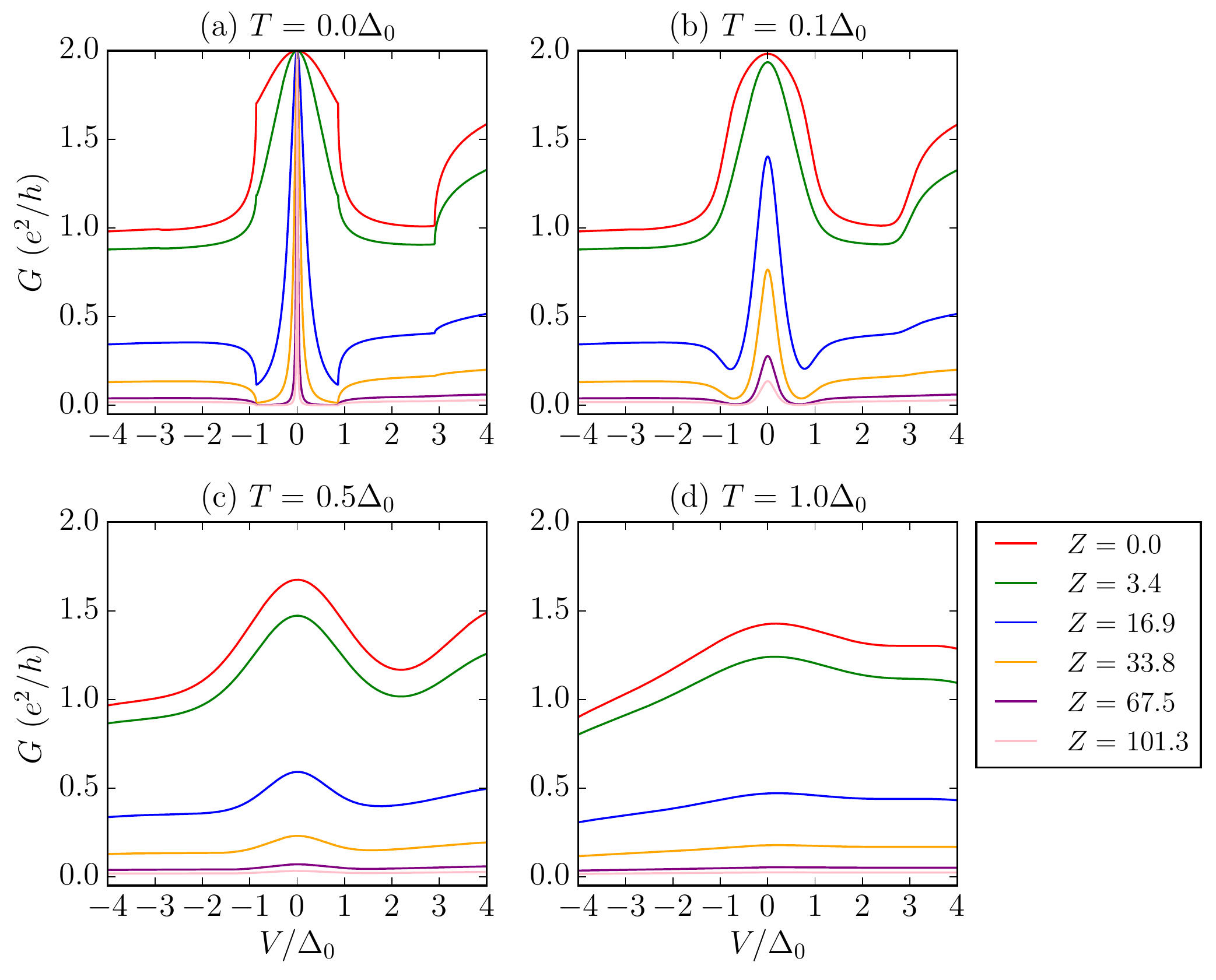}
\end{center}
\caption{(color online) Differential conductance $G$ vs voltage $V$ for different barrier strengths $Z$ at several fixed temperatures $T$: (a) $T = 0$, (b) $T = 0.1 \Delta_0$, (c) $T = 0.5\Delta_0$, and (d) $T = \Delta_0$. The parameters used are $\mu = 5$, $\Delta_0 = 1$, and $V_Z = 8$ (above TQPT). The plots are for $V_Z$-independent bulk gap case.}\label{fig:dIdV_diff_Z_nocollapse} 
\end{figure}

\begin{figure}[h!]
\begin{center}
\includegraphics[width=\linewidth]{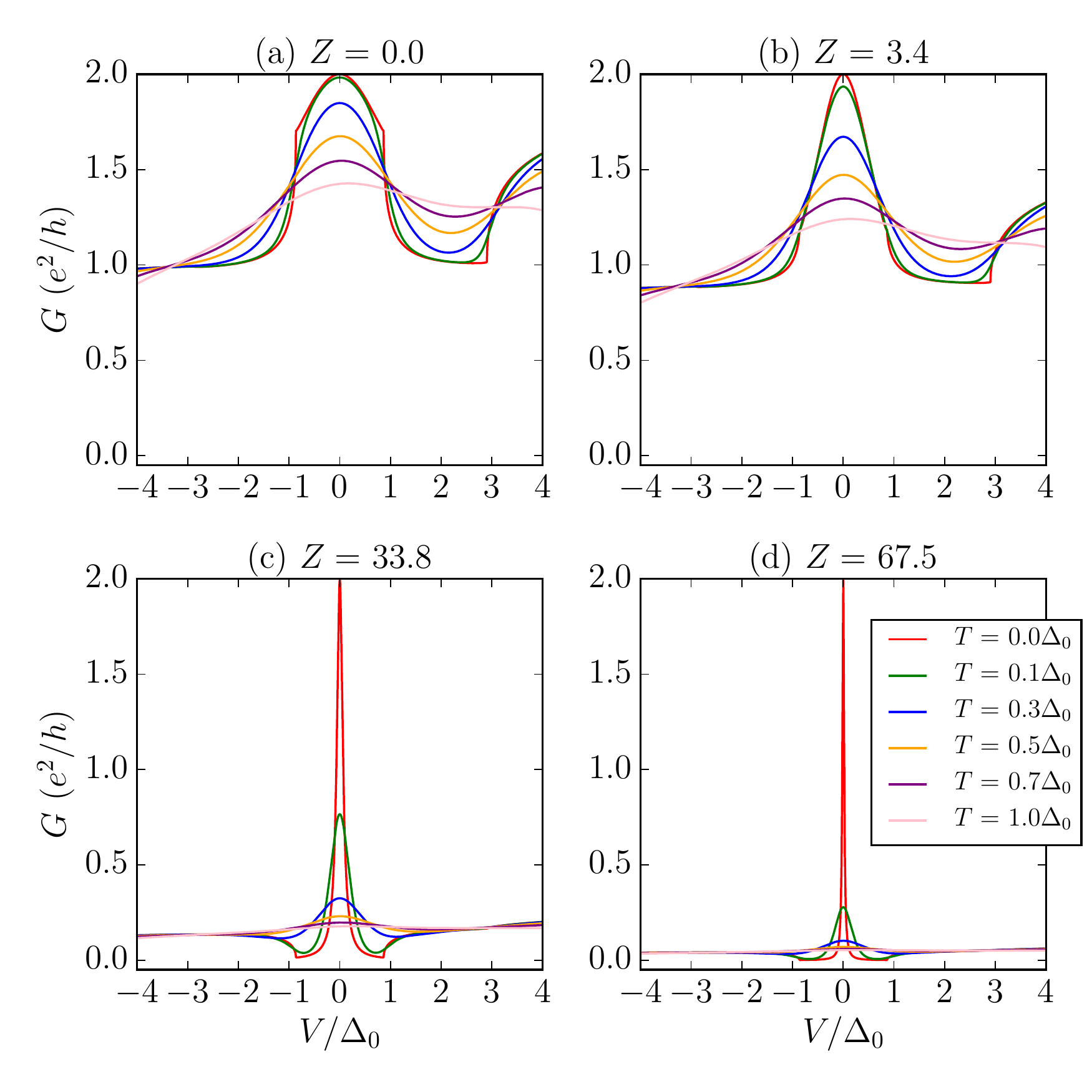}
\end{center}
\caption{(color online) Differential conductance $G$ vs voltage $V$ for different temperatures $T$ at several fixed barrier strengths $Z$: (a) $Z =0$, (b) $Z = 3.4$, (c) $Z = 33.8$, and (d) $Z = 67.5$. The parameters used are $\mu = 5$, $\Delta_0 = 1$, and $V_Z = 8$ (above TQPT). The plots are for $V_Z$-independent bulk gap case.}\label{fig:dIdV_nocollapse_diff_temp} 
\end{figure}

\begin{figure}[h!]
\begin{center}
\includegraphics[width=\linewidth]{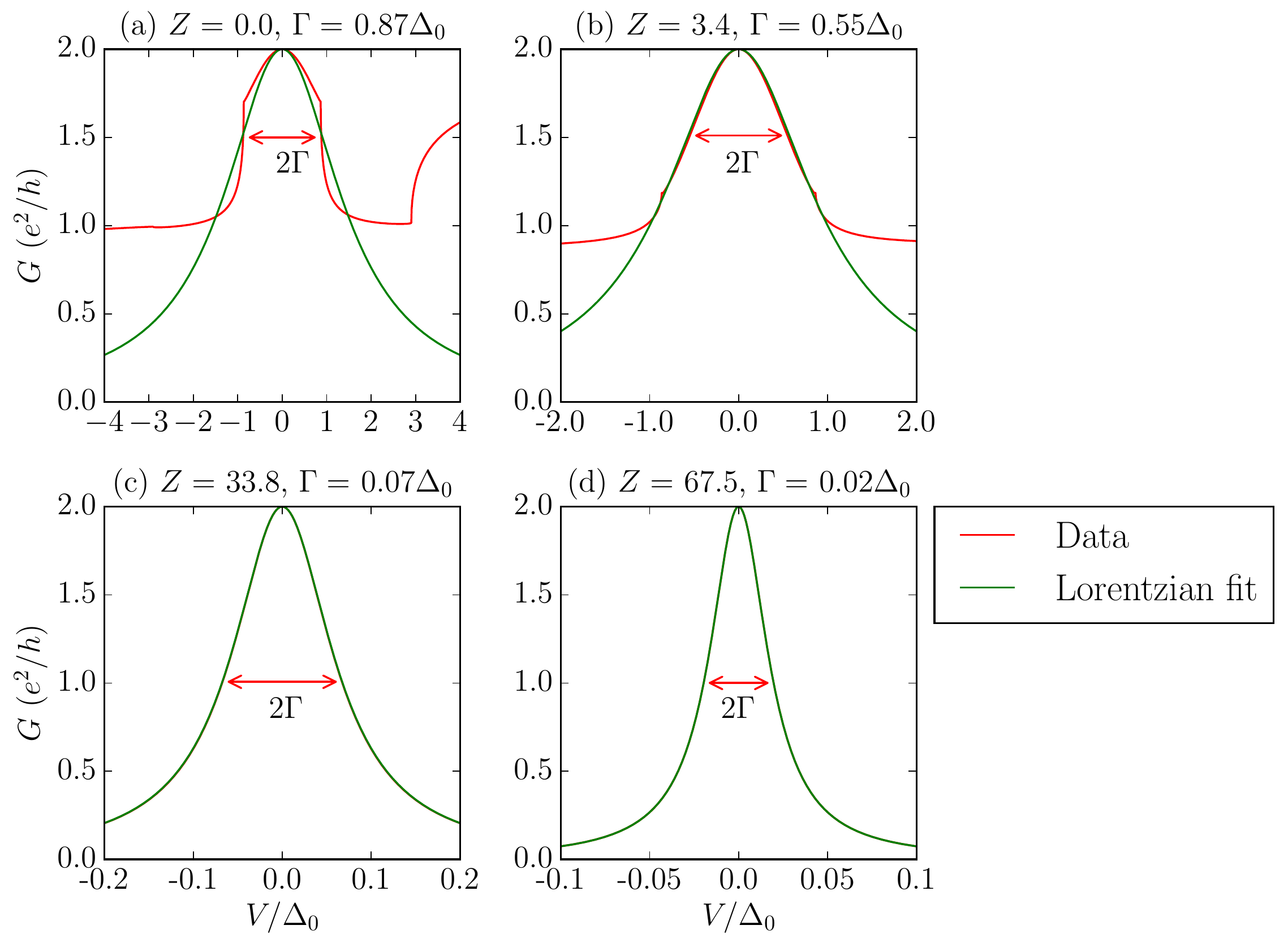}
\end{center}
\caption{(color online) Zero-temperature differential conductance $G$ vs voltage $V$ for different barrier strengths $Z$: (a) $Z =0$, (b) $Z = 3.4$, (c) $Z = 33.8$, and (d) $Z = 67.5$. The width of the zero-temperature conductance peak is characterized by $\Gamma$, which is the half-width at half maximum of the conductance plots shown by the red curves. Note that only in the weak-tunneling or high-barrier-strength limit the conductance peak can be fitted to a Lorentzian curve [$G(V) = 2 \Gamma^2 / (V^2 + \Gamma^2)$] as shown by the green curves. The plots are for $V_Z$-independent bulk gap case.}\label{fig:dIdV_diff_bh_diff_temp} 
\end{figure}

In this subsection, we focus on the case where the induced bulk superconducting gap is independent of the Zeeman splitting, i.e., $\Delta = \Delta_0 \equiv \Delta(V_Z = 0)$. Figures~\ref{fig:dIdV_nocollapse_diff_temp} and \ref{fig:dIdV_diff_bh_diff_temp} show the calculated differential conductance $G$ for the nanowire in the topological regime $(V_Z > \sqrt{\Delta_0^2 + \mu^2})$. The conductance decreases with increasing barrier strength $Z$, except the zero-temperature zero-bias conductance value which is quantized at $2e^2/h$ [Fig.~\ref{fig:dIdV_nocollapse_diff_temp}(a)] for all $Z$ due to perfect resonant Andreev reflection from the MZM~\cite{Sengupta2001Midgap, Law2009Majorana,Flensberg2010Tunneling,Wimmer2011Quantum,Setiawan2015Conductance}.  In the weak-tunneling or high-barrier-strength limit, the zero-bias conductance appears as a peak in the spectrum with its width decreasing with increasing barrier strength. However, in the strong-tunneling limit, depending on the system details (i.e., spin-orbit coupling strength $\alpha$, the chemical potential of the lead and nanowire, Zeeman field, etc.), the zero-bias conductance may or may not manifest as a peak in the conductance spectrum~\cite{Setiawan2015Conductance}.  At finite temperature, the zero-bias conductance value is no longer quantized at $2e^2/h$ but it decreases with increasing barrier strength as shown in Figs.~\ref{fig:dIdV_diff_Z_nocollapse}(b),~\ref{fig:dIdV_diff_Z_nocollapse}(c), and~\ref{fig:dIdV_diff_Z_nocollapse}(d).

\begin{figure}[h!]
\begin{center}
\includegraphics[width=\linewidth]{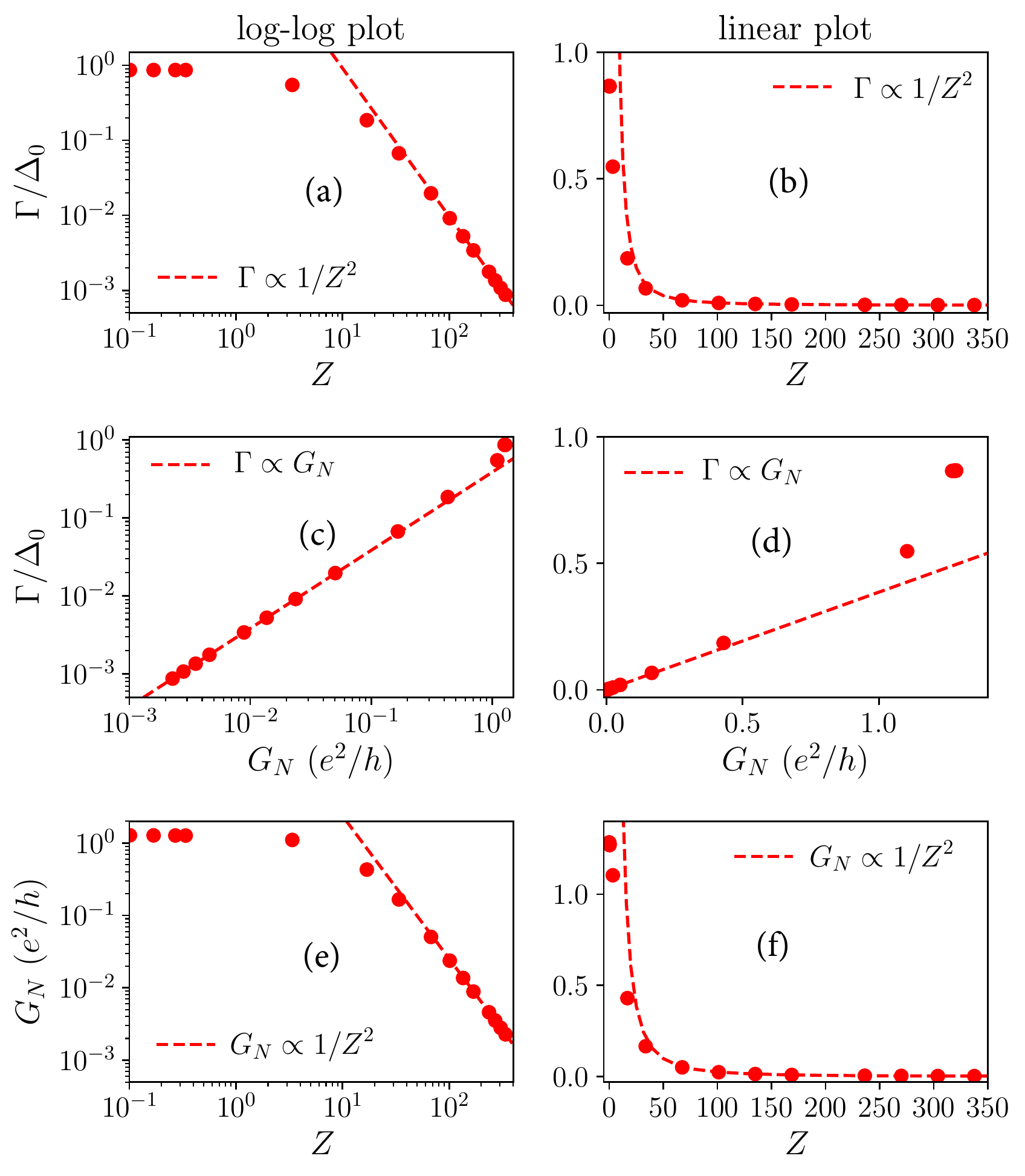}
\end{center}
\caption{(color online) Log-log plot (left) and linear plot (right) of  $\Gamma$ versus $Z$ (top),  $\Gamma$ versus $G_N$ (middle) and $G_N$ vs $Z$ (bottom). In the limit where $Z \gg 1$, we have $\Gamma \propto 1/Z^2$, $\Gamma \propto G_N$ and $G_N \propto 1/Z^2$ as shown by the dashed lines in the top, middle and bottom panels, respectively. The parameters used are $\mu = 5$, $\Delta_0 = 1$, $T = 0$, and $V_Z = 8$ (above TQPT). The plots are for $V_Z$-independent bulk gap case.}\label{fig:Gamma_GN_Z_all} 
\end{figure}

Temperature broadens the zero-bias peak (in addition to its intrinsic broadening due to tunnel coupling $Z$, which is present even at $T=0$)  and lowers its conductance value, keeping the area under the peak constant.  The dependence of the conductance on temperature is shown more explicitly in Fig.~\ref{fig:dIdV_nocollapse_diff_temp}. Temperature smoothes the conductance profile where at high enough temperature (temperature greater than the induced gap), the conductance profile becomes essentially flat. As seen in the figure, the ZBP has a stronger temperature dependence for higher barrier strength. This provides an immediate explanation for why it is difficult to see the $2e^2/h$ quantization in the weak-tunneling limit---finite temperature has a drastic suppression effect on the ZBP strength in the weak-tunneling limit as is obvious in Figs.~\ref{fig:dIdV_nocollapse_diff_temp}(c) and~\ref{fig:dIdV_nocollapse_diff_temp}(d).  At the same time, however, the ZBP is much sharper in the weak-tunneling limit with the peak to valley ratio being much larger.  By contrast, the ZBP remains close to $2e^2/h$ value in the strong-tunneling limit even as the temperature is raised, but the peak to valley ratio is small in the strong-tunneling limit at finite temperatures since the background conductance also increases when the tunneling is strong.  These findings are all consistent with recent observations~\cite{nichele2017scaling}. These comments also imply some qualitative connection between $T$- and $Z$-dependencies of the ZBP.

We quantify the ZBP width or broadening using the full width at half maximum (FWHM) which is defined as the distance between points on the conductance against bias voltage plot at which the conductance reaches the value $(\max(G) + \min(G))/2$ as shown in Fig.~\ref{fig:dIdV_diff_bh_diff_temp}. We note that this definition for the ZBP broadening is unique and does not depend on any assumed form for the ZBP (e.g., Gaussian or Lorentzian). We denote the zero-temperature ZBP width by $2\Gamma$. Only in the weak-tunneling or high-barrier-strength limit the ZBP can be fitted to a Lorentzian curve $G(V) = 2\Gamma^2/(V^2 + \Gamma^2)$ [see Figs.~\ref{fig:dIdV_diff_bh_diff_temp}(c) and~\ref{fig:dIdV_diff_bh_diff_temp}(d)]. In general, for arbitrary values of $Z$, such a simple form for the ZBP does not apply~\cite{Sengupta2001Midgap}. 

\begin{figure}[h!]
\begin{center}
\includegraphics[width=\linewidth]{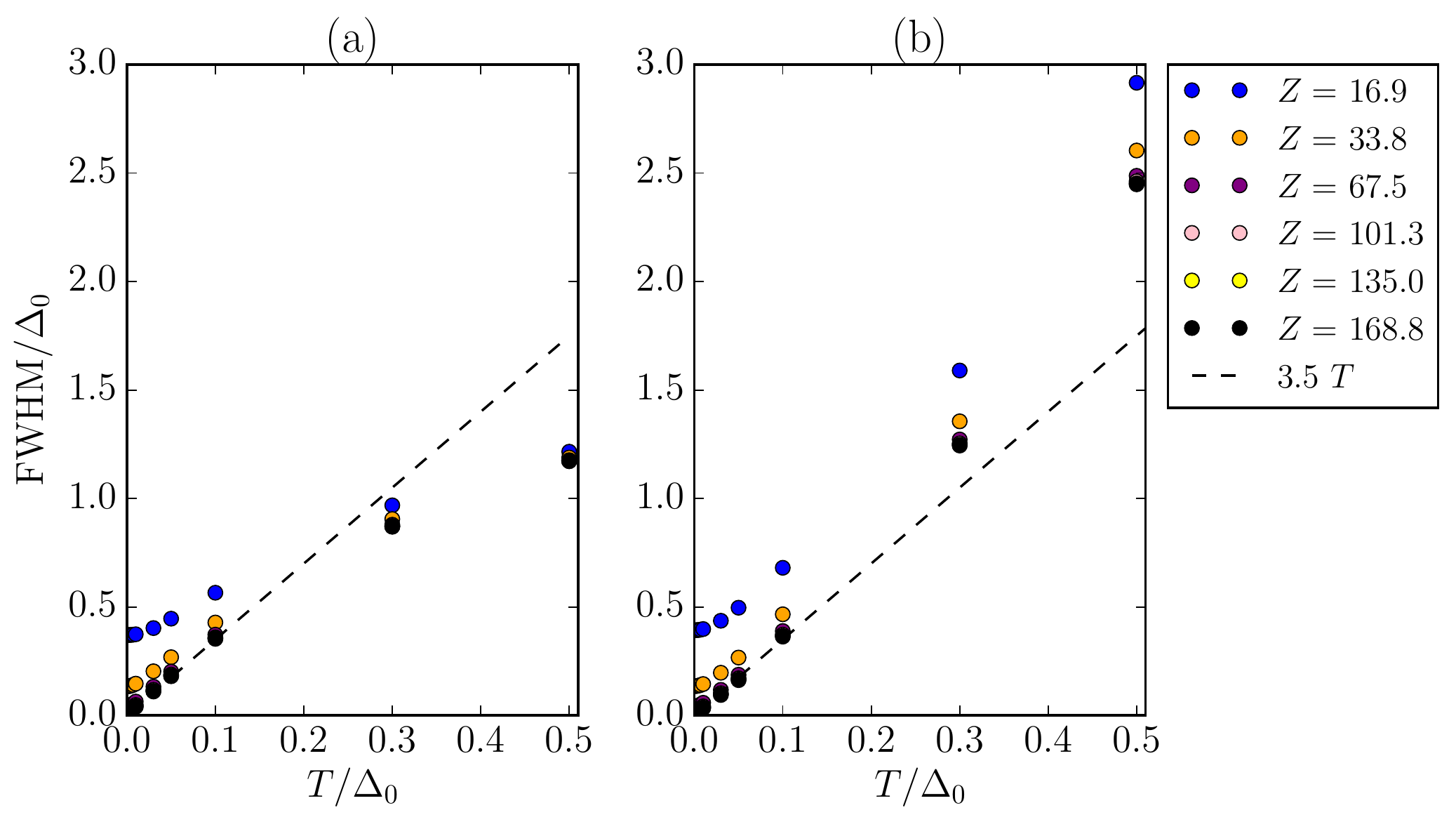}
\end{center}
\caption{(color online) FWHM/$\Delta_0$ vs $T$  for different barrier strengths $Z$ for the case where (a) the FWHM is taken to be the distance between the points where the conductance is $(\max(G)+\min(G))/2$, (b) the FWHM is taken to be the width of the Lorentzian fit. For $\Gamma \ll T \ll \widetilde{\Delta}$, FWHM $= 3.5 T$, which is the width of a Lorentzian resonant peak broadened by temperature (dashed line). The parameters used are $\mu = 5$, $\Delta_0 = 1$, and $V_Z = 8$ (above TQPT), which correspond to the quasiparticle gap $\widetilde{\Delta}  = 0.87$. The plots are for $V_Z$-independent bulk gap case.}\label{fig:FWHM_T_all} 
\end{figure}

The actual dependence of the intrinsic ZBP width on the barrier strength is shown in Figs.~\ref{fig:Gamma_GN_Z_all}(a) and~\ref{fig:Gamma_GN_Z_all}(b). The ZBP width decreases with increasing barrier strength (decreasing junction transparency) as shown in Figs.~\ref{fig:Gamma_GN_Z_all}(a) and~\ref{fig:Gamma_GN_Z_all}(b) [Figs.~\ref{fig:Gamma_GN_Z_all}(c) and~\ref{fig:Gamma_GN_Z_all}(d)]. We note that $\Gamma$ decreases fast with increasing $Z$ [Figs.~\ref{fig:Gamma_GN_Z_all}(a) and~\ref{fig:Gamma_GN_Z_all}(b)] for large $Z$ ($Z>10)$. In the weak-tunneling or high-barrier-strength limit (characterized by a large $Z$, i.e., $Z \gg 1$), $\Gamma \propto G_N$, where $G_N$ is the effective normal-state above-gap conductance, as shown by the dashed lines in Figs.~\ref{fig:Gamma_GN_Z_all}(c) and~\ref{fig:Gamma_GN_Z_all}(d). Here, we characterize the junction transparency by the normal-state conductance $G_N$ (in unit of $e^2/h$), which is the conductance at high voltage. The junction transparency depends not only on the barrier strength $Z$, but is also affected by the parameters mismatch across the junction (e.g., chemical potential, spin-orbit coupling, etc.). Only in the weak-tunneling limit ($G_N \ll e^2/h$), the simple formula $\Gamma \propto G_N$ holds as can be seen in Figs.~\ref{fig:Gamma_GN_Z_all}(c) and~\ref{fig:Gamma_GN_Z_all}(d). As the above-gap conductance can in general be particle-hole asymmetric due to the non-unitarity of the above-gap reflection matrix, we define the junction transparency as 
\begin{equation}
G_N = \frac{G_{N-} + G_{N+}}{2},
\end{equation}
where $G_{N\pm}$ are the conductances at positive and negative voltages above the gap (which are taken to be $V = \pm 4 \Delta_0$ throughout this paper). As seen in Figs.~\ref{fig:Gamma_GN_Z_all}(e) and~\ref{fig:Gamma_GN_Z_all}(f), the junction transparency $G_N$ is inversely related to the barrier strength $Z$.

\begin{figure*}[t]
\begin{center}
\includegraphics[width=0.9\linewidth]{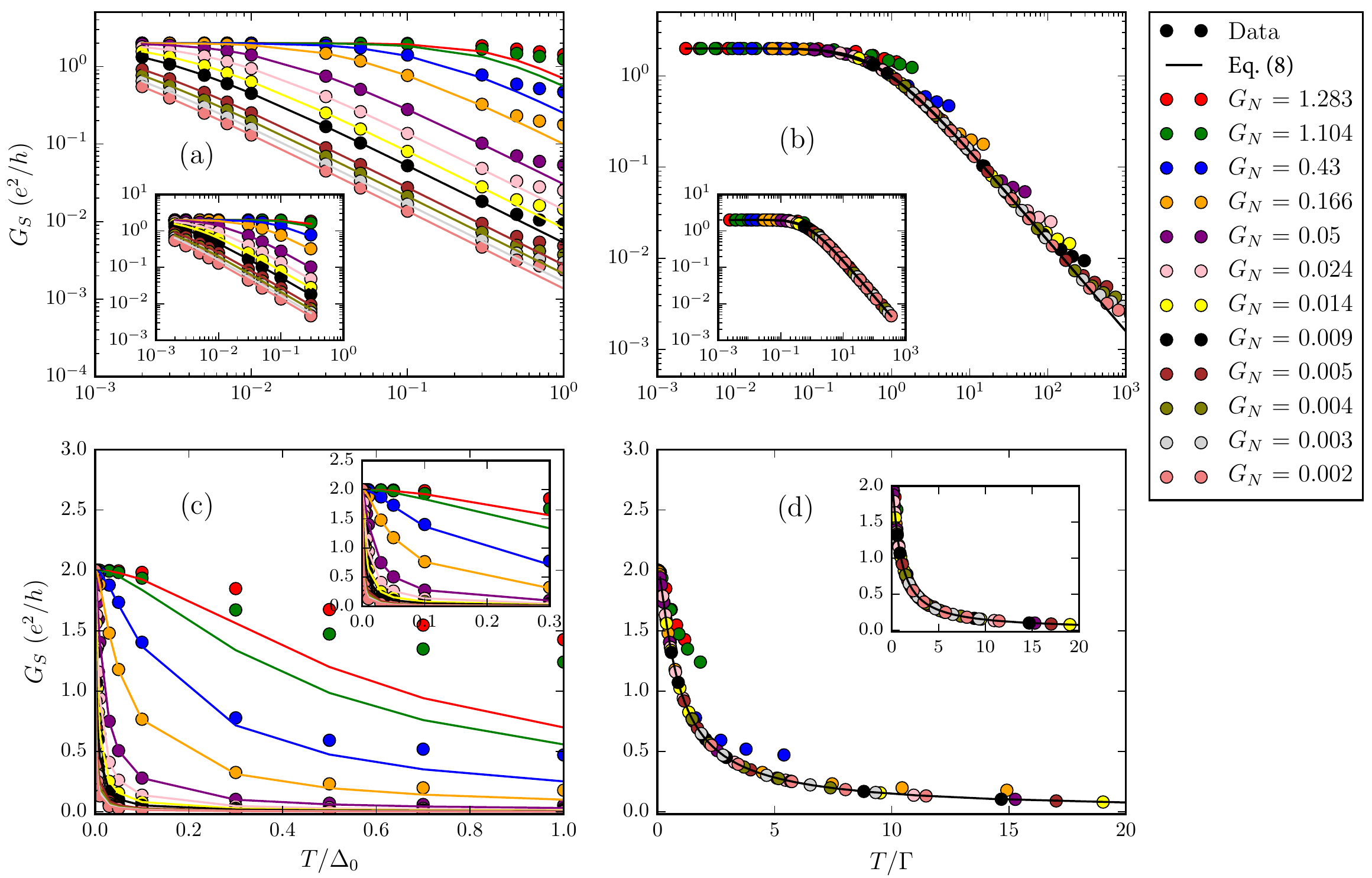}
\end{center}
\caption{(color online) Log-log plot (top) and linear plot (bottom) of  $G_S$ vs $T$ (left) and $G_S$ vs $T/\Gamma$ (right) for different junction transparencies $G_N$. The solid line is Eq.~\eqref{eq:fit}. Note that only the zero-bias conductance values in the weak-tunneling limit ($\Gamma \ll \widetilde{\Delta}$) and for small temperatures ($T \ll \widetilde{\Delta}$) can be fitted into Eq.~\eqref{eq:fit}. (Inset) Plots with high temperature ($T > 0.3 \Delta_0$) data points removed. The parameters used are $\mu = 5$, $\Delta_0 = 1$, and $V_Z$ is $8$ (above TQPT), which correspond to the quasiparticle gap $\widetilde{\Delta}  = 0.87$. The plots are for $V_Z$-independent bulk gap case.}\label{fig:GS_diff_T_all_fit} 
\end{figure*}

It turns out that for small $\Gamma$, i.e., in the weak tunneling $Z \gg 1$ limit when the ZBP itself can be approximated as a Lorentzian [see Eq.~\eqref{eq:fit}], the broadening can be analytically obtained~\cite{Sengupta2001Midgap} to be
\begin{equation}\label{eq:gamma_tau}
\Gamma= \frac{\widetilde{\Delta}\tau}{\sqrt{1-\tau}},
\end{equation}
where $\widetilde{\Delta}$ is the quasiparticle gap and the junction transmission coefficient $\tau$ is given in terms of the barrier strength $Z$ by $\tau= 1/(1+Z^2)$, implying $\tau$ lies between zero (for $Z=\infty$) and one (for $Z=0$). Note that Eq.~\eqref{eq:gamma_tau} holds only in the weak-tunneling $\tau\ll 1$ (and $Z \gg 1$) limit.  In the same weak-tunneling limit, of course, $G_N \propto \tau$ also.  We can rewrite the above analytical formula expressing $\Gamma$ in terms of $Z$ as
\begin{equation}
\Gamma= \frac{\widetilde{\Delta}} {Z\sqrt{1+Z^2}}. 
\end{equation}
The above equation implies that $\Gamma$ has the following limiting scaling forms;  $\Gamma \propto 1/Z^2$ for $Z\gg 1$ and $\Gamma \propto 1/Z$ for $Z \ll 1$.  In Fig.~\ref{fig:Gamma_GN_Z_all}, we show by dashed lines the analytical expectations based on these equations, and indeed, $\Gamma \propto 1/Z^2$ (and $G_N \propto 1/Z^2$) for the weak-tunneling $Z\gg 1$ limit. However, the numerically calculated broadening $\Gamma$ does not obey any power-law scaling for the strong-tunneling $Z<1$ regime.  We note that we do not expect these analytical scaling relations, derived on the basis of weak-tunneling approximation, to apply in the strong-tunneling ($Z<1$) regime, a fact that seems to have been overlooked in Ref.~\cite{nichele2017scaling}. In particular, the experimentally observed $G_S \approx  2e^2/h$ always happens in the $Z<1$ strong-tunneling regime where $\Gamma$ does not have a simple dependence on the junction transparency.  Thus, for the most experimentally interesting situations, the simple analytical formula for $\Gamma$ in terms of $Z$ (or $\tau$) is not of any use.

Figure~\ref{fig:FWHM_T_all} shows the plots of FWHM versus temperature for two different FWHMs obtained from different ways of extracting the FWHMs of the ZBP. Figure~\ref{fig:FWHM_T_all}(a) shows the FWHM obtained by taking the distance between points on the conductance plot where the conductance value is $(\max(G) + \min(G))/2$, while Fig.~\ref{fig:FWHM_T_all}(b)  shows the FWHM of the Lorentzian curve used to fit the conductance plot. In the weak-tunneling limit, both FWHMs are essentially the same as the conductance plots can be fitted exactly to a Lorentzian function. The FWHM of the ZBP increases with increasing temperature since thermal broadening now adds to the intrinsic broadening $\Gamma$. As shown in Fig.~\ref{fig:FWHM_T_all}, in the limit where $\Gamma \ll T \ll \widetilde{\Delta}$ where $\widetilde{\Delta}$ is the induced quasiparticle gap, the FWHM is $3.5 T$ (which is the width of a Lorentzian resonant peak broadened by temperature only~\cite{kramer2012quantum,oda2005silicon}). However, in the strong-tunneling limit, the FWHM obtained from the Lorentzian fit is always greater than the FWHM obtained from taking the distance between the midpoint of the conductance plots. This is simply due to the fact the conductance profile has a minimum at some finite value in the strong-tunneling limit while the Lorentzian fit always has its minimum at zero which makes the half maximum defined for the Lorentzian fit to be lower than that of the midpoint of the conductance plots and hence larger FWHM. Note that the Lorentzian fitting always gives a larger broadening or FWHM than the direct definition in the strong-tunneling regime whereas for weak tunneling the two definitions give essentially the same broadening.  Physically, the difference arises from the fact that the Lorentzian fit automatically entails an effective background subtraction in the extraction of FWHM whereas the direct definition involves no such subtraction---since the background conductance is substantial (negligible) in the strong (weak) tunneling regime, the two definitions provide different (similar) results in the two cases. Unless otherwise stated, throughout this paper we will take the FWHM to be the distance between the midpoint of the conductance plots, which is the correct and direct definition of FWHM.  We also note that the intrinsic (tunnel) broadening parameter $\Gamma$ entering Eq.~\eqref{eq:fit} below is by definition the half width at half maximum at $T=0$. We point out, as is obvious from Fig.~\ref{fig:GS_diff_T_all_fit}(a), that the thermal broadening lower bound of $3.5T$ only applies to the Lorentzian intrinsic broadening---in general, the finite-temperature broadening could be well below the $3.5T$ limit.

In the limit where $\Gamma, T \ll \widetilde{\Delta}$, where $\widetilde{\Delta}$ is the induced quasiparticle gap in the nanowire at a given $V_Z$,  the ZBP can be approximated by a Lorentzian function with the zero-bias tunneling conductance given by~\cite{Sengupta2001Midgap}
\begin{align}\label{eq:fit}
G_S &\approx \frac{e^2}{h}\int_{-\infty}^{\infty} dE \frac{2 \Gamma^2}{E^2 + \Gamma^2} \frac{1}{4 T \cosh^2(E/(2T))}, \nonumber\\
&= \frac{2e^2}{h} g (T/\Gamma),
\end{align}
where $g$ is a scaling function that depends only on the ratio of temperature $T$ to the tunnel broadening $\Gamma$. We emphasize that by definition this formula [Eq.~\eqref{eq:fit}] is valid only in the weak-tunneling and low-temperature limit as was already mentioned in Ref.~\cite{Sengupta2001Midgap}. 
We refer to our discussion above in the context of Fig.~\ref{fig:Gamma_GN_Z_all} to emphasize that this convolution formula for $G_S$ using a Lorentzian form for the zero-temperature ZBP shape [$2\Gamma^2/(E^2 +\Gamma^2)$], which is the basis for the scaling properties discussed in Ref.~\cite{nichele2017scaling}, is valid only in the weak-tunneling regime of $Z\gg 1$, and as such, is of very limited utility in the interpretation of the experimental Majorana data, where the strong-tunneling regime (with $Z<1$) applies making the simple scaling relation above invalid. It is of course also obvious that to the extent Eq.~\eqref{eq:fit} applies, i.e., the $T=0$ conductance obeys the simple resonant Lorentzian form given by the first term inside the convolution integral in Eq.~\eqref{eq:fit} with no energy dependence in the intrinsic broadening $\Gamma$, a scaling of conductance on the dimensionless variable $\Gamma/T$ is guaranteed. Figures~\ref{fig:GS_diff_T_all_fit}(a) and~\ref{fig:GS_diff_T_all_fit}(c) show the plot of the zero-bias conductance as a function of temperature $T$ for different junction transparencies characterized by $G_N$ on a log-log scale and a linear scale, respectively. From these plots, we can see that only the zero-bias conductance values in the weak-tunneling ($\Gamma \ll \widetilde{\Delta}$) and low-temperature ($T \ll \widetilde{\Delta}$) limit can be fitted well by Eq.~\eqref{eq:fit}. We note that a recent experiment~\cite{nichele2017scaling} extracted the values of $\Gamma$ by fitting the low-temperature ZBP values to the scaling relation [Eq.~\eqref{eq:fit}] instead of using the directly measured ZBP width itself. In principle, the values of $\Gamma$ obtained by fitting into the scaling relation are valid only in the weak-tunneling and low-temperature regime since the scaling relation holds only in this asymptotic limit as established in our theoretical work. Figures~\ref{fig:GS_diff_T_all_fit}(b) and~\ref{fig:GS_diff_T_all_fit}(d) show the zero-bias conductance as a function of a single dimensionless paramater, namely, the ratio of temperature $T$ to the tunnel broadening $\Gamma$ on a log-log scale and a linear scale, respectively. Again, we can see that only the zero-bias conductance values for $\Gamma,T \ll \widetilde{\Delta}$ fit into the scaling function, but the scaling fails at larger values of $\Gamma$ and $T$ since Eq.~\eqref{eq:fit} itself is an approximation for small $\Gamma$ and $T$. The inset in Fig.~\ref{fig:GS_diff_T_all_fit} shows the plots with high-temperature data points removed. We also mention that the conductance varies strongly for small values of $T/\Gamma$, where the log scale is somewhat misleading, and therefore, extrapolation of $G_S$ to $T=0$ is problematic. 
\begin{figure}[h!]
\begin{center}
\includegraphics[width=0.7\linewidth]{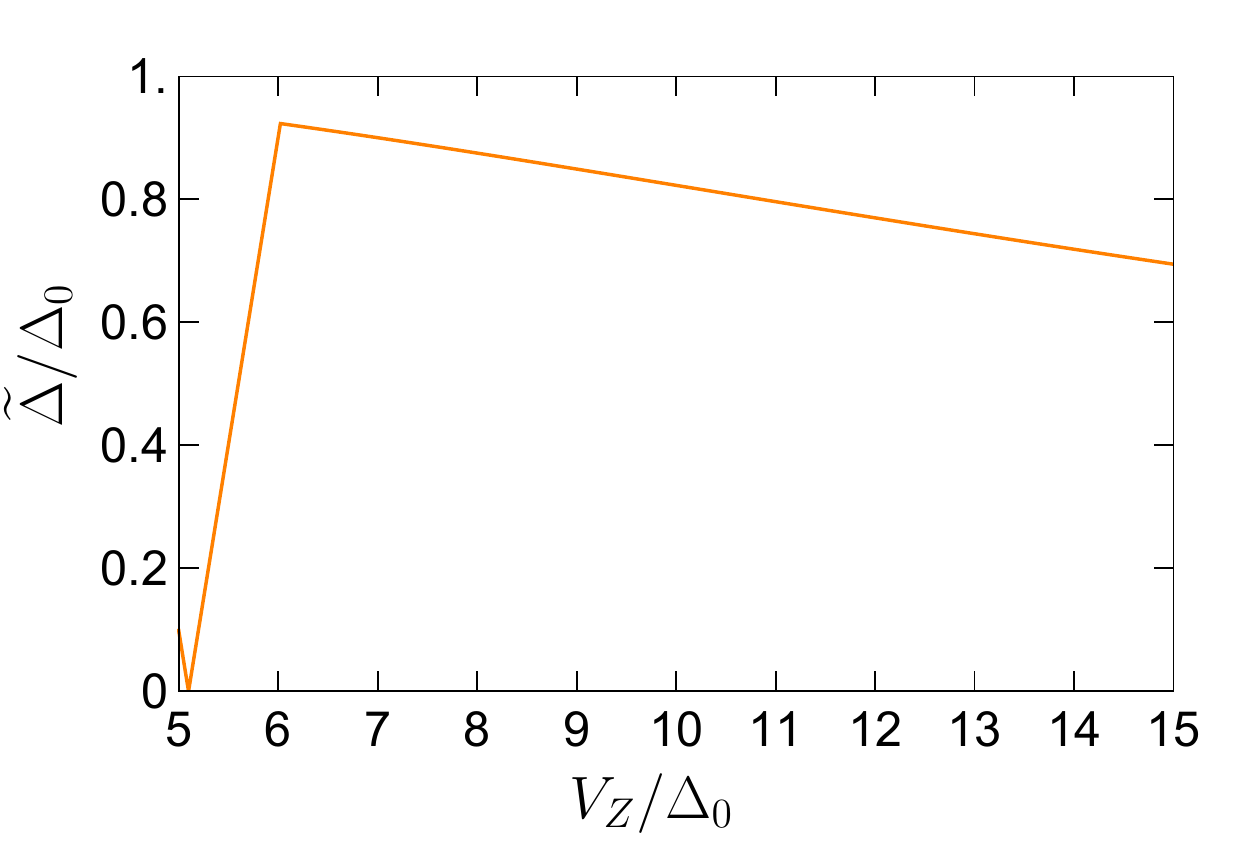}
\end{center} 
\caption{(color online) Quasiparticle gap $\widetilde{\Delta}$ as a function of Zeeman field strength $V_Z$. The parameters used are $\mu = 5$ and $\Delta_0 = 1$ which correspond to the critical Zeeman field $V_{Zc} \equiv \sqrt{\mu^2 + \Delta_0^2}$ = 5.1. The plots are for $V_Z$-independent bulk gap case.}\label{fig:Gap}
\end{figure}

\begin{figure}[h!]
\begin{center}
\includegraphics[width=\linewidth]{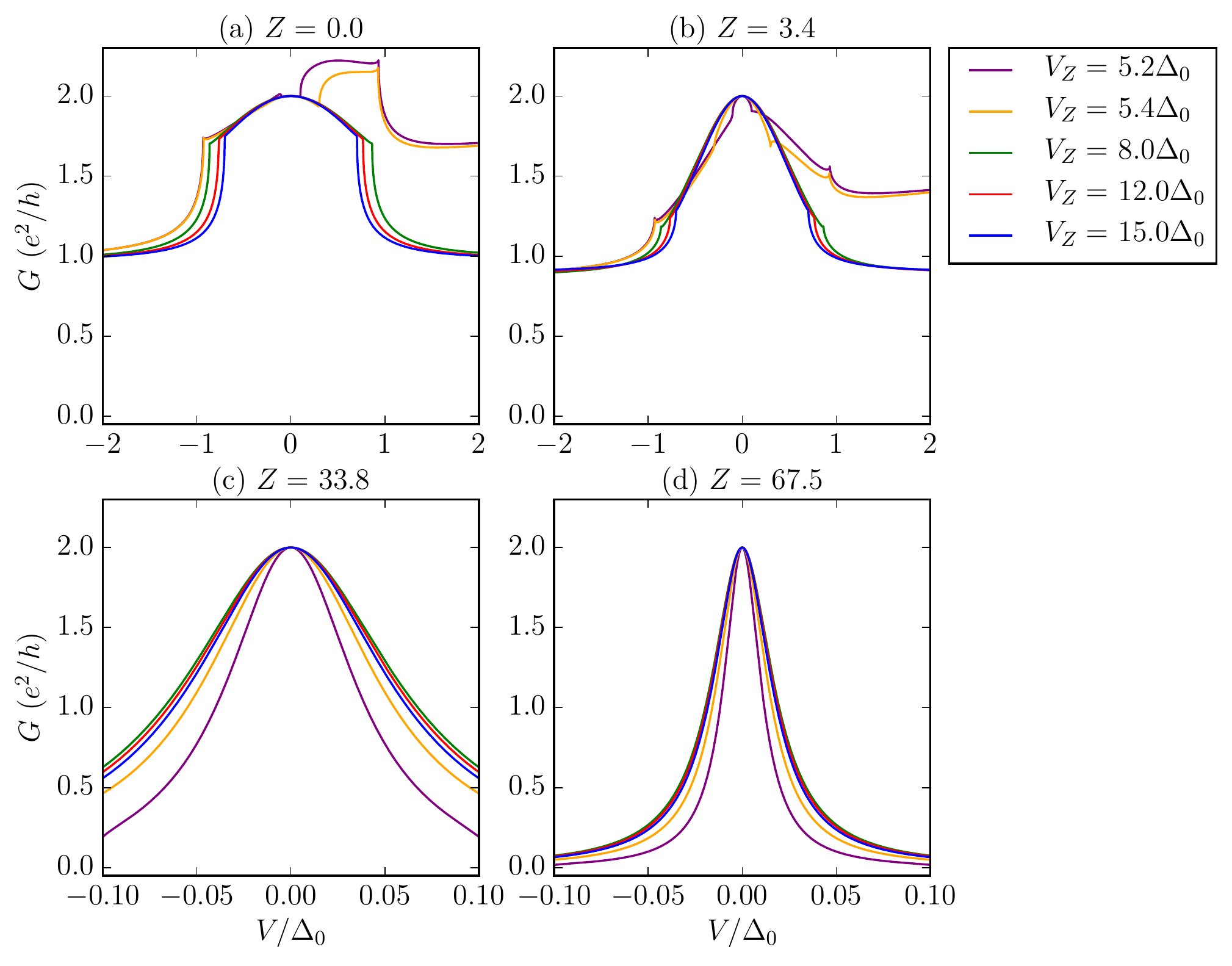}
\end{center}
\caption{(color online) Zero-temperature differential conductance $G$ versus voltage $V$ for several Zeeman fields above the TQPT and for different barrier strengths $Z$: (a) $Z =0$, (b) $Z = 3.4$, (c) $Z = 33.8$, and (d) $Z = 67.5$. The parameters used are $\mu = 5$ and $\Delta_0 = 1$ which correspond to the critical Zeeman field $V_{Zc} \equiv \sqrt{\mu^2 + \Delta_0^2}$ = 5.1. The plots are for $V_Z$-independent bulk gap case.}\label{fig:dIdV_diff_bh_diff_VZ_zeroT} 
\end{figure}

Above the critical Zeeman field at which the TQPT happens ($V_{Zc} = \sqrt{\mu^2 + \Delta^2}$), the quasiparticle gap $\widetilde{\Delta}$ first increases sharply with increasing Zeeman field strength near the TQPT and then decreases slowly with increasing Zeeman field~\cite{Sau2010Non} as shown in Fig.~\ref{fig:Gap}. The quasiparticle gap is given by
\begin{widetext}
\begin{equation}
\widetilde{\Delta} = \min \left(V_Z - \sqrt{\mu^2 + \Delta^2}, \sqrt{2 (V_Z^2 + \alpha^2) + \Delta^2 - 2\sqrt{(V_Z^2 + \alpha^2)^2 + V_Z^2 \Delta^2}}\right),
\end{equation}
\end{widetext}
where near the TQPT, $\widetilde{\Delta} = V_Z -\sqrt{\mu^2+\Delta^2}$ is given by the energy spectrum at $k =0$ which increases with increasing $V_Z$. The subsequent decrease of the quasiparticle gap with a further increase in the Zeeman field is due to the gap at the Fermi momentum which is given by $\widetilde{\Delta} = \sqrt{2 (V_Z^2 + \alpha^2) + \Delta^2 - 2\sqrt{(V_Z^2 + \alpha^2)^2 + V_Z^2 \Delta^2}}$. 
This Zeeman-field dependence of the gap is reflected in the change of the ZBP width in the weak-tunneling limit where it first increases with increasing Zeeman field and then decreases with a further increase in the Zeeman field as shown in Fig.~\ref{fig:dIdV_diff_bh_diff_VZ_zeroT}. We note that in the strong-tunneling limit, the zero-bias conductance may not appear as a peak near the TQPT as shown by Fig.~\ref{fig:dIdV_diff_bh_diff_VZ_zeroT}(a).

\begin{figure}[h!]
\begin{center}
\includegraphics[width=\linewidth]{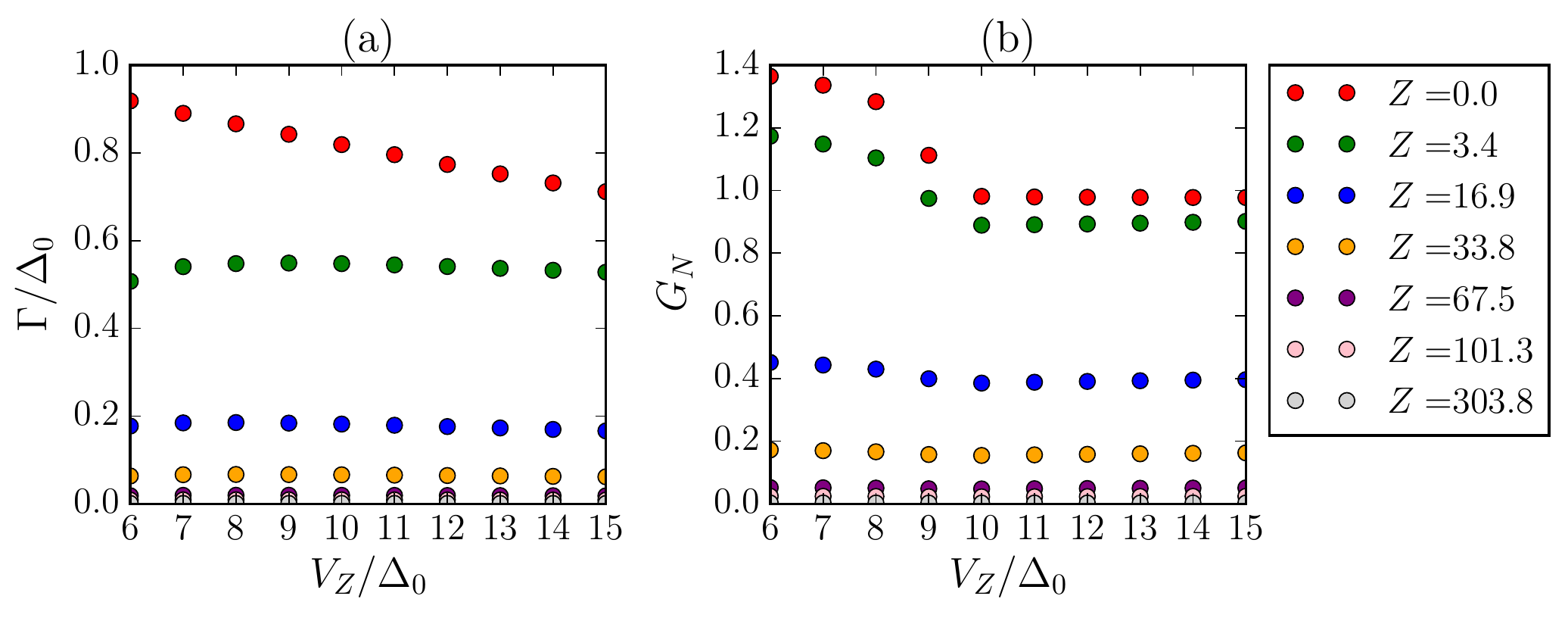}
\end{center}
\caption{(color online) (a) $\Gamma$ versus $V_Z$ and (b) $G_N$ versus $V_Z$ for different barrier strengths $Z$. The parameters used are $\mu = 5$, $\Delta_0 = 1$, and $T =0$. The plots are for $V_Z$-independent bulk gap case.}\label{fig:Gamma_GN_B_fullrange_fit} 
\end{figure}

\begin{figure}[h!]
\begin{center}
\includegraphics[width=\linewidth]{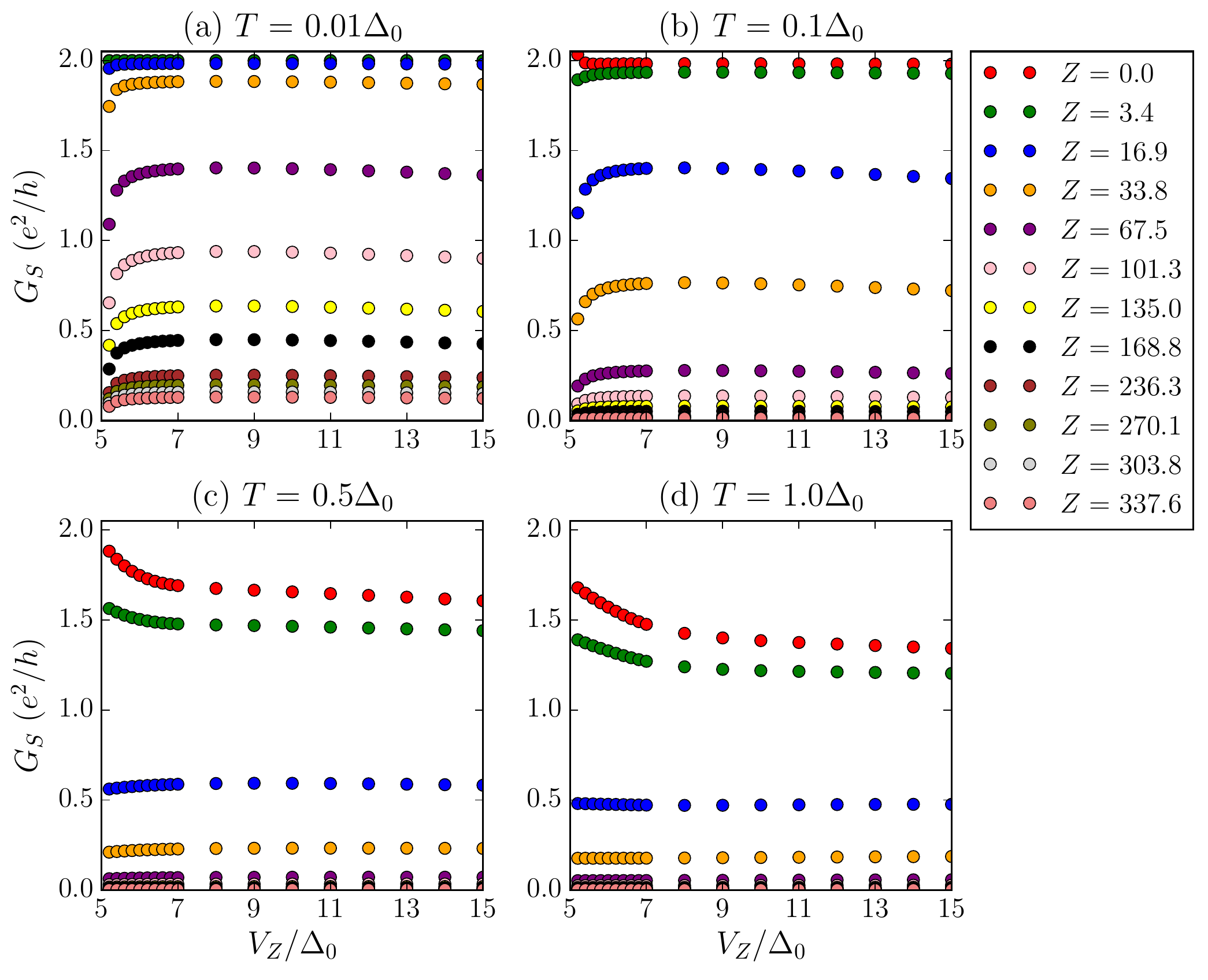}
\end{center}
\caption{(color online) Zero-bias differential conductance $G_S$ versus Zeeman field $V_Z$ above the TQPT for different barrier strengths $Z$ at several fixed temperatures $T$: (a) $T = 0.01 \Delta_0$, (b) $T = 0.1 \Delta_0$, (c) $T = 0.5 \Delta_0$, and (d) $T = \Delta_0$. The parameters used are $\mu = 5$ and $\Delta_0 = 1$ which correspond to the TQPT critical field $V_{Zc} \equiv \sqrt{\mu^2 + \Delta_0^2} = 5.1$. The plots are for $V_Z$-independent bulk gap case.}\label{fig:GS_B} 
\end{figure}

\begin{figure}[h!]
\begin{center}
\includegraphics[width=\linewidth]{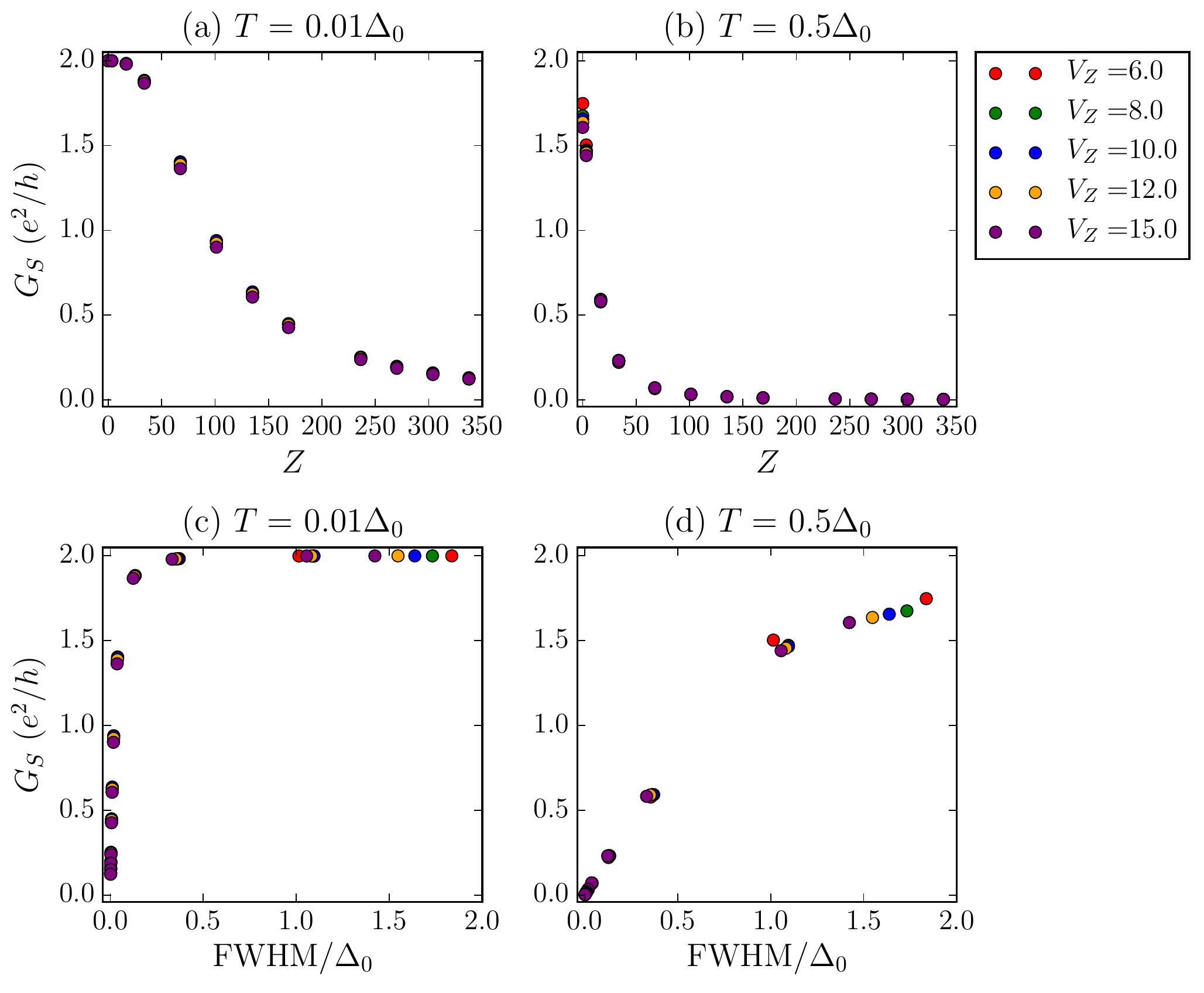}
\end{center}
\caption{(color online) Zero-bias differential conductance $G_S$ vs barrier strength $Z$ (top) and $G_S$ versus FWHM (bottom) for several values of $V_Z$ above the TQPT at fixed temperature: $T = 0.01\Delta_0$ (left) and $T = 0.5\Delta_0$ (right). The parameters used are $\mu = 5$ and $\Delta_0 = 1$, which correspond to the TQPT critical field $V_{Zc} \equiv \sqrt{\mu^2 + \Delta_0^2} = 5.1$. The plots are for $V_Z$-independent bulk gap case.}\label{fig:GS_Z_Gamma} 
\end{figure}

In the weak-tunneling limit ($\Gamma \ll \widetilde{\Delta}$), $\Gamma$ is related to the imaginary part of the normal-lead self-energy~\cite{datta1997electronic}, which is given by
\begin{equation}
\Gamma \sim \tau |\phi(0)|^2,
\end{equation}
where $\tau$ is the junction transmission coefficient and $\phi(0)$ is the nanowire wave function at the NS interface. Since the isolated MZM wave function is approximately given by
\begin{equation}
\phi(x) \approx \frac{1}{\sqrt{\xi}}e^{-x/\xi},
\end{equation}
where $\xi = v_F/\widetilde{\Delta}$ is the superconducting coherence length, we get~\cite{DasSarma2016How}
\begin{equation}\label{eq:GammaDelta}
\Gamma \propto \tau \widetilde{\Delta}, 
\end{equation}
which implies that in the weak-tunneling limit (i.e., $\tau\ll 1$) the tunnel broadening $\Gamma$ depends on the Zeeman field in the same way that $\widetilde{\Delta}$ depends on $V_Z$ (for $\Gamma\ll \widetilde{\Delta}$). We note that Eq.~\eqref{eq:GammaDelta} is essentially the same as Eq.~\eqref{eq:gamma_tau} in the limit where $\tau \ll 1$. As shown in Fig.~\ref{fig:Gamma_GN_B_fullrange_fit}(a), $\Gamma$ decreases with increasing $V_Z$. For the case of Zeeman-independent bulk superconducting gap, beyond the TQPT the quasiparticle gap decreases slowly with increasing Zeeman field. As a result, the dependence of $\Gamma$ on $V_Z$ is barely noticeable in Fig.~\ref{fig:Gamma_GN_B_fullrange_fit}. This dependence will become more apparent in the following subsection when we consider the case of Zeeman-dependent bulk superconducting gap. Figure~\ref{fig:Gamma_GN_B_fullrange_fit}(b) shows the plot of $G_N$ versus $V_Z$. We can see from the plot that $G_N$ only depends on $V_Z$ in the strong-tunneling limit where it decreases with increasing $V_Z$ near the TQPT and becomes independent of $V_Z$ for sufficiently large $V_Z$. 

Now let us look at the dependence of the finite-temperature zero-bias conductance on the Zeeman field (as shown in Fig.~\ref{fig:GS_B}). This dependence arises from the dependence of conductance on the quasiparticle gap $\widetilde{\Delta}$. We note three different limiting cases:
\begin{align}
G_S 
\begin{cases}
 \approx 2 e^2/h, & \text{for\hspace{0.3cm}} T \ll \Gamma \ll \widetilde{\Delta}, \\
\propto \Gamma \propto \tau\widetilde{\Delta}, & \text{for\hspace{0.3cm}} \Gamma \ll T \ll \widetilde{\Delta}, \\
 \sim G_N, &\text{for\hspace{0.3cm}} \widetilde{\Delta} \ll T.
\end{cases}
\end{align}
In Figs.~\ref{fig:GS_B}(a) and~\ref{fig:GS_B}(b), we can see that for small $Z$, $G_S \approx 2 e^2/h$ which corresponds to the regime where $T \ll \Gamma \ll \widetilde{\Delta}$, and for large $Z$, which corresponds to the regime $\Gamma \ll T \ll \widetilde{\Delta}$, $G_S \propto \widetilde{\Delta}$ where it increases with increasing $V_Z$ near the TQPT, and beyond a certain value of $V_Z$, it decreases with increasing $V_Z$. As depicted in Figs.~\ref{fig:GS_B}(c) and~\ref{fig:GS_B}(d), for sufficiently large temperature the zero-bias conductance in the strong-tunneling limit first decreases with increasing $V_Z$ near the TQPT and approaches $G_N$ in the large $V_Z$ limit. In the weak-tunneling limit, the zero-bias conductance $G_S$ at large temperature is independent of $V_Z$. 

We show the plot of $G_S$ versus barrier strength $Z$ and $G_S$ versus FWHM of the ZBP in Fig.~\ref{fig:GS_Z_Gamma} for several Zeeman fields $V_Z$ above the TQPT. The finite-temperature zero-bias conductance decreases with increasing barrier strength $Z$ or decreasing ZBP FWHM.  Since for the Zeeman-independent bulk gap case, the quasiparticle gap decreases slowly with increasing $V_Z$ above the TQPT, the zero-bias conductance does not vary greatly with $V_Z$ (at least in this long-wire limit).

\subsection{Zeeman-dependent bulk gap}

\begin{figure}[h]
\begin{center}
\includegraphics[width=\linewidth]{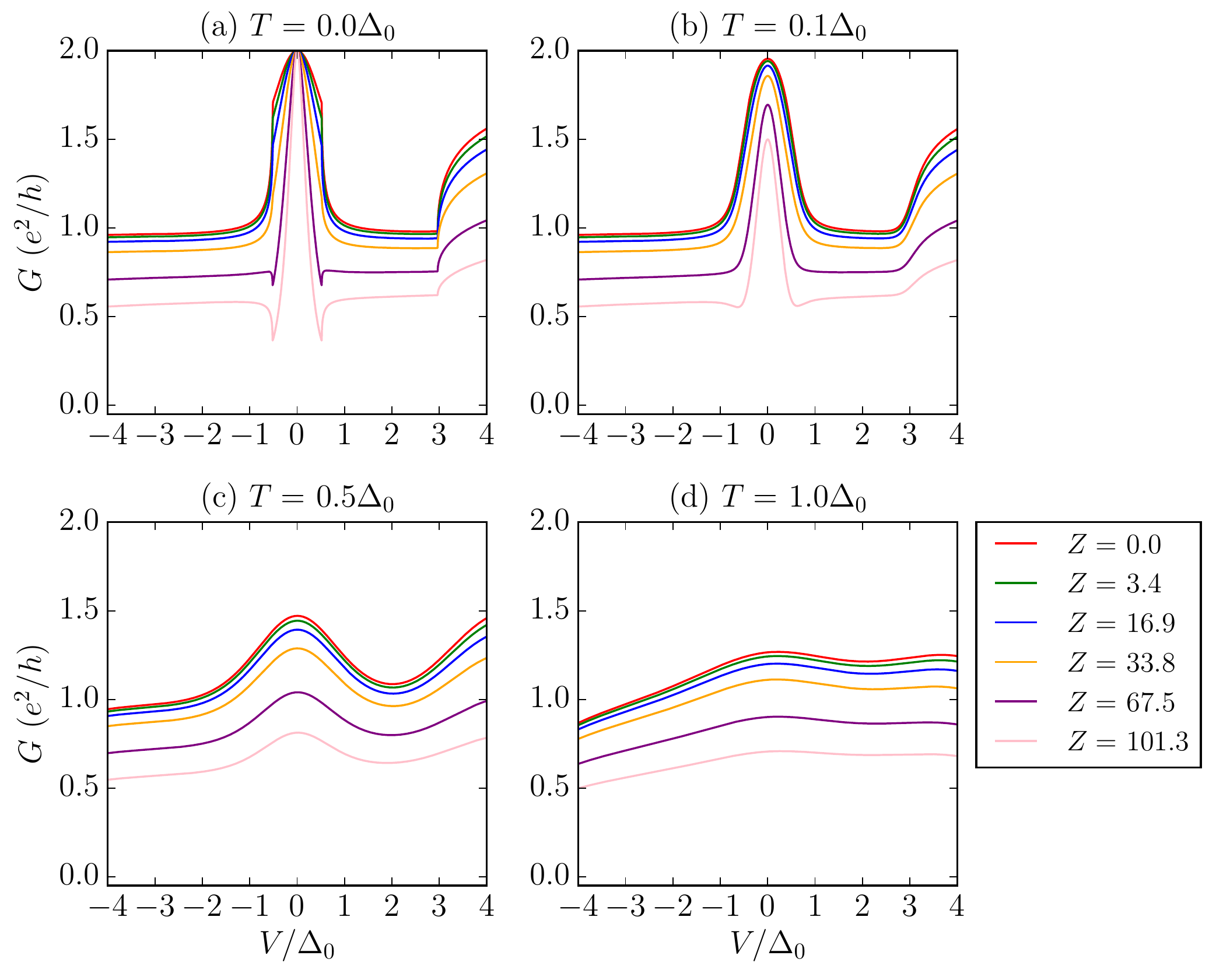}
\end{center}
\caption{(color online) Differential conductance $G$ vs voltage $V$ for different barrier strengths $Z$ at several fixed temperatures $T$: (a) $T = 0$, (b) $T = 0.1\Delta_0$, (c) $T = 0.5 \Delta_0$, and (d) $T = \Delta_0$. The parameters used are $\mu = 5$, $\Delta_0 = 1$, and $V_Z = 8$ (above TQPT). The plots are for $V_Z$-dependent bulk gap case where the bulk gap collapses at $V_Z^* = 10$.}\label{fig:dIdV_diff_Z_collapse} 
\end{figure}

\begin{figure}[h!]
\begin{center}
\includegraphics[width=\linewidth]{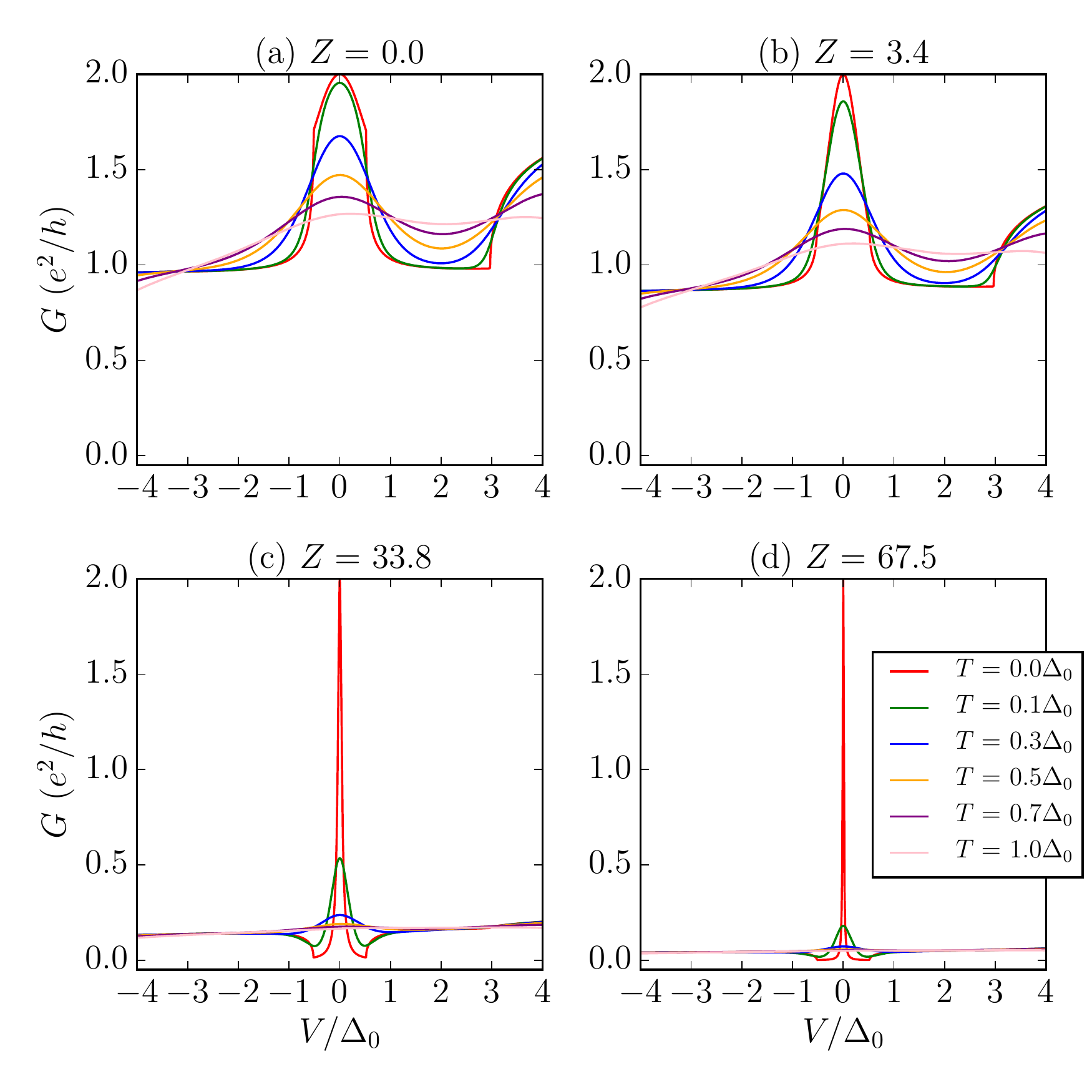}
\end{center}
\caption{(color online) Differential conductance $G$ vs voltage $V$ for different temperatures $T$ at several fixed barrier strengths $Z$: (a) $Z =0$, (b) $Z = 3.4$, (c) $Z = 33.8$, and (d) $Z = 67.5$. The parameters used are $\mu = 5$, $\Delta_0 = 1$, and $V_Z = 8$ (above TQPT). The plots are for $V_Z$-dependent bulk gap case where the bulk gap collapses at $V_Z^* = 10$.}\label{fig:dIdV_collapse_diff_temp} 
\end{figure}

\begin{figure}[h!]
\begin{center}
\includegraphics[width=0.95\linewidth]{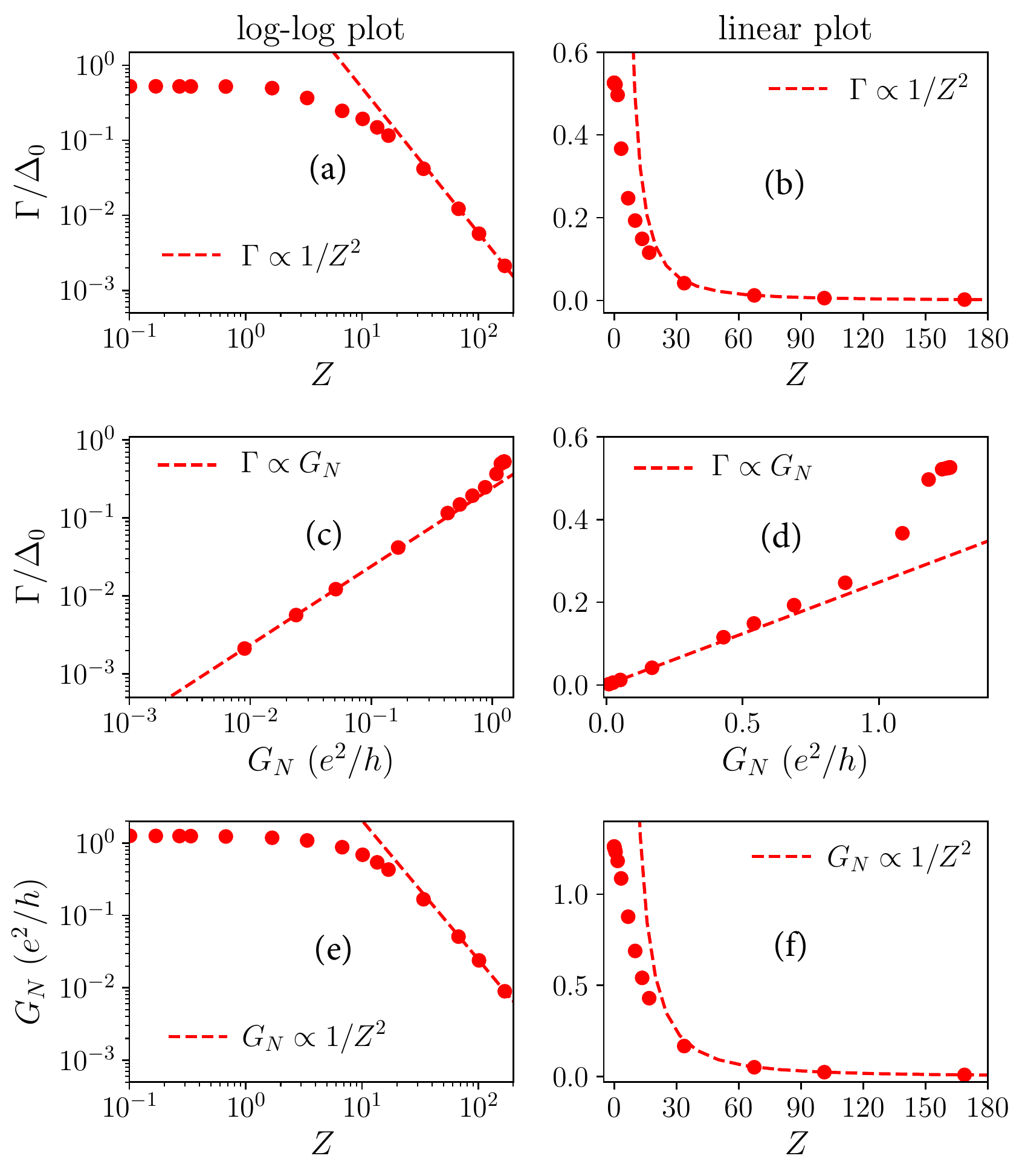}
\end{center}
\caption{(color online) Log-log plot (left) and linear plot (right) of  $\Gamma$ versus $Z$ (top),  $\Gamma$ versus $G_N$ (middle) and $G_N$ versus $Z$ (bottom). In the limit where $Z \gg 1$, we have $\Gamma \propto 1/Z^2$, $\Gamma \propto G_N$ and $G_N \propto 1/Z^2$ as shown by the dashed lines in the top, middle, and bottom panels, respectively. The parameters used are $\mu = 5$, $\Delta_0 = 1$, $T = 0$, and $V_Z$ is $8$ (above TQPT). The plots are for $V_Z$-dependent bulk gap case where the bulk gap collapses at $V_Z^* = 10$.}\label{fig:Gamma_GN_Z_collapse_all} 
\end{figure}

\begin{figure}[h!]
\begin{center}
\includegraphics[width=0.95\linewidth]{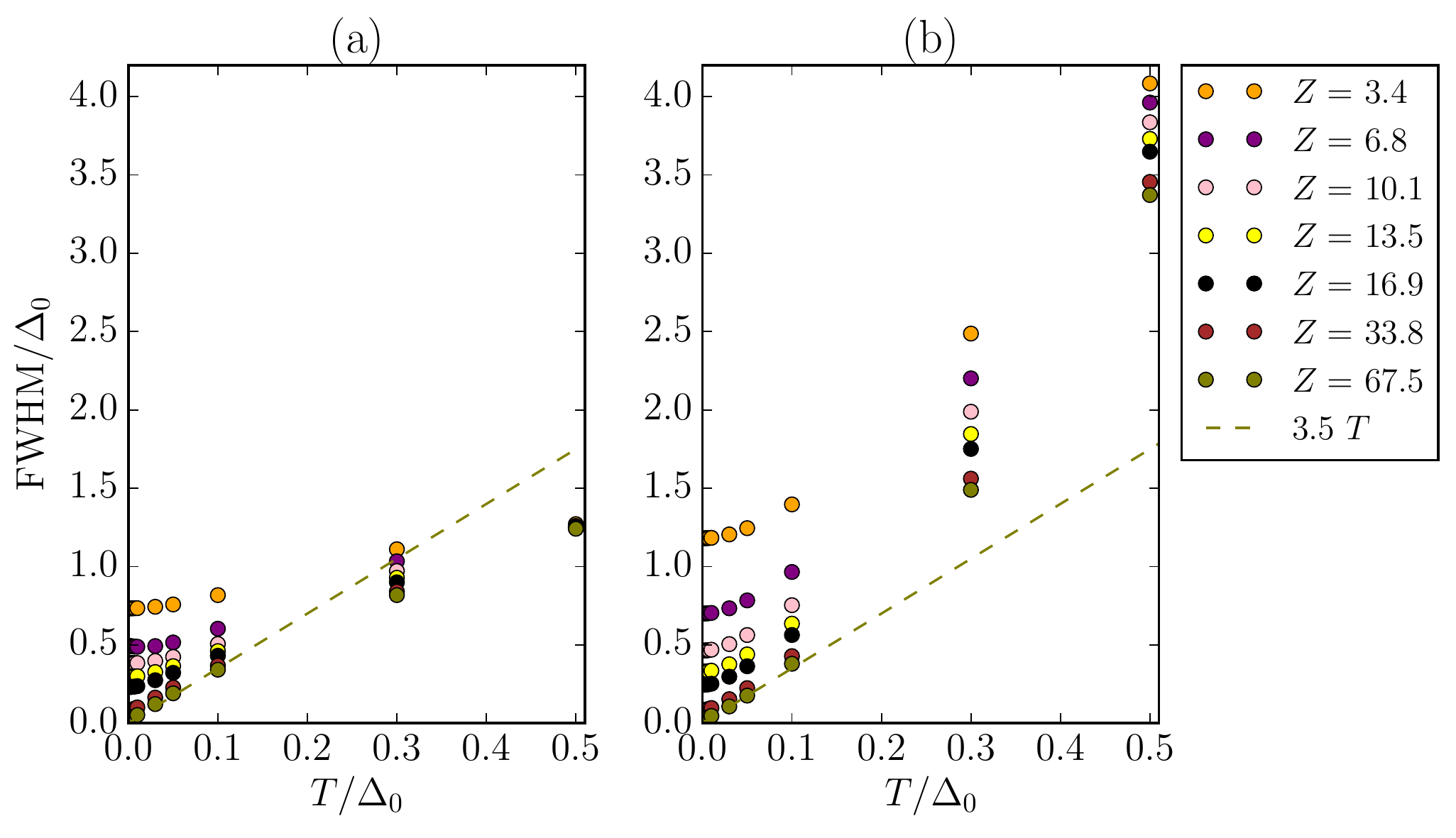}
\end{center}
\caption{(color online) FWHM/$\Delta_0$ vs $T$  for different barrier strengths $Z$ for the case where the FWHM is (a) the distance between the points where the conductance is $(\max(G)+\min(G))/2$, (b) the width of the Lorentzian fit. For $\Gamma \ll T \ll \widetilde{\Delta}$, FWHM $= 3.5 T$, which is the width of a Lorentzian resonant peak broadened by temperature (dashed line). The parameters used are $\mu = 5$, $\Delta_0 = 1$, and $V_Z = 8$ (above the TQPT). The plots are for $V_Z$-dependent bulk gap case where the bulk gap collapses at $V_Z^* = 10$.}\label{fig:FWHM_collapse_T_all} 
\end{figure}

\begin{figure*}[t]
\begin{center}
\includegraphics[width=0.9\linewidth]{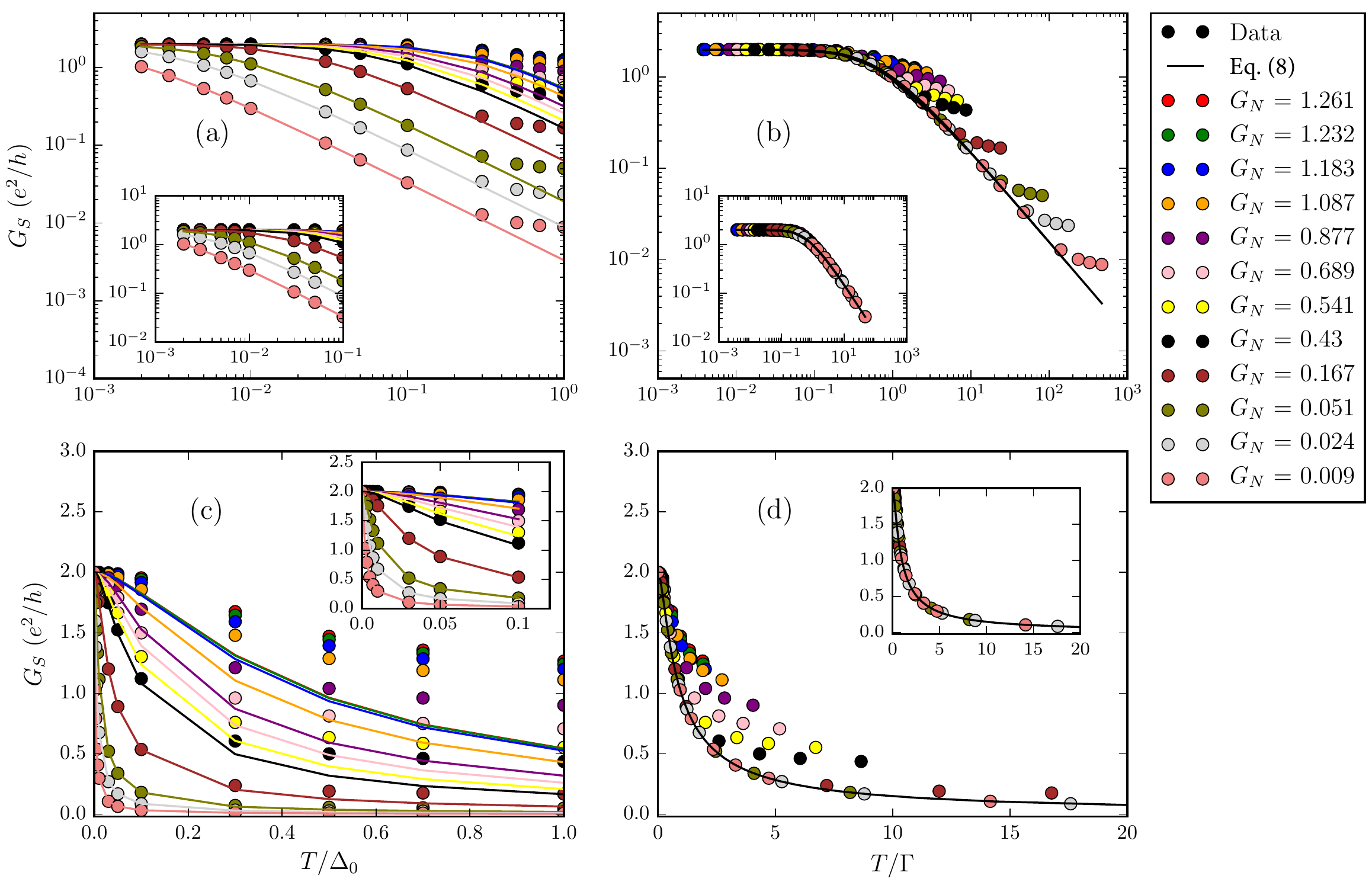}
\end{center}
\caption{(color online) Log-log plot (top) and linear plot (bottom) of  $G_S$ vs $T$ (left) and $G_S$ versus $T/\Gamma$ (right) for different junction transparencies $G_N$. Solid line is Eq.~\eqref{eq:fit}. Note that only the zero-bias conductance values in the weak-tunneling limit ($\Gamma \ll \widetilde{\Delta}$) and for small temperature ($T \ll \widetilde{\Delta}$) can be fitted into Eq.~\eqref{eq:fit}. (Inset) Plots with high temperature ($T > 0.1 \Delta_0$) data points removed. The parameters used are $\mu = 5$, $\Delta_0 = 1$, and $V_Z$ is $8$ (above TQPT), which correspond to the quasiparticle gap $\widetilde{\Delta}  = 0.52$. The plots are for $V_Z$-dependent bulk gap case where the bulk gap collapses at $V_Z^* = 10$.}\label{fig:GS_collapse_diff_T_all_fit} 
\end{figure*}

In this subsection, we consider a more experimentally realistic scenario~\cite{Deng2016Majorana,nichele2017scaling} where the bulk gap shrinks with increasing Zeeman field, i.e., 
\begin{equation}\label{eq:gapcollapse}
\Delta(V_Z) = \Delta_0 \sqrt{1- (V_Z/V_{Z}^*)^2},
\end{equation}
where $V_{Z}^*$ is the critical field at which the bulk superconducting gap collapses. This appears to be a common experimental scenario~\cite{Deng2016Majorana,nichele2017scaling}. The results for the case of Zeeman-dependent bulk gap are qualitatively similar to the case of Zeeman-independent bulk gap considered in the previous subsection, but there are important quantitative differences since the effective topological gap is considerably suppressed in the presence of a $V_Z$-dependent bulk gap, and consequently, all the energy scales are suppressed. 

Figure~\ref{fig:dIdV_diff_Z_collapse} shows the plot of differential conductance of the nanowire in the topological regime [$V_Z > \sqrt{\mu^2 + \Delta(V_Z)^2}$] for different barrier strengths at several values of temperatures. It can be seen that the conductance decreases with increasing barrier strength except the zero-bias conductance at zero temperature which is always fixed at $2e^2/h$ [Fig.~\ref{fig:dIdV_diff_Z_collapse}(a)]. The dependence of the conductance on the temperature for fixed barrier strength is shown in Fig.~\ref{fig:dIdV_collapse_diff_temp}. As for the case of Zeeman-independent bulk gap, the decrease of the zero-bias conductance with increasing temperature is more pronounced for higher barrier strength.

\begin{figure}[h!]
\includegraphics[width=0.7\linewidth]{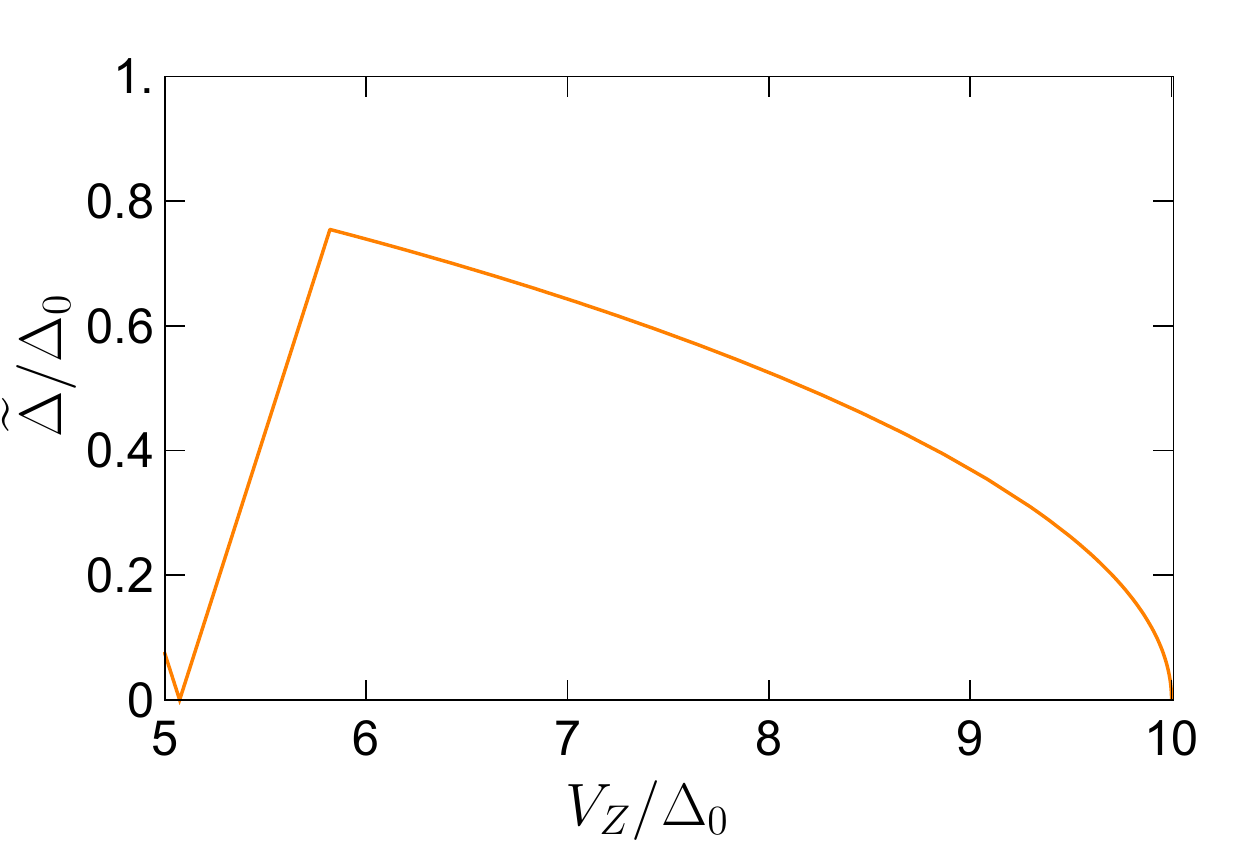}
\caption{(color online) Quasiparticle gap $\widetilde{\Delta}$ as a function of Zeeman field $V_Z$. The parameters used are $\mu = 5$ and $\Delta_0 = 1$. The critical field at which the TQPT happens is $V_{Zc} \equiv \sqrt{\mu^2 + \Delta(V_{Zc})^2} = 5.1$. The plots are for $V_Z$-dependent bulk gap case where the bulk gap collapses at $V_Z^* = 10$.}\label{fig:Gap_collapse} 
\end{figure}

\begin{figure}[h!]
\begin{center}
\includegraphics[width=\linewidth]{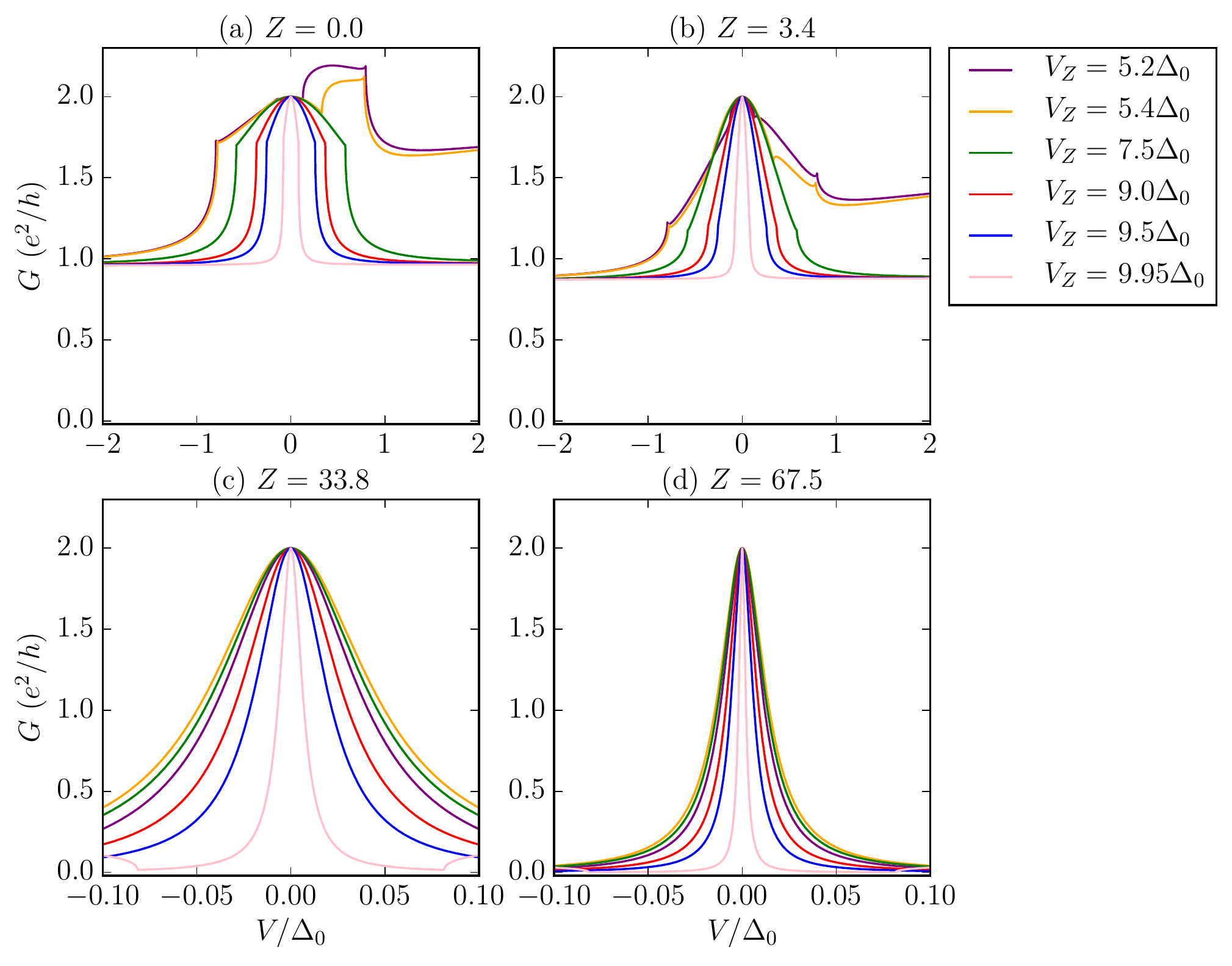}
\end{center}
\caption{(color online) Zero-temperature differential conductance $G$ vs voltage $V$ for several Zeeman fields above the TQPT with different barrier strengths $Z$: (a) $Z =0$, (b) $Z = 3.4$, (c) $Z = 33.8$, and (d) $Z = 67.5$. The parameters used are $\mu = 5$ and $\Delta_0 = 1$ which correspond to  $V_{Zc} \equiv \sqrt{\mu^2 + \Delta(V_{Zc})^2} = 5.1$. The plots are for $V_Z$-dependent bulk gap case where the bulk gap collapses at $V_Z^* = 10$.}\label{fig:dIdV_diff_bh_diff_VZ_collapse_zeroT} 
\end{figure}

\begin{figure}[h!]
\begin{center}
\includegraphics[width=\linewidth]{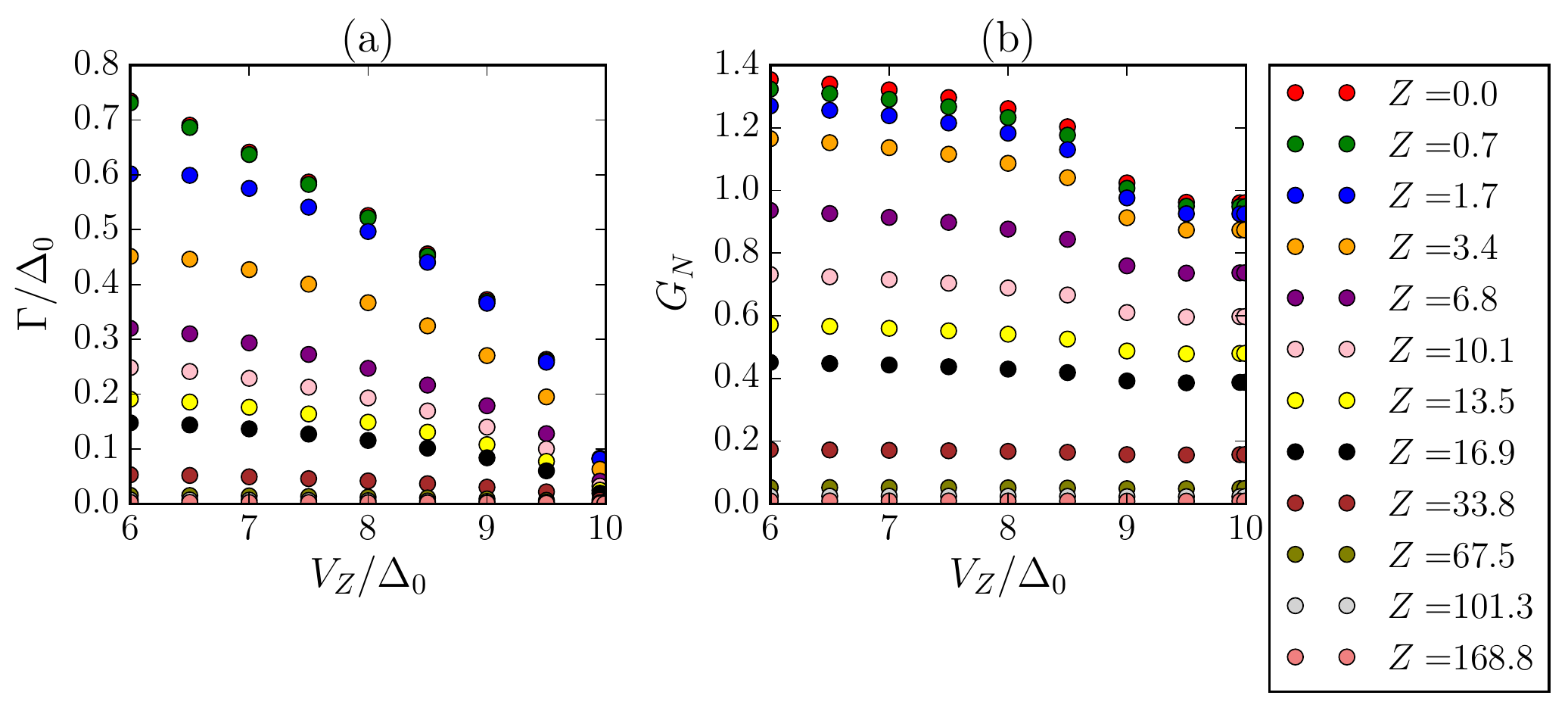}
\end{center}
\caption{(color online) (a) $\Gamma$ vs $V_Z$ and (b) $G_N$ vs $V_Z$ for different barrier strengths $Z$. The parameters used are $T = 0$, $\mu = 5$ and $\Delta_0 = 1$, which correspond to $V_{Zc} \equiv \sqrt{\mu^2 + \Delta(V_{Zc})^2} = 5.1$. The plots are for $V_Z$-dependent bulk gap case where the bulk gap collapses at $V_Z^* = 10$.}\label{fig:Gamma_GN_collapse_B_fit} 
\end{figure}

The ZBP width for the case of Zeeman-dependent bulk gap is smaller than that for the Zeeman-independent case with the same parameters, simply because the induced gap is smaller, which can be understood from Eq.~\eqref{eq:GammaDelta}. The width of the ZBP decreases with increasing barrier strength (decreasing junction transparency) as shown in Fig.~\ref{fig:Gamma_GN_Z_collapse_all}. Similar to the case of Zeeman-independent bulk superconducting gap (see Fig.~\ref{fig:Gamma_GN_Z_all}), in the weak-tunneling $Z\gg 1$ limit, the scaling behavior $\Gamma \propto 1/Z^2$ still holds as shown by the dashed lines in the upper panel of Fig.~\ref{fig:Gamma_GN_Z_collapse_all}. However, there is no simple power-law dependence of $\Gamma$ on $Z$ in the strong-tunneling $Z<1$ regime. We note that in the weak-tunneling limit, the relation $\Gamma \propto G_N$ and $G_N \propto 1/Z^2$ also hold as seen in the middle and lower panels of Fig.~\ref{fig:Gamma_GN_Z_collapse_all}. 

Figure~\ref{fig:FWHM_collapse_T_all} shows the dependence of FWHM of ZBP on temperature where the FWHM shown in Fig.~\ref{fig:FWHM_collapse_T_all}(a) corresponds to the FWHM obtained from taking the distance between the midpoint in the conductance plots and the FWHM in Fig.~\ref{fig:FWHM_collapse_T_all}(b) corresponds to the width of the Lorentzian fit to  the conductance plot. In the limit where $\Gamma \ll T \ll \widetilde{\Delta}$, the FWHM has a width $3.5 T$ as shown by the dashed lines in the figure. We note again that this thermal broadening constraint of FWHM $> 3.5T$ applies to the specific assumption of a Lorentzian broadening, and is in general inapplicable to MZM broadening for larger $T$ and $\Gamma$---see, e.g., Figs.~\ref{fig:FWHM_T_all}(a) and \ref{fig:FWHM_collapse_T_all}(a).

In Fig.~\ref{fig:GS_collapse_diff_T_all_fit}, we show the scaling fit of the zero-bias conductance to Eq.~\eqref{eq:fit}. Only the zero-bias conductance values in the weak-tunneling ($\Gamma \ll \widetilde{\Delta}$) and small-temperature ($T \ll \widetilde{\Delta}$) limit follow the scaling of Eq.~\eqref{eq:fit}. Figures~\ref{fig:GS_collapse_diff_T_all_fit}(b) and~\ref{fig:GS_collapse_diff_T_all_fit}(d) show the fit to the scaling function with a dimensionless parameter $T/\Gamma$ on a log-log scale and a linear scale, respectively. Obviously, scaling is poor except for small $T$ and $\Gamma$.

\begin{figure}[h!]
\begin{center}
\includegraphics[width=\linewidth]{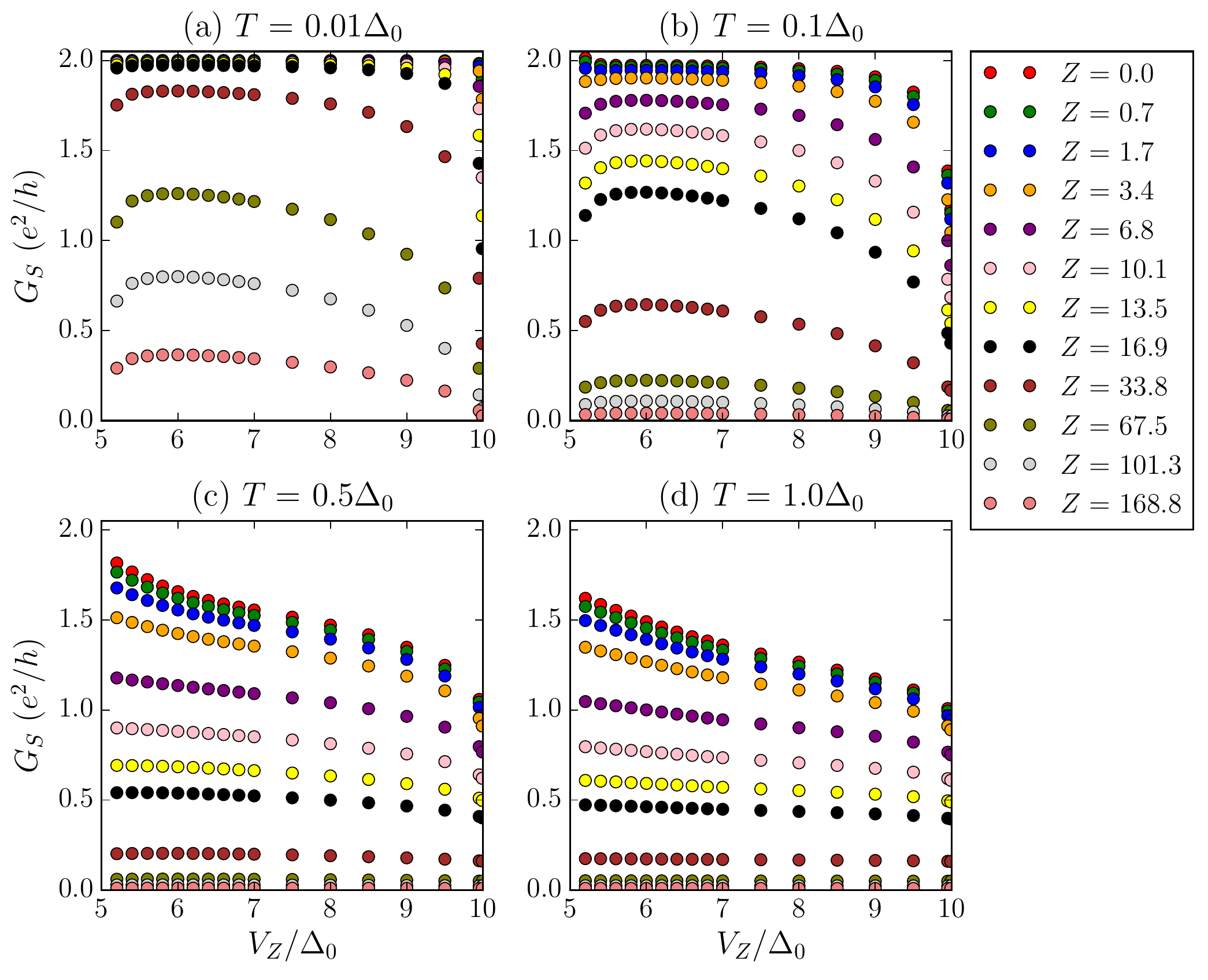}
\end{center}
\caption{(color online) Zero-bias differential conductance $G_S$ vs Zeeman field $V_Z$ above the TQPT for different barrier strengths $Z$ at several fixed temperatures $T$: (a) $T = 0.01\Delta_0$, (b) $T = 0.1 \Delta_0$, (c) $T = 0.5\Delta_0$, and (d) $T = \Delta_0$. The parameters used are $\mu = 5$ and $\Delta_0 = 1$, which correspond to $V_{Zc} \equiv \sqrt{\mu^2 + \Delta(V_{Zc})^2} = 5.1$. The plots are for $V_Z$-dependent bulk gap case where the bulk gap collapses at $V_Z^* = 10$.}\label{fig:GS_B_collapse} 
\end{figure}

\begin{figure}[h!]
\begin{center}
\includegraphics[width=\linewidth]{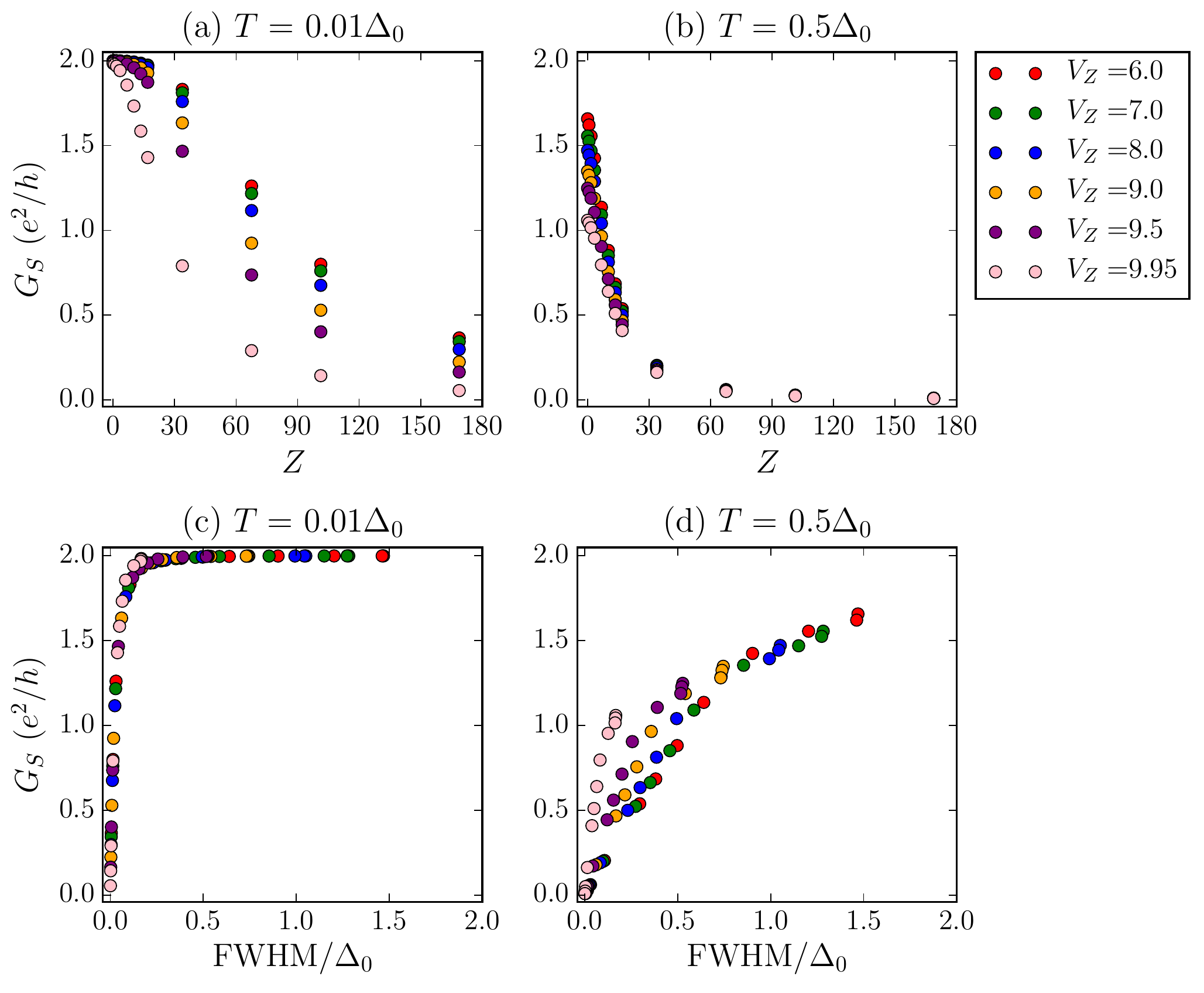}
\end{center}
\caption{(color online) Zero-bias differential conductance $G_S$ vs barrier strength $Z$ (top)  and $G_S$ versus FWHM (bottom) for several values of $V_Z$ above the TQPT at fixed temperature: $T = 0.01\Delta_0$ (left) and $T = 0.5\Delta_0$ (right). The parameters used are $\mu = 5$ and $\Delta_0 = 1$, which correspond to the TQPT critical field $V_{Zc} \equiv \sqrt{\mu^2 + \Delta(V_{Zc})^2} = 5.1$. The plots are for $V_Z$-dependent bulk gap case where the bulk gap collapses at $V_Z^* = 10$.}\label{fig:GS_Z_Gamma_collapse} 
\end{figure}

The quasiparticle gap $\widetilde{\Delta}$ is shown as a function of $V_Z$ in Fig.~\ref{fig:Gap_collapse}.  Here we have the bulk superconducting gap collapsing at $V_Z = V_Z^* = 10 \Delta_0$. Compared to the case of Zeeman-independent bulk gap, the quasiparticle gap here has a stronger dependence on the Zeeman field $V_Z$ resulting in a more pronounced change in the ZBP width with the Zeeman field (compare Fig.~\ref{fig:dIdV_diff_bh_diff_VZ_collapse_zeroT} with Fig.~\ref{fig:dIdV_diff_bh_diff_VZ_zeroT}). We note that since the quasiparticle gap changes sharply near the critical bulk-gap closing field $V_Z^*$, a more pronounced change in the ZBP width with $V_Z$ can be observed near $V_Z^*$.

The zero-temperature differential conductance plots for different Zeeman fields above the TQPT at several fixed tunnel barrier strengths are shown in Fig.~\ref{fig:dIdV_diff_bh_diff_VZ_collapse_zeroT}. Similar to the case of Zeeman-independent bulk gap (cf. Fig.~\ref{fig:dIdV_diff_bh_diff_VZ_zeroT}), for the Zeeman fields just above the TQPT the zero-bias conductance in the strong-tunneling limit may not develop as a peak in the spectra [see Fig.~\ref{fig:dIdV_diff_bh_diff_VZ_collapse_zeroT}(a)]. As shown in Figs.~\ref{fig:dIdV_diff_bh_diff_VZ_collapse_zeroT}(b)--\ref{fig:dIdV_diff_bh_diff_VZ_collapse_zeroT}(d), the width of the ZBP first increases with increasing Zeeman field and then decreases with a further increase in the Zeeman field.   

Figure~\ref{fig:Gamma_GN_collapse_B_fit}(a) shows more explicitly that sufficiently above the TQPT, the ZBP width decreases with increasing Zeeman field where it vanishes at the bulk-gap critical closing field $V_Z^* = 10 \Delta_0$. The plot of $G_N$ versus $V_Z$ is shown in Fig.~\ref{fig:Gamma_GN_collapse_B_fit}(b). As seen from the plot, $G_N$ first decreases with increasing $V_Z$ and then saturates to a constant value  at large $V_Z$.

We plot the zero-bias conductance $G_S$ as a function of $V_Z$ in Fig.~\ref{fig:GS_B_collapse}. In the limit where $ \Gamma \ll T \ll \widetilde{\Delta}$, the zero-bias conductance is $G_S\propto G_N \widetilde{\Delta}$. The dependence of $G_S$ on $\widetilde{\Delta}$ is reflected in the zero-bias conductance values shown in Figs.~\ref{fig:GS_B_collapse}(a) and~\ref{fig:GS_B_collapse}(b) for sufficiently large barrier strength ($Z \geq 6.8$), where it first increases with increasing $V_Z$ and beyond a certain value of $V_Z$, it decreases with increasing $V_Z$.

The plots of finite-temperature $G_S$ versus barrier strength $Z$ and $G_S$ versus FWHM of the ZBP are depicted in Fig.~\ref{fig:GS_Z_Gamma_collapse}. The value of $G_S$ decreases with increasing barrier strength (decreasing ZBP width). For the case of Zeeman-dependent bulk gap considered here, the variation of $G_S$ on $V_Z$ is more pronounced than the Zeeman-independent bulk gap case (cf. Fig.~\ref{fig:GS_Z_Gamma}). 

\section{Andreev Bound State}\label{sec:ABS}

In Sec.~\ref{sec:majorana}, we examined the temperature and tunnel coupling properties of the MZM-induced ZBP, e.g., how the conductance varies with respect to temperature, tunnel barrier strength, and Zeeman field. In particular, we find that only in the low-temperature and weak-tunneling limit, i.e., $T,~\Gamma \ll \widetilde{\Delta}$, the zero-bias conductance for the MZM-induced ZBP scales with a single variable $T/\Gamma$, i.e., the ratio of temperature to zero-temperature half width at half maximum (i.e., intrinsic ZBP broadening $\Gamma$), according to Eq.~\eqref{eq:fit}. Here in this section, we investigate how ABS-induced ZBPs~\cite{Liu2017Andreev} change with respect to $T$, $\Gamma$, and $V_Z$. In contrast to the MZMs, the ABSs are topologically trivial states, which can appear below the TQPT in class-D superconducting nanowires with spin-orbit coupling and Zeeman field in the absence of any disorder as long as there is a smooth quantum dot potential in the system~\cite{Liu2017Andreev}. The smooth potential is a crucial ingredient for the formation of the nontopological near-zero-energy ABSs. Effectively, we can think of the nanowire being composed of two spinless $p$-wave superconductors holding a pair of MZMs at the ends. Only under smooth potential circumstances, the two MZMs at one end from both of the channels are spatially separated and couple weakly with each other, giving rise to near-zero-energy ABSs~\cite{moore2016majorana}. Whether ABSs form or not in specific experimental nanowires is not known \textit{a priori} and there is no easy way to distinguish ZBPs arising from ABSs from those arising from MZMs in experimental systems since the TQPT point is not known experimentally~\cite{Liu2017Andreev}.  We numerically simulate such ABS-induced ZBPs at different temperature and tunnel barrier strength, and investigate how the peak and width of the ZBP change with respect to such external parameters. The key question, of course, is whether the temperature and tunnel coupling dependence of ABS-induced trivial ZBP would manifest any qualitative difference with that of the MZM-induced topological ZBP, thus enabling an experimental technique to distinguish them without any \textit{a priori} knowledge of the TQPT point. It is important to emphasize here that Eq.~\eqref{eq:fit}, with an explicit scaling of the conductance on $T/\Gamma$, is simply the finite-temperature convolution of a resonant scattering peak with the derivative of the Fermi function.  The integrand of Eq.~\eqref{eq:fit} has nothing whatsoever to do with topology, and is valid (for small $T$ and $\Gamma$) equally well for the subgap conductance contributed by MZM or ABS.  In fact, Eq.~\eqref{eq:fit} was first used in this context for spin-polarized Andreev transport in semiconductor-superconductor hybrid structures~\cite{Zutic1999Spin} before any discussion of MZMs in the literature~\cite{Sengupta2001Midgap}. Thus, we expect any scaling behavior of $G_S$ on $T/\Gamma$ to be equally valid for both MZMs and ABSs with the only difference being that the $T=0$ conductance associated with the MZM is always strictly $2e^2/h$.  Thus, if the experimental ZBP is already $> 2e^2/h$ at a finite temperature, the likelihood that the corresponding subgap state is an ABS rather than an MZM is rather high~\cite{nichele2017scaling}.

\begin{figure}[h!]
\includegraphics[width=0.45\textwidth]{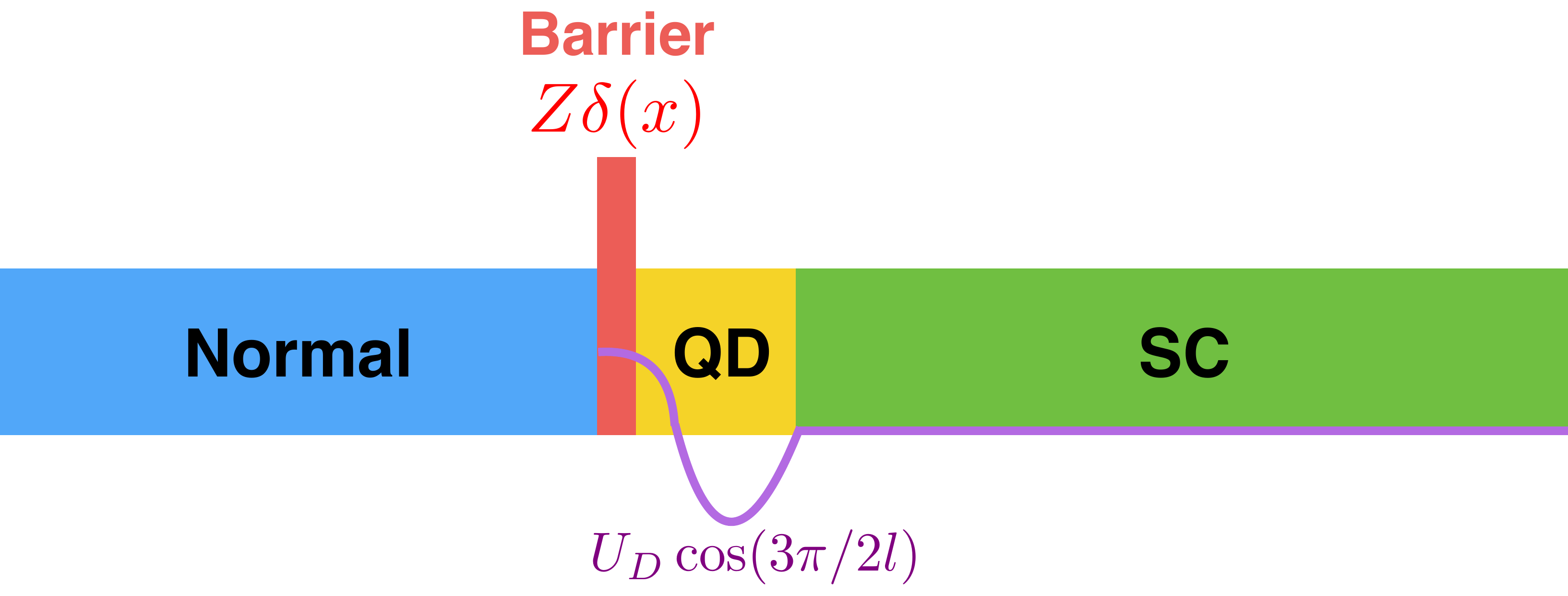}
\caption{(color online) Schematic of the NS junction for the ABS-induced ZBP. An extra and crucial ingredient in the NS junction here is the quantum dot between the normal lead and the proximitized nanowire. The quantum dot is a smooth confinement potential and is not proximitized by the parent superconductor.}
\label{fig:schematic}
\end{figure}

For the simulation of the ABS-induced ZBPs, we follow Ref.~\cite{Liu2017Andreev} and use a finite-length superconducting nanowire with a smooth potential quantum dot and a delta-function barrier lying in between the lead and the nanowire, as shown by the schematic in Fig.~\ref{fig:schematic}. The smooth potential confinement in the quantum dot gives rise to a near-zero-energy localized ABS (sometimes mimicking the MZM). The BdG Hamiltonian of the lead and nanowire are the same as Eq.~\eqref{eq:hamiltonian} with the BdG Hamiltonian for the quantum dot given by
\begin{equation}
\mathcal{H}_{\text{QD}} = \left( -\frac{\partial^2_x}{2m^*}  -i \alpha \partial_x \sigma_y + U(x) - \mu + Z\delta(x) \right)\tau_z,
\label{eq:hamiltonianQD}
\end{equation}
where the BdG Hamiltonian is given in the same basis as Eq.~\eqref{eq:hamiltonian}. We use the following parameters for the simulation in this section:  $\mu = 3.8$, $\Delta_0=1$, $U(x)=U_D \cos\left(3\pi x/2l\right)$ is the smooth confinement potential with the dot depth $U_D= 4$, and dot length $l=0.3$ (in unit of $\ell_{\mathrm{SO}}$ where $\ell_{\mathrm{SO}}$ is the spin-orbit length). The total length of the semiconductor nanowire is $L=13$. Here we assume the bulk gap to be independent of the Zeeman field, i.e., $\Delta (V_Z) = \Delta_0 \equiv \Delta (V_Z = 0)$. The tunnel barrier at the NS interface is modeled as an onsite energy with strength $Z$ (exactly as in Sec.~\ref{sec:majorana}) at the first site of the quantum dot next to the lead. Unless otherwise stated, for the ABS-induced ZBPs, we choose the Zeeman field to be $V_Z \simeq 1.7$, where the energy of the ABS is zero. Such an ABS is topologically trivial, because $V_Z \simeq 1.7 < V_{Zc} = \sqrt{\mu^2 + \Delta^2_0} \simeq 3.9$. The precise value of $V_Z$ at which the ABS energy is zero varies slightly with respect to the tunnel barrier strength, since the barrier is also part of the quantum dot. However, the variation is tiny, and often it is less than $5\%$ of the average $V_Z$. (This also indicates that it may be tricky, although feasible as a matter of principle, to use any sensitivity to the tunnel barrier strength as a diagnostic tool to distinguish between MZM and ABS.)

\begin{figure}[h!]
\includegraphics[width=0.48\textwidth]{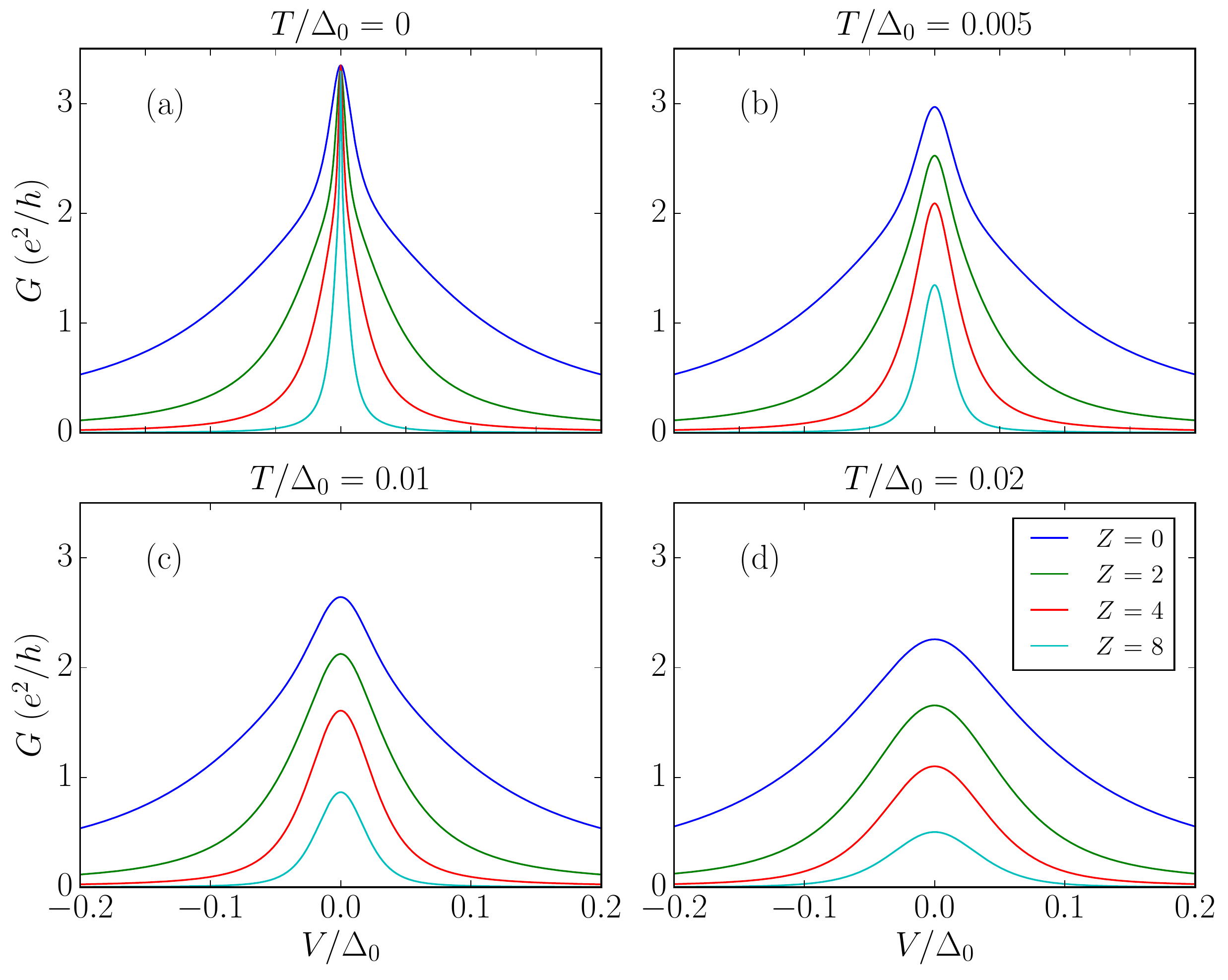}
\caption{(color online) Plot of ABS-induced conductance $G$ as a function of voltage $V$ for different tunnel barrier strengths $Z$ at various temperatures $T$: (a) $T = 0$, (b) $T = 0.005\Delta_0$, (c) $T = 0.01\Delta_0$, and (d) $T = 0.02\Delta_0$.}
\label{fig:TforZ}
\end{figure}

First, we examine how the profile of the ABS-induced ZBP varies with the tunnel barrier strength $Z$ at fixed temperature, as shown in Fig.~\ref{fig:TforZ}. Figure~\ref{fig:TforZ}(a) shows the zero-temperature ZBP profiles at four different values of tunnel barrier strength. When the tunnel barrier strength is lowered, the ZBP gets broadened, with the ZBP conductance at zero temperature around $3e^2/h$, insensitive to the variation of the tunnel barrier strength. The stability of the ABS-induced ZBPs is related to the origin of the formation of the ABS. The ABS is formed by two MZMs weakly coupled with each other, with each MZM contributing almost perfect Andreev reflection, independent of the tunnel barrier strength, to the conductance. The ZBP conductance, however, is not robust against temperature, as indicated by Figs.~\ref{fig:TforZ}(b)--(d). For the finite temperature case, with the lowering of the tunnel barrier strength, both the ZBP conductance and the broadening increase. These features are qualitatively similar to our findings on MZM-induced ZBPs as presented in Sec.~\ref{sec:majorana}.

\begin{figure}[h]
\includegraphics[width=0.48\textwidth]{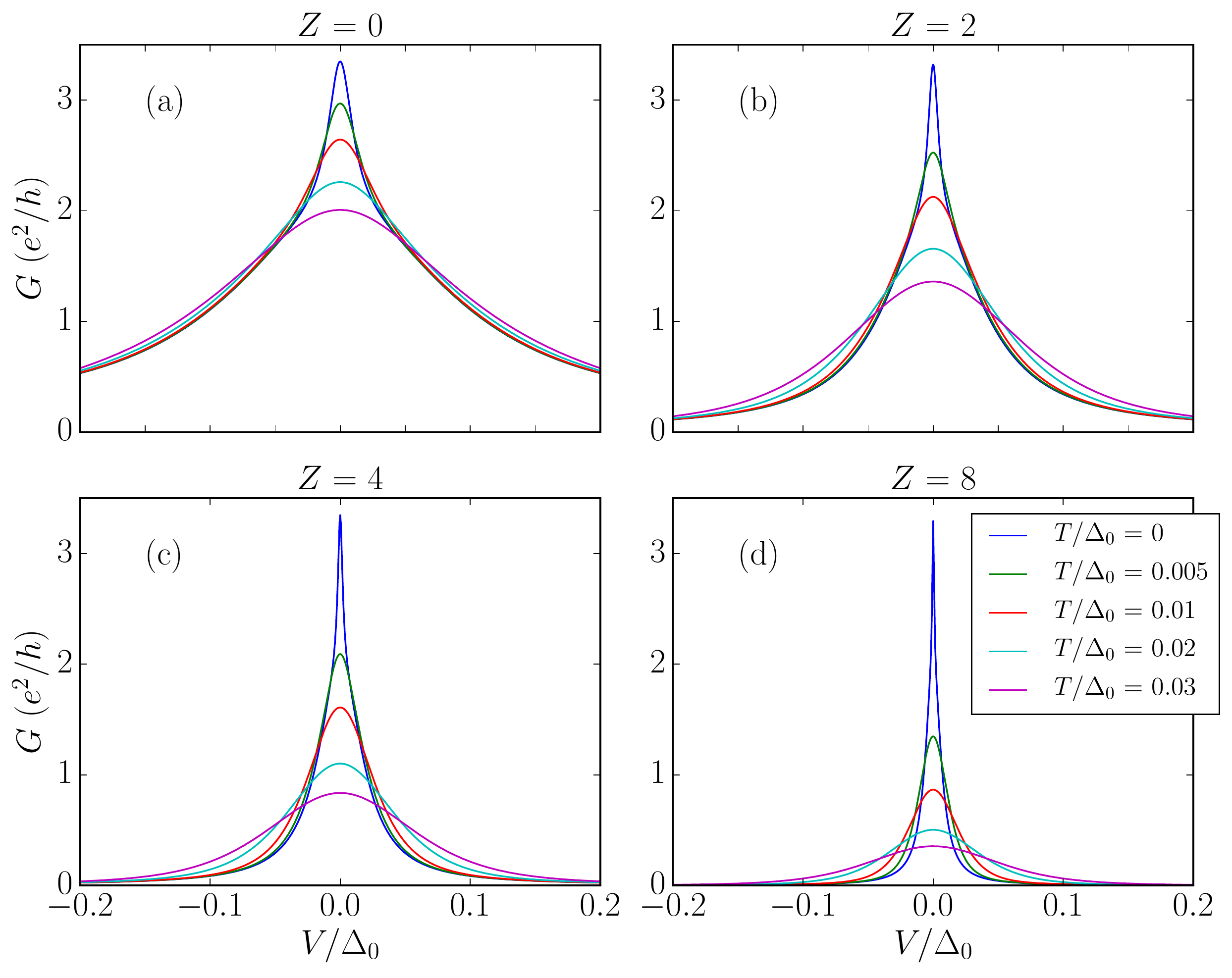}
\caption{(color online) Plot of ABS-induced conductance $G$ as a function of voltage $V$ for different temperatures $T$ at various tunnel barrier strengths $Z$: (a) $Z = 0$, (b) $Z = 2$, (c) $Z = 4$, and (d) $Z = 8$.}
\label{fig:ZforT}
\end{figure}

Figure.~\ref{fig:ZforT} shows more explicitly the effect of temperature on the ABS-induced ZBPs. In general, raising temperature would lower the peak conductance and broaden the width simultaneously. Furthermore, at higher tunnel barrier strength, the peak suppressing effect by the temperature is even more dramatic than that at lower tunnel barrier strength [compare the blue and green curves in both Figs.~\ref{fig:ZforT}(a) and ~\ref{fig:ZforT}(d)].

\begin{figure}[h!]
\includegraphics[width=0.48\textwidth]{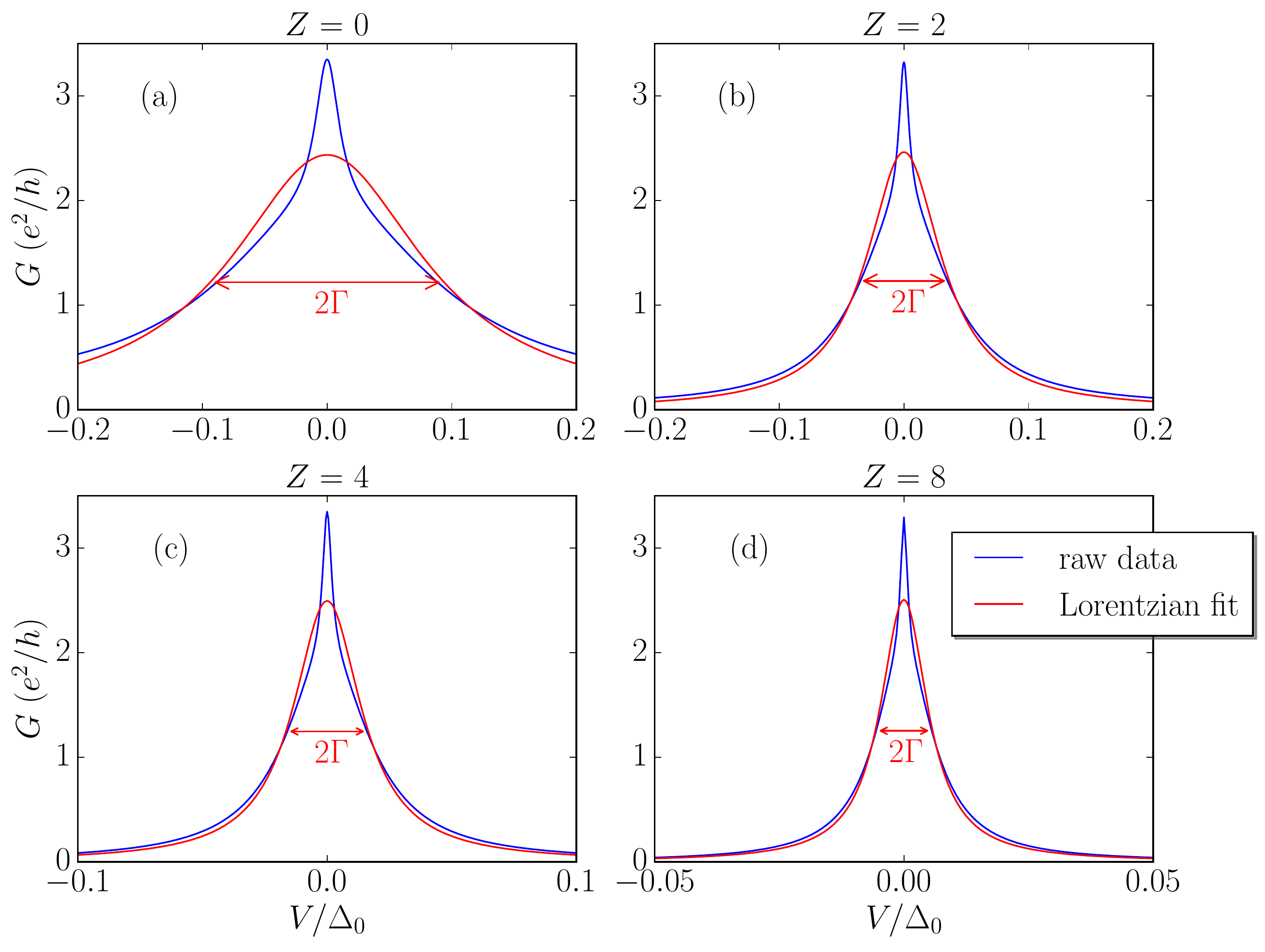}
\caption{(color online) Zero-temperature ABS-induced differential conductance $G$ as a function of bias voltage $V$ (blue curve) fitted by a Lorentzian function (red curve) for different tunnel barrier strengths: (a) $Z = 0$, (b) $Z = 2$, (c) $Z = 4$, and (d) $Z = 8$. Since the ZBP is a superposition of a thin peak at the top and a broad peak at the bottom, the width of the peak $2\Gamma$ is taken to be the full width at half maximum of the Lorentzian fit shown by the red curve. Note that only the ZBP in the high-barrier-strength limit can be fitted nicely to a Lorentzian curve.}
\label{fig:fit}
\end{figure}

\begin{figure}[h!]
\includegraphics[width=0.48\textwidth]{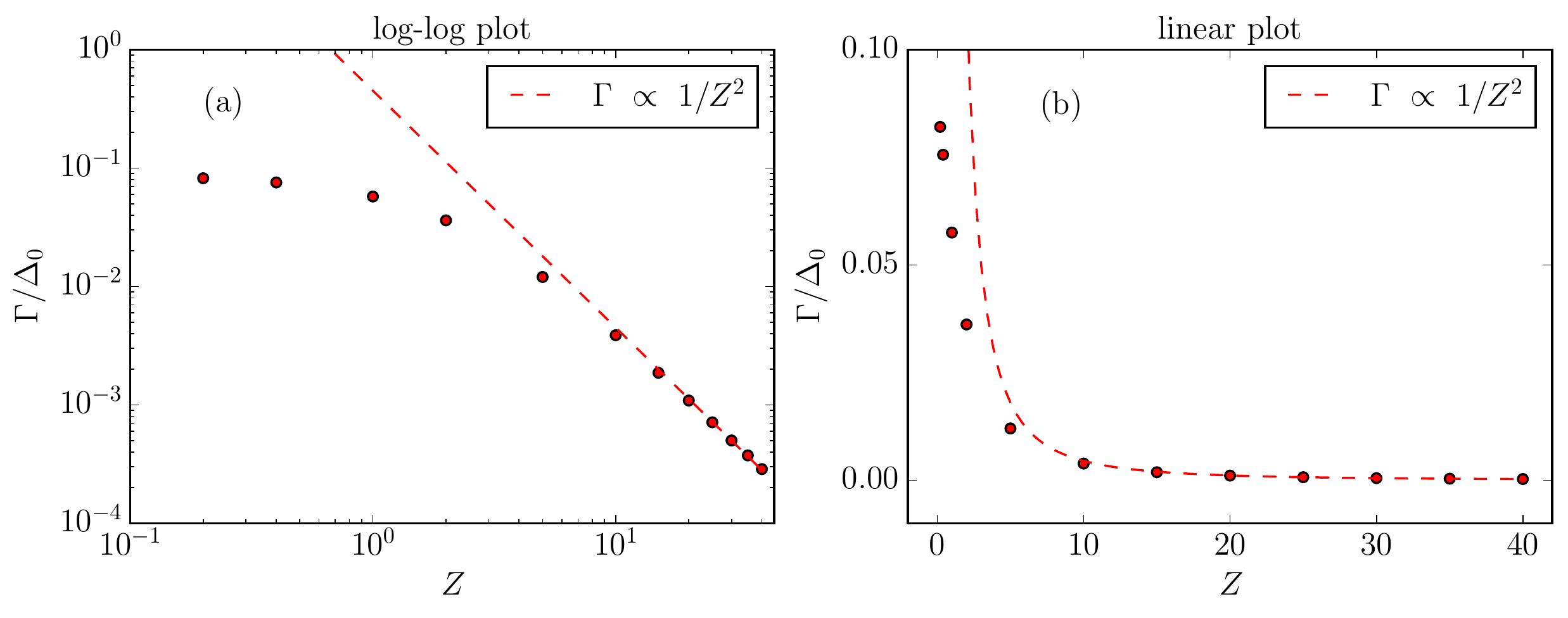}
\caption{(color online) (a) Log-log plot and (b) Linear plot of $\Gamma$ vs barrier strength $Z$ for ABS. In the limit where $Z \gg 1$, $\Gamma \propto 1/Z^2$ as shown by the dashed lines.}
\label{fig:Gamma_Z_ABS}
\end{figure}

\begin{figure}[h!]
\includegraphics[width=0.48\textwidth]{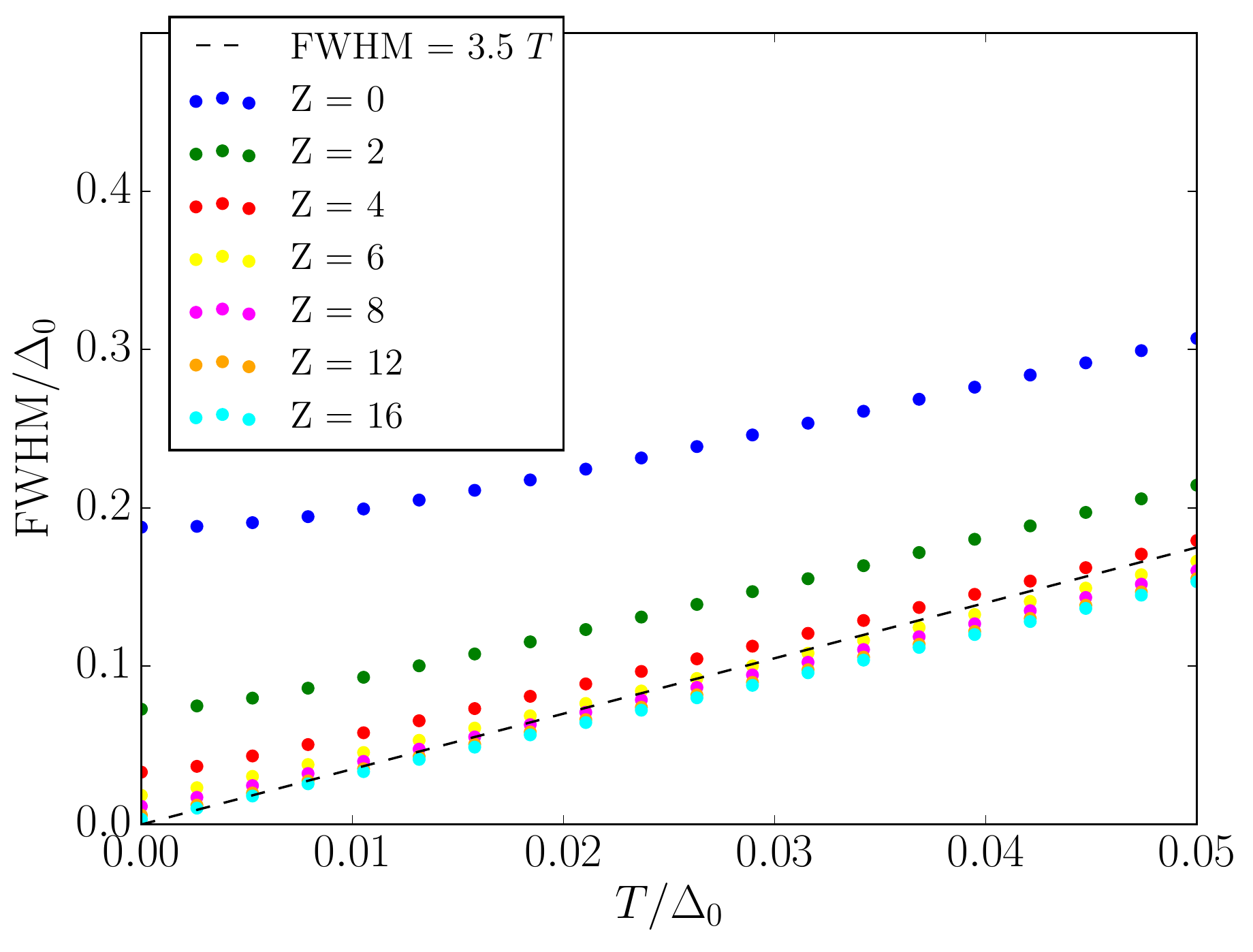}
\caption{(color online) FWHM of the ABS-induced ZBP as a function of temperature for various tunnel barrier strengths $Z$. The FWHM is the width of the Lorentzian function used to fit the conductance plots. Dashed line ($3.5T$) denotes the width of a Lorentzian resonant peak broadened by temperature only.}
\label{fig:35T}
\end{figure}

\begin{figure*}[t]
\includegraphics[width=\textwidth]{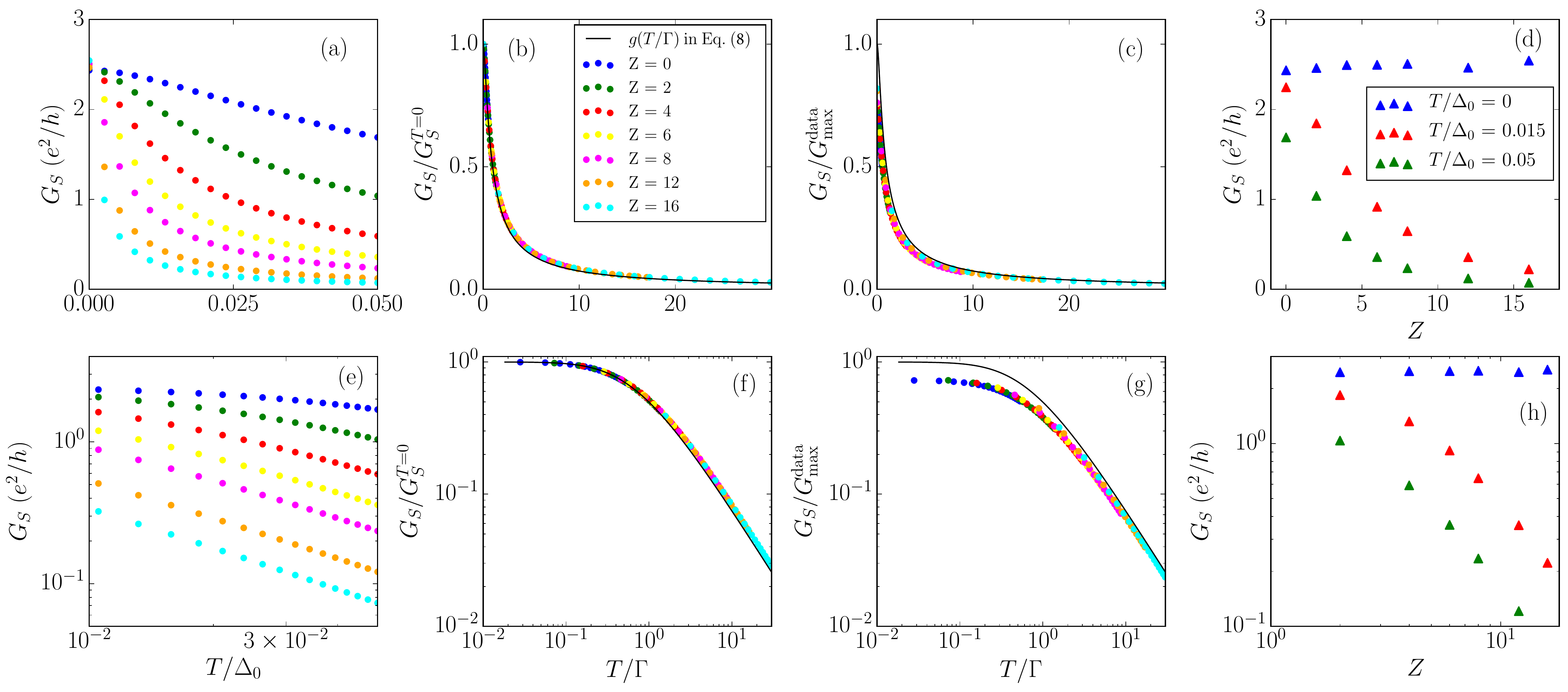}
\caption{(color online) Linear plot (top) and $\log$-$\log$ plot (bottom) of zero-bias conductance $G_S$ as a function of [(a), (e)] temperature $T$, [(b),(f),(c),(g)] $G_S/G_S^{T=0}$ as a function of $T/\Gamma$, and [(d),(h)] $G_S$ vs $Z$. The zero-bias conductance $G_S$ is the peak value of the Lorentzian fit for the ABS-induced ZBP. In (b) and (f), $G_S$ is normalized by the zero-temperature zero-bias value of the Lorentzian fit while in (c) and (g), $G_S$ is normalized by the zero-temperature zero-bias conductance value of the raw data $G_{\max}^{\mathrm{data}}$. Note that the scaling function (solid line), which is given by $g(T/\Gamma)$ in Eq.~\eqref{eq:fit}, where $\Gamma$ is the half width at half maximum of the Lorentzian fit, fits the data points only when $G_S$ is normalized by the peak value of the Lorentzian fit [(b) and (f)].}
\label{fig:scaling}
\end{figure*}

\begin{figure}[h!]
\includegraphics[width=0.48\textwidth]{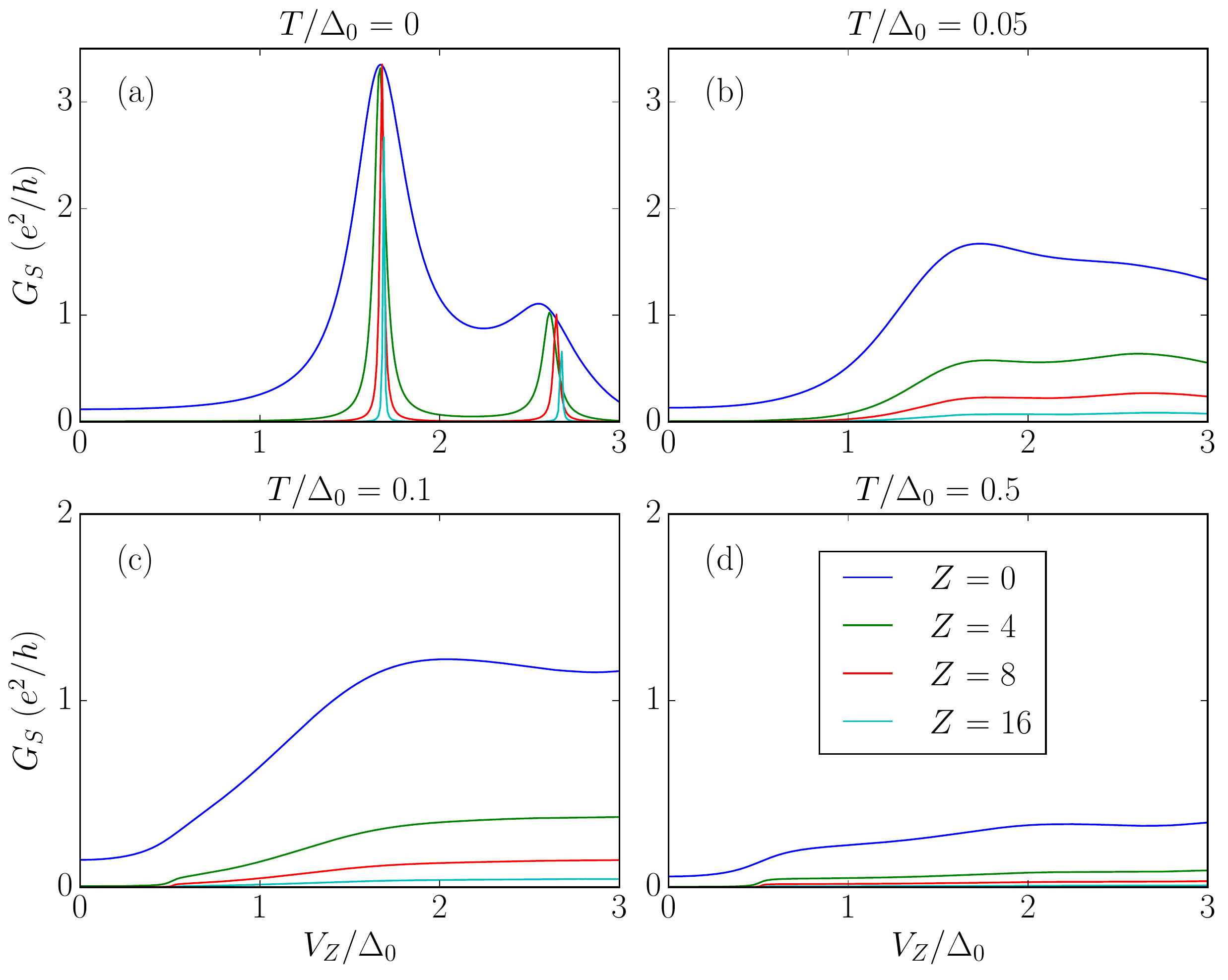}
\caption{(color online) ABS zero-bias conductance $G$ vs Zeeman field $V_Z$ for different tunnel barrier strength $Z$ at various temperatures $T$: (a) $T = 0$, (b) $T = 0.05\Delta_0$, (c) $T = 0.1\Delta_0$, and (d) $T = 0.5\Delta_0$.}
\label{fig:Gs}
\end{figure}

It is important to realize that although the trivial ZBP arising from the ABS has an arbitrary conductance value ($>2e^2/h$) at $T=0$ for all values of $Z$, the ZBP conductance decreases quickly to $2e^2/h$ or below already at rather low temperatures. This is particularly true when the tunnel barrier is not particularly weak, but even for $Z= 4$, the ZBP in Fig.~\ref{fig:ZforT} is $\sim 2e^2/h$ for $T/\Delta_0$ = 0.005 (which is an extremely low temperature experimentally).  Thus, the ABS-induced ZBP at finite temperature may not be that different from the MZM-induced ZBP although at $T=0$, the MZM ZBP is necessarily quantized at $2e^2/h$ and the ABS ZBP not.

In order to characterize some intrinsic properties of the ABS-induced ZBP, e.g., the intrinsic broadening width and the intrinsic peak value, we now focus on the zero-temperature ABS-induced ZBP. Here we obtain the width and peak value of the ZBP by fitting the zero-temperature conductance to the Lorentzian function:
\begin{align}
G(V) = G_{\text{max}} \frac{\Gamma^2}{V^2 + \Gamma^2},
\label{eq:lorentzian}
\end{align}
with two fitting parameters, i.e., the peak value $G_{\text{max}}$ at zero-bias voltage and half width at half maximum $\Gamma$. We note that in contrast to the MZM case where $G_{\mathrm{max}} = 2 e^2/h$, here the ABS-induced ZBP has a nonuniversal peak value $G_{\mathrm{max}}$ which decreases with increasing barrier strength $Z$. The fitting result is shown in Fig.~\ref{fig:fit}, where the blue curves are the zero-temperature differential conductance at various tunnel barrier strengths, and the red curves are the Lorentzian fitting curves. Few remarks are in order here. First, as the blue curves in Figs.~\ref{fig:fit}(a) and \ref{fig:fit}(b) show, the zero-temperature ABS-induced ZBP is a sum of two peaks: one wider peak at the bottom plus one thinner peak on top. The thin top peak is very fragile and is easily suppressed by temperature and dissipation, leaving mostly the broader lower peak behind, leading to the approximate $2e^2/h$ value of the ZBP conductance associated with the ABS~\cite{Liu2017Andreev}.  This phenomenon is consistent with the two weakly coupled Majorana picture for the ABS, where one of the MZMs couples with the lead more strongly, giving rise to a wider ZBP at the bottom, while the other MZM couples with the lead weakly, giving rise to a thin ZBP at the top. Second, all the Lorentzian fitting have their $G_{\text{max}}$ less than the peak value of the raw data. Thus, the Lorentzian fits lead to ABS ZBPs, which are very similar to the MZM ZBPs. Following the picture of two Majoranas forming the ABS, the combined peak is the sum of a wide peak and a thin peak, the wide one has a larger weight since it arises from the MZM located nearer to the NS junction. The Lorentzian function is thus mainly fitting the ZBP from the MZM, which couples more strongly to the lead, while ignoring the small peak on top of the wide ZBP, which is why the peak value of the fitting function is always smaller than the peak value of the raw data. This also has the immediate experimental implication that the thin and weak sharp peak may simply not be observable in the experimental tunneling data at finite temperature and finite energy resolution. Third, Figs.~\ref{fig:fit} shows that the ABS-induced ZBP is more Lorentzian-like in the weak-tunneling limit (large $Z$), while not Lorentzian-like in the strong-tunneling limit (small $Z$). This phenomenon of more Lorentzian-like profile of ZBP at weak-tunneling limit is quite similar to the MZM-induced ZBP situation. Thus, there is very strong qualitative similarity between ABS- and MZM-induced ZBPs.

In Fig.~\ref{fig:Gamma_Z_ABS}, we plot our calculated ABS-induced ZBP broadening (or width) $\Gamma$, i.e., ZBP half width at half maximum at $T=0$, as a function of barrier strength $Z$ (similar to what was done in Fig.~\ref{fig:Gamma_GN_Z_all} for the corresponding MZM-induced ZBP).  It is clear that the dependence of $\Gamma$ on $Z$ for the ABS ZBP is very similar to that of MZM ZBP.  In particular, $\Gamma \propto 1/Z^2$ in the weak-tunneling $Z \gg 1$ limit, and the strong-tunneling regime ($Z<1$) does not manifest any power-law behavior on $Z$.  Thus, the ZBP broadening for ABS and MZM are very similar.

We now investigate how the width of the ABS-induced ZBP gets broadened by temperature (i.e., in addition to the intrinsic $T=0$ width $\Gamma$ arising from tunnel coupling). As discussed in the previous section, finite temperature is introduced by the convolution between the zero-temperature differential conductance and the derivative of the Fermi-Dirac distribution [Eq.~\eqref{eq:GT}]. Since the ABS ZBPs have a thin sharp peak on top of a wide peak, we define the FWHM of the peak to be the FWHM of the Lorentzian curve [Eq.~\eqref{eq:lorentzian}] used to fit the conductance plot instead of taking the FWHM to be the midpoint of the conductance plots as in the previous section. We plot FWHM as a function of temperature for various tunnel barrier strengths in Fig.~\ref{fig:35T}. In addition, there is a dashed line in Fig.~\ref{fig:35T} denoting FWHM = 3.5$T$, which is the thermal width of a Lorentzian resonance curve broadened by the derivative of the Fermi-Dirac distribution. Generically, for all values of the tunnel barrier strength, increasing temperature would increase the FWHM. In particular, in the strong-tunneling limit, the finite-temperature broadened FWHM always lies well above 3.5$T$, while in the weak-tunneling limit, the broadened FWHM would go below 3.5$T$ as temperature increases.

We now examine the scaling behavior of the ABS-induced ZBP. In particular, we study how the zero-bias conductance $G_S$ depends on the temperature and broadening. We also investigate whether $G_S$ could be written as a scaling function of a single variable $T/\Gamma$ as implied by Eq.~\eqref{eq:fit}. The scaling results are shown in Fig.~\ref{fig:scaling}, where in the upper panels a linear scale is used, while in the lower panels a $\log$-$\log$ scale is used. In Figs.~\ref{fig:scaling}(a) and ~\ref{fig:scaling}(e), we show the zero-bias conductance as a function of temperature at various tunnel barrier strengths. Generically, the zero-bias conductance $G_S$ decreases with increasing tunnel barrier and increasing temperature. But note that at zero temperature, the conductance peak $G_S$ is quite stable against the variation of the tunnel barrier strength $Z$. Here the zero-bias conductance $G_S$ is the peak value of the Lorentzian fit to the conductance plot. The reason of using the peak value of the Lorentzian fit instead of the peak value of the original conductance plot is because the ABS-induced ZBP is a superposition of two conductance peaks from two weakly coupled MZMs, which by no means have Lorentzian shapes. In Figs.~\ref{fig:scaling}(b) and~\ref{fig:scaling}(f), the zero-bias conductance $G_S$ divided by the zero-bias conductance from the Lorentzian fit at zero temperature $G_S^{T = 0}$ is plotted as a function of the scaling variable [according to Eq.~\eqref{eq:fit}] defined by temperature over intrinsic broadening width. Most of the data points fall on the same scaling function defined by Eq.~\eqref{eq:fit}. In Figs.~\ref{fig:scaling}(c) and~\ref{fig:scaling}(g), we show the zero-bias conductance divided by the zero-bias conductance of the raw data $G_\mathrm{max}^\mathrm{data}$ without any Lorentzian fitting. Although the data seem to present an approximate scaling on the linear plot [Fig.~\ref{fig:scaling}(c)], the scaling behavior actually no longer follows that described by Eq.~\eqref{eq:fit} as clearly shown by the solid line in the log-log plot [Fig.~\ref{fig:scaling}(f)].  Thus, without Lorentzian fits, the scaling properties are quite different. Although the data do collapse, the scaling function obtained from Eq.~\eqref{eq:fit} for the case where $G_S$ is normalized by $G_\mathrm{max}^\mathrm{data}$ can no longer fit the data points. Figures~\ref{fig:scaling}(d) and~\ref{fig:scaling}(h) show the plot of the zero-bias conductance as a function of barrier strength $Z$ for different temperatures, in a linear and log-log scale, respectively. As shown in the figures, at zero temperature the zero-bias conductance peak value is stable over a huge range of barrier strength. The stability of the zero-temperatue ABS ZBP value against the barrier strength is strikingly similar to the robustness of the zero-temperature MZM ZBP against the barrier strength with the difference being that the MZM ZBP is always quantized exactly at $2e^2/h$. At finite temperature the zero-bias conductance values decrease with increasing barrier strength as seen in Figs.~\ref{fig:scaling}(b)-\ref{fig:scaling}(d).

Figure~\ref{fig:Gs} shows the ABS zero-bias conductance as a function of Zeeman field for different tunnel barrier strength at various temperatures. Since at zero temperature, the ABS energy crosses zero only at specific values of $V_Z$~\cite{Liu2017Andreev}, the zero-temperature zero-bias conductance develops peaks only at specific value of $V_Z$ [see Fig.~\ref{fig:Gs}(a)]. We note that the zero-temperature zero-bias conductance peak value is stable over quite a wide range of tunnel barrier strength $Z$ although the exact value of $V_Z$ at which the peak is located varies slightly with $Z$. Finite temperature broadens the zero-temperature conductance peaks and lowers their peak values. As a result, the peaks in the $G_S$ versus $V_Z$ plot are broadened at finite temperature [see Figs.~\ref{fig:Gs}(b)--\ref{fig:Gs}(d)]. For finite temperature, the peak value is no longer stable against the change in the barrier strength, but its value quickly falls to $2e^2/h$ or below, mimicking the MZM situation. 

\section{Conclusion}\label{sec:conclusion}

We have described in depth the properties of the zero-bias tunneling conductance for Majorana and Andreev bound states in proximity-induced superconducting semiconductor nanowires in the presence of spin-orbit coupling and Zeeman splitting, focusing on the temperature and tunnel coupling dependence of the subgap conductance.  We establish that the general qualitative dependence of the zero-bias peak on temperature and tunnel coupling is similar for ABS and MZM, thus making a distinction between the two difficult based just on tunneling spectroscopy.  In particular, both ZBPs show approximate scaling of the ZBP conductance at low temperature and tunnel coupling, but there is no scaling for larger temperature and tunnel coupling (and indeed, there is no reason to expect such a scaling).  The MZM-induced ZBP has a quantized value $2e^2/h$ at $T=0$ for all tunnel couplings whereas the ABS-induced ZBP could have a conductance up to $4e^2/h$ since the ABS is composed essentially of two strongly overlapping MZMs.  Although this zero-temperature theoretical distinction is crucial between the trivial ABS and the topological MZM ZBPs, at finite temperature this distinction may be without any difference since the ABS ZBP quickly decreases to $2e^2/h$ or below with increasing temperature, and obviously a strict zero-temperature experiment is impossible.

In this context, it is crucial to emphasize that if a ZBP is manifesting a conductance value $2e^2/h$ or above already at a finite temperature, it is doubtful that the corresponding origin could be the MZM (unless multichannel effects are substantial) simply because the likely zero-temperature extrapolated value of the corresponding conductance is likely to be $>2e^2/h$, making it an ABS candidate. The $T$ dependence of the ZBP height is weak only for $T \ll \Gamma \ll \widetilde{\Delta}$, a regime almost never reached experimentally. We do find the interesting property that the MZM ZBP is more likely to reflect the $2e^2/h$ value even at finite temperatures at stronger tunnel coupling, where the overall background conductance is high and the ratio of the peak conductance to the valley conductance is rather small.  This should be contrasted with the weak-tunneling behavior, where the ZBP is very sharp with low background conductance (so that the peak to valley subgap conductance ratio is large), but in this case the corresponding ZBP conductance is likely to be much lower than $2e^2/h$ even at rather low temperature. Thus, the ZBP conductance of $\sim 2e^2/h$ may occur experimentally with a rather small peak over a broad background (i.e., strong tunneling) whereas a sharp ZBP with a pronounced peak over the background may reflect a low-conductance value (i.e., weak tunneling). Unfortunately, the ABS ZBP also reflects similar behavior since this behavior is rather generic for a peak with a finite width convoluted by the derivative of a Fermi function.

Multichannel effects further complicate the comparison of conductances between theory and experiment. Indeed, the theoretical prediction of quantized conductance into an MZM is only valid for a single-channel or two-channel quantum point contact~\cite{Wimmer2011Quantum}. This is one of the reasons for focusing on the single band wires for the simple models considered in this manuscript. Away from this case, the conductance into an MZM can have any value higher than $2e^2/h$ and would therefore be rather difficult to distinguish from ABSs or other nontopological resonances. Therefore, an experimental determination of an MZM by ZBP requires ensuring that the quantum point contact used in the tunneling is single channel.  Note that this has nothing to do with whether the topological nanowire itself is single channel but only whether the tunnel barrier is single channel, i.e., has transmission dominated by one transmission eigenvalue. This is rather difficult to ensure in the present experiment where the structure of the potential barrier is complicated by the electrostatics of the nearby metallic lead and superconductor. In fact, one could say that for high transparency barriers where the near quantized conductance is seen with low-temperature width ($\sim \Gamma$) comparable to the gap that the barrier is more than likely to be transmitting several channels. This issue can be resolved in experiment by increasing the separation between the metallic lead and superconductor to leave a larger purely semiconductor region which can be depleted to produce a well-controlled single-channel point contact. The observation of a quantized conductance in this limit would constitute a rather definitive test for an MZM~\cite{Wimmer2011Quantum}.

Many of the above ZBP properties have recently been observed experimentally~\cite{Marcus2017PrivateCom,Kouwenhoven2017PrivateCom,HaoZhang2017PrivateCom,nichele2017scaling}, but a firm conclusion on whether the corresponding underlying cause is MZM or ABS remains problematic because extrapolation to $T=0$ from the experimental data is difficult, and to the extent it can be done (see, e.g., the lowest $T/\Gamma$ regime, $T/\Gamma<0.1$, in Fig. 4 of Ref.~\cite{nichele2017scaling}), the extrapolation to $T=0$ is more consistent with the ZBP having a zero-temperature strength of $4e^2/h$ rather than $2e^2/h$, indicating perhaps an ABS origin. We mention, as is obvious from our results in Figs.~\ref{fig:GS_diff_T_all_fit} and~\ref{fig:GS_collapse_diff_T_all_fit}, log-log scaling plots near small values of $T/\Gamma$ are dangerous for the extraction of zero-temperature ZBP value since the scaled conductance varies rapidly for small values of $T/\Gamma$ and distinguishing 1 from 2 on the ordinate is extremely difficult based on such log-log plots.  We caution against any conclusion based on log-log scaling plots of experimental conductance data.

We mention that dissipation has been left out of the theory whereas experimental systems are expected to have some dissipative broadening from unknown sources as discussed in Sec.~\ref{sec:introduction} and in Refs.~\cite{Liu2017Role,DasSarma2016How}.  Dissipation will add to the ZBP broadening for both ABS and MZM, and will further suppress the ZBP height (i.e., the conductance value).  Dissipation effects will be quantitatively most important when the electron temperature and tunnel coupling are small, precisely the regime where scaling of subgap conductance is most accurate.  This may further suppress any scaling behavior in the nanowire conductance, even in the low-temperature and weak-tunneling limit.  For explicit comparison with experiments, dissipation, neglected here and in the analysis of Ref.~\cite{nichele2017scaling}, should be taken into account.

One serious problem in quantitatively comparing experiment and theory in Majorana nanowires is that many relevant parameters are unknown in the experiment.  In fact, except for temperature (even ignoring the complication of the possibly unknown electron temperature) and the measured tunnel conductance itself, no other theoretical parameters of even the minimal model [Eq.~\eqref{eq:H_NW}] are really experimentally known.  Even quantities as basic as the spin splitting ($V_Z$), the induced gap ($\Delta$), and the spin-orbit coupling ($\alpha$) are unknown experimentally in the actual experimental structures.  The tunnel transmission parameter ($Z$) is certainly unknown, and one could even question whether a single tunneling parameter is adequate in describing the tunnel barrier (as in the minimal model).  The ZBP width or broadening is known at finite temperature, but its zero-temperature value (i.e., $\Gamma$, which is a key parameter in the theory describing the tunnel coupling itself) is unknown experimentally.  One can measure the experimental ZBP FWHM at finite temperature, and obtain $\Gamma$ by an extrapolation to $T=0$, but the situation is subtle.  In particular, the experimental scaling behavior in Ref.~\cite{nichele2017scaling} is established by numerically fitting the subgap conductance to Eq.~\eqref{eq:fit} to extract $\Gamma$, but a direct comparison of this ``fitted" $\Gamma$ to the actual experimentally measured FWHM at low temperature shows a very large discrepancy.  In fact, the extraction of $\Gamma$ at finite temperature is nontrivial from the ZBP since fitting it to an arbitrary Lorentzian automatically implies a background subtraction, and as emphasized in the current work, such a Lorentzian approximation is only valid in the very weak tunneling situation of large  (small) $Z$ ($\Gamma$).  We do not believe that connecting $\Gamma$ and the above-gap normal conductance $G_N$ is quantitatively compelling for reasons discussed in this work (and in more details elsewhere~\cite{liu2017phenomenology}) mainly because $G_N$, which is typically non-unique (and has some energy dependence),  also depends on parameters other than the tunnel barrier strength.  Finally, there is no independent information at all on the applicable chemical potential in the Majorana nanowires, which is a parameter of great theoretical significance because it directly defines the TQPT and hence the $V_Z$-extent of the topological regime.

It may still be possible to distinguish ABS and MZM from ZBP behavior provided a great deal of data are available for many different sets of values of $T$, $Z$, $V_Z$, and $\Delta_0$ in many different wires of various lengths through very careful quantitative analysis of the type we have done in the current theoretical work where the nature of the ZBP (i.e., ABS or MZM) is known by construction. However, such extensive data over large parameter sets are currently unavailable for drawing any firm conclusion.  We believe that many more results are necessary, particularly in wires with much larger induced gap so that the dimensionless low-temperature regime (with very small temperature/gap values) can be explored quantitatively, before we can reach a firm conclusion on the nature of the ZBP which is ubiquitous in Majorana nanowires.  Eliminating disorder from the wires has been a tremendous experimental accomplishment over the last five years, and hopefully, achieving large induced proximity gap even at relatively high applied magnetic field will be the next key accomplishment in order to make progress in the field. We believe that any eventual topological quantum computation carried out using Majorana nanowires necessitates substantial more materials science development leading to much larger robust induced superconducting gap in the nanowire at high applied magnetic field.


\begin{acknowledgements}
This work is supported by Microsoft, Laboratory for Physical Sciences, and NSF Physics Frontier Center at JQI. J.D.S. acknowledges the funding from Sloan Research Fellowship and NSF-DMR-1555135 (CAREER).
\end{acknowledgements}




\bibliography{scaling1}

\begin{thebibliography}{61}%
\makeatletter
\providecommand \@ifxundefined [1]{%
 \@ifx{#1\undefined}
}%
\providecommand \@ifnum [1]{%
 \ifnum #1\expandafter \@firstoftwo
 \else \expandafter \@secondoftwo
 \fi
}%
\providecommand \@ifx [1]{%
 \ifx #1\expandafter \@firstoftwo
 \else \expandafter \@secondoftwo
 \fi
}%
\providecommand \natexlab [1]{#1}%
\providecommand \enquote  [1]{``#1''}%
\providecommand \bibnamefont  [1]{#1}%
\providecommand \bibfnamefont [1]{#1}%
\providecommand \citenamefont [1]{#1}%
\providecommand \href@noop [0]{\@secondoftwo}%
\providecommand \href [0]{\begingroup \@sanitize@url \@href}%
\providecommand \@href[1]{\@@startlink{#1}\@@href}%
\providecommand \@@href[1]{\endgroup#1\@@endlink}%
\providecommand \@sanitize@url [0]{\catcode `\\12\catcode `\$12\catcode
  `\&12\catcode `\#12\catcode `\^12\catcode `\_12\catcode `\%12\relax}%
\providecommand \@@startlink[1]{}%
\providecommand \@@endlink[0]{}%
\providecommand \url  [0]{\begingroup\@sanitize@url \@url }%
\providecommand \@url [1]{\endgroup\@href {#1}{\urlprefix }}%
\providecommand \urlprefix  [0]{URL }%
\providecommand \Eprint [0]{\href }%
\providecommand \doibase [0]{http://dx.doi.org/}%
\providecommand \selectlanguage [0]{\@gobble}%
\providecommand \bibinfo  [0]{\@secondoftwo}%
\providecommand \bibfield  [0]{\@secondoftwo}%
\providecommand \translation [1]{[#1]}%
\providecommand \BibitemOpen [0]{}%
\providecommand \bibitemStop [0]{}%
\providecommand \bibitemNoStop [0]{.\EOS\space}%
\providecommand \EOS [0]{\spacefactor3000\relax}%
\providecommand \BibitemShut  [1]{\csname bibitem#1\endcsname}%
\let\auto@bib@innerbib\@empty
\bibitem [{\citenamefont {Kitaev}(2001)}]{Kitaev2001Unpaired}%
  \BibitemOpen
  \bibfield  {author} {\bibinfo {author} {\bibfnamefont {A.~Y.}\ \bibnamefont
  {Kitaev}},\ }\href@noop {} {\bibfield  {journal} {\bibinfo  {journal}
  {Physics-Uspekhi}\ }\textbf {\bibinfo {volume} {44}},\ \bibinfo {pages} {131}
  (\bibinfo {year} {2001})}\BibitemShut {NoStop}%
\bibitem [{\citenamefont {Sengupta}\ \emph {et~al.}(2001)\citenamefont
  {Sengupta}, \citenamefont {\ifmmode \check{Z}\else
  \v{Z}\fi{}uti\ifmmode~\acute{c}\else \'{c}\fi{}}, \citenamefont {Kwon},
  \citenamefont {Yakovenko},\ and\ \citenamefont
  {Das~Sarma}}]{Sengupta2001Midgap}%
  \BibitemOpen
  \bibfield  {author} {\bibinfo {author} {\bibfnamefont {K.}~\bibnamefont
  {Sengupta}}, \bibinfo {author} {\bibfnamefont {I.}~\bibnamefont {\ifmmode
  \check{Z}\else \v{Z}\fi{}uti\ifmmode~\acute{c}\else \'{c}\fi{}}}, \bibinfo
  {author} {\bibfnamefont {H.-J.}\ \bibnamefont {Kwon}}, \bibinfo {author}
  {\bibfnamefont {V.~M.}\ \bibnamefont {Yakovenko}}, \ and\ \bibinfo {author}
  {\bibfnamefont {S.}~\bibnamefont {Das~Sarma}},\ }\href {\doibase
  10.1103/PhysRevB.63.144531} {\bibfield  {journal} {\bibinfo  {journal} {Phys.
  Rev. B}\ }\textbf {\bibinfo {volume} {63}},\ \bibinfo {pages} {144531}
  (\bibinfo {year} {2001})}\BibitemShut {NoStop}%
\bibitem [{\citenamefont {Read}\ and\ \citenamefont
  {Green}(2000)}]{Read2000Paired}%
  \BibitemOpen
  \bibfield  {author} {\bibinfo {author} {\bibfnamefont {N.}~\bibnamefont
  {Read}}\ and\ \bibinfo {author} {\bibfnamefont {D.}~\bibnamefont {Green}},\
  }\href {\doibase 10.1103/PhysRevB.61.10267} {\bibfield  {journal} {\bibinfo
  {journal} {Phys. Rev. B}\ }\textbf {\bibinfo {volume} {61}},\ \bibinfo
  {pages} {10267} (\bibinfo {year} {2000})}\BibitemShut {NoStop}%
\bibitem [{\citenamefont {Nayak}\ \emph {et~al.}(2008)\citenamefont {Nayak},
  \citenamefont {Simon}, \citenamefont {Stern}, \citenamefont {Freedman},\ and\
  \citenamefont {Das~Sarma}}]{Nayak2008Non-Abelian}%
  \BibitemOpen
  \bibfield  {author} {\bibinfo {author} {\bibfnamefont {C.}~\bibnamefont
  {Nayak}}, \bibinfo {author} {\bibfnamefont {S.~H.}\ \bibnamefont {Simon}},
  \bibinfo {author} {\bibfnamefont {A.}~\bibnamefont {Stern}}, \bibinfo
  {author} {\bibfnamefont {M.}~\bibnamefont {Freedman}}, \ and\ \bibinfo
  {author} {\bibfnamefont {S.}~\bibnamefont {Das~Sarma}},\ }\href {\doibase
  10.1103/RevModPhys.80.1083} {\bibfield  {journal} {\bibinfo  {journal} {Rev.
  Mod. Phys.}\ }\textbf {\bibinfo {volume} {80}},\ \bibinfo {pages} {1083}
  (\bibinfo {year} {2008})}\BibitemShut {NoStop}%
\bibitem [{\citenamefont {Alicea}(2012)}]{Alicea2012New}%
  \BibitemOpen
  \bibfield  {author} {\bibinfo {author} {\bibfnamefont {J.}~\bibnamefont
  {Alicea}},\ }\href {http://stacks.iop.org/0034-4885/75/i=7/a=076501}
  {\bibfield  {journal} {\bibinfo  {journal} {Rep. Prog. Phys.}\ }\textbf
  {\bibinfo {volume} {75}},\ \bibinfo {pages} {076501} (\bibinfo {year}
  {2012})}\BibitemShut {NoStop}%
\bibitem [{\citenamefont {Sarma}\ \emph {et~al.}(2015)\citenamefont {Sarma},
  \citenamefont {Freedman},\ and\ \citenamefont
  {Nayak}}]{DasSarma2015Majorana}%
  \BibitemOpen
  \bibfield  {author} {\bibinfo {author} {\bibfnamefont {S.~D.}\ \bibnamefont
  {Sarma}}, \bibinfo {author} {\bibfnamefont {M.}~\bibnamefont {Freedman}}, \
  and\ \bibinfo {author} {\bibfnamefont {C.}~\bibnamefont {Nayak}},\ }\href
  {http://dx.doi.org/10.1038/npjqi.2015.1} {\bibfield  {journal} {\bibinfo
  {journal} {Npj Quantum Information}\ }\textbf {\bibinfo {volume} {1}},\
  \bibinfo {pages} {15001 EP } (\bibinfo {year} {2015})}\BibitemShut {NoStop}%
\bibitem [{\citenamefont {Sau}\ \emph {et~al.}(2010{\natexlab{a}})\citenamefont
  {Sau}, \citenamefont {Lutchyn}, \citenamefont {Tewari},\ and\ \citenamefont
  {Das~Sarma}}]{Sau2010Generic}%
  \BibitemOpen
  \bibfield  {author} {\bibinfo {author} {\bibfnamefont {J.~D.}\ \bibnamefont
  {Sau}}, \bibinfo {author} {\bibfnamefont {R.~M.}\ \bibnamefont {Lutchyn}},
  \bibinfo {author} {\bibfnamefont {S.}~\bibnamefont {Tewari}}, \ and\ \bibinfo
  {author} {\bibfnamefont {S.}~\bibnamefont {Das~Sarma}},\ }\href {\doibase
  10.1103/PhysRevLett.104.040502} {\bibfield  {journal} {\bibinfo  {journal}
  {Phys. Rev. Lett.}\ }\textbf {\bibinfo {volume} {104}},\ \bibinfo {pages}
  {040502} (\bibinfo {year} {2010}{\natexlab{a}})}\BibitemShut {NoStop}%
\bibitem [{\citenamefont {Lutchyn}\ \emph {et~al.}(2010)\citenamefont
  {Lutchyn}, \citenamefont {Sau},\ and\ \citenamefont
  {Das~Sarma}}]{Lutchyn2010Majorana}%
  \BibitemOpen
  \bibfield  {author} {\bibinfo {author} {\bibfnamefont {R.~M.}\ \bibnamefont
  {Lutchyn}}, \bibinfo {author} {\bibfnamefont {J.~D.}\ \bibnamefont {Sau}}, \
  and\ \bibinfo {author} {\bibfnamefont {S.}~\bibnamefont {Das~Sarma}},\ }\href
  {\doibase 10.1103/PhysRevLett.105.077001} {\bibfield  {journal} {\bibinfo
  {journal} {Phys. Rev. Lett.}\ }\textbf {\bibinfo {volume} {105}},\ \bibinfo
  {pages} {077001} (\bibinfo {year} {2010})}\BibitemShut {NoStop}%
\bibitem [{\citenamefont {Oreg}\ \emph {et~al.}(2010)\citenamefont {Oreg},
  \citenamefont {Refael},\ and\ \citenamefont {von Oppen}}]{Oreg2010Helical}%
  \BibitemOpen
  \bibfield  {author} {\bibinfo {author} {\bibfnamefont {Y.}~\bibnamefont
  {Oreg}}, \bibinfo {author} {\bibfnamefont {G.}~\bibnamefont {Refael}}, \ and\
  \bibinfo {author} {\bibfnamefont {F.}~\bibnamefont {von Oppen}},\ }\href
  {\doibase 10.1103/PhysRevLett.105.177002} {\bibfield  {journal} {\bibinfo
  {journal} {Phys. Rev. Lett.}\ }\textbf {\bibinfo {volume} {105}},\ \bibinfo
  {pages} {177002} (\bibinfo {year} {2010})}\BibitemShut {NoStop}%
\bibitem [{\citenamefont {Sau}\ \emph {et~al.}(2010{\natexlab{b}})\citenamefont
  {Sau}, \citenamefont {Tewari}, \citenamefont {Lutchyn}, \citenamefont
  {Stanescu},\ and\ \citenamefont {Das~Sarma}}]{Sau2010Non}%
  \BibitemOpen
  \bibfield  {author} {\bibinfo {author} {\bibfnamefont {J.~D.}\ \bibnamefont
  {Sau}}, \bibinfo {author} {\bibfnamefont {S.}~\bibnamefont {Tewari}},
  \bibinfo {author} {\bibfnamefont {R.~M.}\ \bibnamefont {Lutchyn}}, \bibinfo
  {author} {\bibfnamefont {T.~D.}\ \bibnamefont {Stanescu}}, \ and\ \bibinfo
  {author} {\bibfnamefont {S.}~\bibnamefont {Das~Sarma}},\ }\href {\doibase
  10.1103/PhysRevB.82.214509} {\bibfield  {journal} {\bibinfo  {journal} {Phys.
  Rev. B}\ }\textbf {\bibinfo {volume} {82}},\ \bibinfo {pages} {214509}
  (\bibinfo {year} {2010}{\natexlab{b}})}\BibitemShut {NoStop}%
\bibitem [{\citenamefont {Mourik}\ \emph {et~al.}(2012)\citenamefont {Mourik},
  \citenamefont {Zuo}, \citenamefont {Frolov}, \citenamefont {Plissard},
  \citenamefont {Bakkers},\ and\ \citenamefont
  {Kouwenhoven}}]{Mourik2012Signatures}%
  \BibitemOpen
  \bibfield  {author} {\bibinfo {author} {\bibfnamefont {V.}~\bibnamefont
  {Mourik}}, \bibinfo {author} {\bibfnamefont {K.}~\bibnamefont {Zuo}},
  \bibinfo {author} {\bibfnamefont {S.~M.}\ \bibnamefont {Frolov}}, \bibinfo
  {author} {\bibfnamefont {S.}~\bibnamefont {Plissard}}, \bibinfo {author}
  {\bibfnamefont {E.~P. A.~M.}\ \bibnamefont {Bakkers}}, \ and\ \bibinfo
  {author} {\bibfnamefont {L.~P.}\ \bibnamefont {Kouwenhoven}},\ }\href
  {\doibase 10.1126/science.1222360} {\bibfield  {journal} {\bibinfo  {journal}
  {Science}\ }\textbf {\bibinfo {volume} {336}},\ \bibinfo {pages} {1003}
  (\bibinfo {year} {2012})}\BibitemShut {NoStop}%
\bibitem [{\citenamefont {Deng}\ \emph {et~al.}(2012)\citenamefont {Deng},
  \citenamefont {Yu}, \citenamefont {Huang}, \citenamefont {Larsson},
  \citenamefont {Caroff},\ and\ \citenamefont {Xu}}]{Deng2012Anomalous}%
  \BibitemOpen
  \bibfield  {author} {\bibinfo {author} {\bibfnamefont {M.~T.}\ \bibnamefont
  {Deng}}, \bibinfo {author} {\bibfnamefont {C.~L.}\ \bibnamefont {Yu}},
  \bibinfo {author} {\bibfnamefont {G.~Y.}\ \bibnamefont {Huang}}, \bibinfo
  {author} {\bibfnamefont {M.}~\bibnamefont {Larsson}}, \bibinfo {author}
  {\bibfnamefont {P.}~\bibnamefont {Caroff}}, \ and\ \bibinfo {author}
  {\bibfnamefont {H.~Q.}\ \bibnamefont {Xu}},\ }\href {\doibase
  10.1021/nl303758w} {\bibfield  {journal} {\bibinfo  {journal} {Nano Lett.}\
  }\textbf {\bibinfo {volume} {12}},\ \bibinfo {pages} {6414} (\bibinfo {year}
  {2012})}\BibitemShut {NoStop}%
\bibitem [{\citenamefont {Das}\ \emph {et~al.}(2012)\citenamefont {Das},
  \citenamefont {Ronen}, \citenamefont {Most}, \citenamefont {Oreg},
  \citenamefont {Heiblum},\ and\ \citenamefont {Shtrikman}}]{Das2012Zero}%
  \BibitemOpen
  \bibfield  {author} {\bibinfo {author} {\bibfnamefont {A.}~\bibnamefont
  {Das}}, \bibinfo {author} {\bibfnamefont {Y.}~\bibnamefont {Ronen}}, \bibinfo
  {author} {\bibfnamefont {Y.}~\bibnamefont {Most}}, \bibinfo {author}
  {\bibfnamefont {Y.}~\bibnamefont {Oreg}}, \bibinfo {author} {\bibfnamefont
  {M.}~\bibnamefont {Heiblum}}, \ and\ \bibinfo {author} {\bibfnamefont
  {H.}~\bibnamefont {Shtrikman}},\ }\href {http://dx.doi.org/10.1038/nphys2479}
  {\bibfield  {journal} {\bibinfo  {journal} {Nat. Phys.}\ }\textbf {\bibinfo
  {volume} {8}},\ \bibinfo {pages} {887} (\bibinfo {year} {2012})}\BibitemShut
  {NoStop}%
\bibitem [{\citenamefont {Churchill}\ \emph {et~al.}(2013)\citenamefont
  {Churchill}, \citenamefont {Fatemi}, \citenamefont {Grove-Rasmussen},
  \citenamefont {Deng}, \citenamefont {Caroff}, \citenamefont {Xu},\ and\
  \citenamefont {Marcus}}]{Churchill2013Superconductor}%
  \BibitemOpen
  \bibfield  {author} {\bibinfo {author} {\bibfnamefont {H.~O.~H.}\
  \bibnamefont {Churchill}}, \bibinfo {author} {\bibfnamefont {V.}~\bibnamefont
  {Fatemi}}, \bibinfo {author} {\bibfnamefont {K.}~\bibnamefont
  {Grove-Rasmussen}}, \bibinfo {author} {\bibfnamefont {M.~T.}\ \bibnamefont
  {Deng}}, \bibinfo {author} {\bibfnamefont {P.}~\bibnamefont {Caroff}},
  \bibinfo {author} {\bibfnamefont {H.~Q.}\ \bibnamefont {Xu}}, \ and\ \bibinfo
  {author} {\bibfnamefont {C.~M.}\ \bibnamefont {Marcus}},\ }\href {\doibase
  10.1103/PhysRevB.87.241401} {\bibfield  {journal} {\bibinfo  {journal} {Phys.
  Rev. B}\ }\textbf {\bibinfo {volume} {87}},\ \bibinfo {pages} {241401}
  (\bibinfo {year} {2013})}\BibitemShut {NoStop}%
\bibitem [{\citenamefont {Finck}\ \emph {et~al.}(2013)\citenamefont {Finck},
  \citenamefont {Van~Harlingen}, \citenamefont {Mohseni}, \citenamefont
  {Jung},\ and\ \citenamefont {Li}}]{Finck2013Anomalous}%
  \BibitemOpen
  \bibfield  {author} {\bibinfo {author} {\bibfnamefont {A.~D.~K.}\
  \bibnamefont {Finck}}, \bibinfo {author} {\bibfnamefont {D.~J.}\ \bibnamefont
  {Van~Harlingen}}, \bibinfo {author} {\bibfnamefont {P.~K.}\ \bibnamefont
  {Mohseni}}, \bibinfo {author} {\bibfnamefont {K.}~\bibnamefont {Jung}}, \
  and\ \bibinfo {author} {\bibfnamefont {X.}~\bibnamefont {Li}},\ }\href
  {\doibase 10.1103/PhysRevLett.110.126406} {\bibfield  {journal} {\bibinfo
  {journal} {Phys. Rev. Lett.}\ }\textbf {\bibinfo {volume} {110}},\ \bibinfo
  {pages} {126406} (\bibinfo {year} {2013})}\BibitemShut {NoStop}%
\bibitem [{\citenamefont {Zhang}\ \emph {et~al.}(2016)\citenamefont {Zhang},
  \citenamefont {G{\"u}l} \emph {et~al.}}]{Zhang2016Ballistic}%
  \BibitemOpen
  \bibfield  {author} {\bibinfo {author} {\bibfnamefont {H.}~\bibnamefont
  {Zhang}}, \bibinfo {author} {\bibfnamefont {{\"O}.}~\bibnamefont {G{\"u}l}},
  \emph {et~al.},\ }\href {https://arxiv.org/abs/1603.04069} {\bibfield
  {journal} {\bibinfo  {journal} {arXiv:1603.04069}\ } (\bibinfo {year}
  {2016})}\BibitemShut {NoStop}%
\bibitem [{\citenamefont {Deng}\ \emph {et~al.}(2016)\citenamefont {Deng},
  \citenamefont {Vaitiekenas}, \citenamefont {Hansen}, \citenamefont {Danon},
  \citenamefont {Leijnse}, \citenamefont {Flensberg}, \citenamefont
  {Nyg{\aa}rd}, \citenamefont {Krogstrup},\ and\ \citenamefont
  {Marcus}}]{Deng2016Majorana}%
  \BibitemOpen
  \bibfield  {author} {\bibinfo {author} {\bibfnamefont {M.~T.}\ \bibnamefont
  {Deng}}, \bibinfo {author} {\bibfnamefont {S.}~\bibnamefont {Vaitiekenas}},
  \bibinfo {author} {\bibfnamefont {E.~B.}\ \bibnamefont {Hansen}}, \bibinfo
  {author} {\bibfnamefont {J.}~\bibnamefont {Danon}}, \bibinfo {author}
  {\bibfnamefont {M.}~\bibnamefont {Leijnse}}, \bibinfo {author} {\bibfnamefont
  {K.}~\bibnamefont {Flensberg}}, \bibinfo {author} {\bibfnamefont
  {J.}~\bibnamefont {Nyg{\aa}rd}}, \bibinfo {author} {\bibfnamefont
  {P.}~\bibnamefont {Krogstrup}}, \ and\ \bibinfo {author} {\bibfnamefont
  {C.~M.}\ \bibnamefont {Marcus}},\ }\href {\doibase 10.1126/science.aaf3961}
  {\bibfield  {journal} {\bibinfo  {journal} {Science}\ }\textbf {\bibinfo
  {volume} {354}},\ \bibinfo {pages} {1557} (\bibinfo {year}
  {2016})}\BibitemShut {NoStop}%
\bibitem [{\citenamefont {Chen}\ \emph {et~al.}(2017)\citenamefont {Chen},
  \citenamefont {Yu}, \citenamefont {Stenger}, \citenamefont {Hocevar},
  \citenamefont {Car}, \citenamefont {Plissard}, \citenamefont {Bakkers},
  \citenamefont {Stanescu},\ and\ \citenamefont
  {Frolov}}]{Chen2016Experimental}%
  \BibitemOpen
  \bibfield  {author} {\bibinfo {author} {\bibfnamefont {J.}~\bibnamefont
  {Chen}}, \bibinfo {author} {\bibfnamefont {P.}~\bibnamefont {Yu}}, \bibinfo
  {author} {\bibfnamefont {J.}~\bibnamefont {Stenger}}, \bibinfo {author}
  {\bibfnamefont {M.}~\bibnamefont {Hocevar}}, \bibinfo {author} {\bibfnamefont
  {D.}~\bibnamefont {Car}}, \bibinfo {author} {\bibfnamefont {S.~R.}\
  \bibnamefont {Plissard}}, \bibinfo {author} {\bibfnamefont {E.~P.}\
  \bibnamefont {Bakkers}}, \bibinfo {author} {\bibfnamefont {T.~D.}\
  \bibnamefont {Stanescu}}, \ and\ \bibinfo {author} {\bibfnamefont {S.~M.}\
  \bibnamefont {Frolov}},\ }\href@noop {} {\bibfield  {journal} {\bibinfo
  {journal} {Science Advances}\ }\textbf {\bibinfo {volume} {3}},\ \bibinfo
  {pages} {e1701476} (\bibinfo {year} {2017})}\BibitemShut {NoStop}%
\bibitem [{Mar()}]{Marcus2017PrivateCom}%
  \BibitemOpen
  \href@noop {} {}\bibinfo {note} {C. Marcus, APS March Meeting talk at New
  Orleans, March 2017}\BibitemShut {NoStop}%
\bibitem [{Kou()}]{Kouwenhoven2017PrivateCom}%
  \BibitemOpen
  \href@noop {} {}\bibinfo {note} {L. Kouwenhoven, APS March Meeting talk at
  New Orleans, March 2017}\BibitemShut {NoStop}%
\bibitem [{\citenamefont {Liu}\ \emph {et~al.}(2017{\natexlab{a}})\citenamefont
  {Liu}, \citenamefont {Setiawan}, \citenamefont {Sau},\ and\ \citenamefont
  {Das~Sarma}}]{liu2017phenomenology}%
  \BibitemOpen
  \bibfield  {author} {\bibinfo {author} {\bibfnamefont {C.-X.}\ \bibnamefont
  {Liu}}, \bibinfo {author} {\bibfnamefont {F.}~\bibnamefont {Setiawan}},
  \bibinfo {author} {\bibfnamefont {J.~D.}\ \bibnamefont {Sau}}, \ and\
  \bibinfo {author} {\bibfnamefont {S.}~\bibnamefont {Das~Sarma}},\ }\href
  {\doibase 10.1103/PhysRevB.96.054520} {\bibfield  {journal} {\bibinfo
  {journal} {Phys. Rev. B}\ }\textbf {\bibinfo {volume} {96}},\ \bibinfo
  {pages} {054520} (\bibinfo {year} {2017}{\natexlab{a}})}\BibitemShut
  {NoStop}%
\bibitem [{\citenamefont {Beenakker}(2013)}]{Beenakker2013Search}%
  \BibitemOpen
  \bibfield  {author} {\bibinfo {author} {\bibfnamefont {C.}~\bibnamefont
  {Beenakker}},\ }\href {\doibase 10.1146/annurev-conmatphys-030212-184337}
  {\bibfield  {journal} {\bibinfo  {journal} {Annu. Rev. Condens. Matter
  Phys.}\ }\textbf {\bibinfo {volume} {4}},\ \bibinfo {pages} {113} (\bibinfo
  {year} {2013})}\BibitemShut {NoStop}%
\bibitem [{\citenamefont {Leijnse}\ and\ \citenamefont
  {Flensberg}(2012)}]{Leijnse2012Introduction}%
  \BibitemOpen
  \bibfield  {author} {\bibinfo {author} {\bibfnamefont {M.}~\bibnamefont
  {Leijnse}}\ and\ \bibinfo {author} {\bibfnamefont {K.}~\bibnamefont
  {Flensberg}},\ }\href {http://stacks.iop.org/0268-1242/27/i=12/a=124003}
  {\bibfield  {journal} {\bibinfo  {journal} {Semicond. Sci. Technol.}\
  }\textbf {\bibinfo {volume} {27}},\ \bibinfo {pages} {124003} (\bibinfo
  {year} {2012})}\BibitemShut {NoStop}%
\bibitem [{\citenamefont {Stanescu}\ and\ \citenamefont
  {Tewari}(2013)}]{Stanescu2013Majorana}%
  \BibitemOpen
  \bibfield  {author} {\bibinfo {author} {\bibfnamefont {T.~D.}\ \bibnamefont
  {Stanescu}}\ and\ \bibinfo {author} {\bibfnamefont {S.}~\bibnamefont
  {Tewari}},\ }\href
  {http://iopscience.iop.org/article/10.1088/0953-8984/25/23/233201/meta}
  {\bibfield  {journal} {\bibinfo  {journal} {J. Phys.: Condens. Matter}\
  }\textbf {\bibinfo {volume} {25}},\ \bibinfo {pages} {233201} (\bibinfo
  {year} {2013})}\BibitemShut {NoStop}%
\bibitem [{\citenamefont {Elliott}\ and\ \citenamefont
  {Franz}(2015)}]{Elliott2015Colloquium}%
  \BibitemOpen
  \bibfield  {author} {\bibinfo {author} {\bibfnamefont {S.~R.}\ \bibnamefont
  {Elliott}}\ and\ \bibinfo {author} {\bibfnamefont {M.}~\bibnamefont
  {Franz}},\ }\href {\doibase 10.1103/RevModPhys.87.137} {\bibfield  {journal}
  {\bibinfo  {journal} {Rev. Mod. Phys.}\ }\textbf {\bibinfo {volume} {87}},\
  \bibinfo {pages} {137} (\bibinfo {year} {2015})}\BibitemShut {NoStop}%
\bibitem [{\citenamefont {Stanescu}(2016)}]{stanescu2016introduction}%
  \BibitemOpen
  \bibfield  {author} {\bibinfo {author} {\bibfnamefont {T.~D.}\ \bibnamefont
  {Stanescu}},\ }\href@noop {} {\emph {\bibinfo {title} {Introduction to
  Topological Quantum Matter \& Quantum Computation}}}\ (\bibinfo  {publisher}
  {CRC Press},\ \bibinfo {year} {2016})\BibitemShut {NoStop}%
\bibitem [{\citenamefont {Sato}\ and\ \citenamefont
  {Fujimoto}(2016)}]{sato2016majorana}%
  \BibitemOpen
  \bibfield  {author} {\bibinfo {author} {\bibfnamefont {M.}~\bibnamefont
  {Sato}}\ and\ \bibinfo {author} {\bibfnamefont {S.}~\bibnamefont
  {Fujimoto}},\ }\href@noop {} {\bibfield  {journal} {\bibinfo  {journal}
  {Journal of the Physical Society of Japan}\ }\textbf {\bibinfo {volume}
  {85}},\ \bibinfo {pages} {072001} (\bibinfo {year} {2016})}\BibitemShut
  {NoStop}%
\bibitem [{\citenamefont {Lutchyn}\ \emph {et~al.}(2017)\citenamefont
  {Lutchyn}, \citenamefont {Bakkers}, \citenamefont {Kouwenhoven},
  \citenamefont {Krogstrup}, \citenamefont {Marcus},\ and\ \citenamefont
  {Oreg}}]{lutchyn2017realizing}%
  \BibitemOpen
  \bibfield  {author} {\bibinfo {author} {\bibfnamefont {R.}~\bibnamefont
  {Lutchyn}}, \bibinfo {author} {\bibfnamefont {E.}~\bibnamefont {Bakkers}},
  \bibinfo {author} {\bibfnamefont {L.}~\bibnamefont {Kouwenhoven}}, \bibinfo
  {author} {\bibfnamefont {P.}~\bibnamefont {Krogstrup}}, \bibinfo {author}
  {\bibfnamefont {C.}~\bibnamefont {Marcus}}, \ and\ \bibinfo {author}
  {\bibfnamefont {Y.}~\bibnamefont {Oreg}},\ }\href@noop {} {\bibfield
  {journal} {\bibinfo  {journal} {arXiv:1707.04899}\ } (\bibinfo {year}
  {2017})}\BibitemShut {NoStop}%
\bibitem [{\citenamefont {Lee}\ \emph {et~al.}(2012)\citenamefont {Lee},
  \citenamefont {Jiang}, \citenamefont {Aguado}, \citenamefont {Katsaros},
  \citenamefont {Lieber},\ and\ \citenamefont {De~Franceschi}}]{Lee2012Zero}%
  \BibitemOpen
  \bibfield  {author} {\bibinfo {author} {\bibfnamefont {E.~J.~H.}\
  \bibnamefont {Lee}}, \bibinfo {author} {\bibfnamefont {X.}~\bibnamefont
  {Jiang}}, \bibinfo {author} {\bibfnamefont {R.}~\bibnamefont {Aguado}},
  \bibinfo {author} {\bibfnamefont {G.}~\bibnamefont {Katsaros}}, \bibinfo
  {author} {\bibfnamefont {C.~M.}\ \bibnamefont {Lieber}}, \ and\ \bibinfo
  {author} {\bibfnamefont {S.}~\bibnamefont {De~Franceschi}},\ }\href {\doibase
  10.1103/PhysRevLett.109.186802} {\bibfield  {journal} {\bibinfo  {journal}
  {Phys. Rev. Lett.}\ }\textbf {\bibinfo {volume} {109}},\ \bibinfo {pages}
  {186802} (\bibinfo {year} {2012})}\BibitemShut {NoStop}%
\bibitem [{\citenamefont {Liu}\ \emph {et~al.}(2012)\citenamefont {Liu},
  \citenamefont {Potter}, \citenamefont {Law},\ and\ \citenamefont
  {Lee}}]{Liu2012Zero}%
  \BibitemOpen
  \bibfield  {author} {\bibinfo {author} {\bibfnamefont {J.}~\bibnamefont
  {Liu}}, \bibinfo {author} {\bibfnamefont {A.~C.}\ \bibnamefont {Potter}},
  \bibinfo {author} {\bibfnamefont {K.~T.}\ \bibnamefont {Law}}, \ and\
  \bibinfo {author} {\bibfnamefont {P.~A.}\ \bibnamefont {Lee}},\ }\href
  {\doibase 10.1103/PhysRevLett.109.267002} {\bibfield  {journal} {\bibinfo
  {journal} {Phys. Rev. Lett.}\ }\textbf {\bibinfo {volume} {109}},\ \bibinfo
  {pages} {267002} (\bibinfo {year} {2012})}\BibitemShut {NoStop}%
\bibitem [{\citenamefont {Bagrets}\ and\ \citenamefont
  {Altland}(2012)}]{Dmitry2012Class}%
  \BibitemOpen
  \bibfield  {author} {\bibinfo {author} {\bibfnamefont {D.}~\bibnamefont
  {Bagrets}}\ and\ \bibinfo {author} {\bibfnamefont {A.}~\bibnamefont
  {Altland}},\ }\href {\doibase 10.1103/PhysRevLett.109.227005} {\bibfield
  {journal} {\bibinfo  {journal} {Phys. Rev. Lett.}\ }\textbf {\bibinfo
  {volume} {109}},\ \bibinfo {pages} {227005} (\bibinfo {year}
  {2012})}\BibitemShut {NoStop}%
\bibitem [{\citenamefont {Pikulin}\ \emph {et~al.}(2012)\citenamefont
  {Pikulin}, \citenamefont {Dahlhaus}, \citenamefont {Wimmer}, \citenamefont
  {Schomerus},\ and\ \citenamefont {Beenakker}}]{Pikulin2012Zero}%
  \BibitemOpen
  \bibfield  {author} {\bibinfo {author} {\bibfnamefont {D.}~\bibnamefont
  {Pikulin}}, \bibinfo {author} {\bibfnamefont {J.}~\bibnamefont {Dahlhaus}},
  \bibinfo {author} {\bibfnamefont {M.}~\bibnamefont {Wimmer}}, \bibinfo
  {author} {\bibfnamefont {H.}~\bibnamefont {Schomerus}}, \ and\ \bibinfo
  {author} {\bibfnamefont {C.}~\bibnamefont {Beenakker}},\ }\href
  {http://iopscience.iop.org/article/10.1088/1367-2630/14/12/125011/meta}
  {\bibfield  {journal} {\bibinfo  {journal} {New J. Phys.}\ }\textbf {\bibinfo
  {volume} {14}},\ \bibinfo {pages} {125011} (\bibinfo {year}
  {2012})}\BibitemShut {NoStop}%
\bibitem [{\citenamefont {Kells}\ \emph {et~al.}(2012)\citenamefont {Kells},
  \citenamefont {Meidan},\ and\ \citenamefont {Brouwer}}]{Kells2012Near}%
  \BibitemOpen
  \bibfield  {author} {\bibinfo {author} {\bibfnamefont {G.}~\bibnamefont
  {Kells}}, \bibinfo {author} {\bibfnamefont {D.}~\bibnamefont {Meidan}}, \
  and\ \bibinfo {author} {\bibfnamefont {P.~W.}\ \bibnamefont {Brouwer}},\
  }\href {\doibase 10.1103/PhysRevB.86.100503} {\bibfield  {journal} {\bibinfo
  {journal} {Phys. Rev. B}\ }\textbf {\bibinfo {volume} {86}},\ \bibinfo
  {pages} {100503} (\bibinfo {year} {2012})}\BibitemShut {NoStop}%
\bibitem [{\citenamefont {Liu}\ \emph {et~al.}(2017{\natexlab{b}})\citenamefont
  {Liu}, \citenamefont {Sau}, \citenamefont {Stanescu},\ and\ \citenamefont
  {Das~Sarma}}]{Liu2017Andreev}%
  \BibitemOpen
  \bibfield  {author} {\bibinfo {author} {\bibfnamefont {C.-X.}\ \bibnamefont
  {Liu}}, \bibinfo {author} {\bibfnamefont {J.~D.}\ \bibnamefont {Sau}},
  \bibinfo {author} {\bibfnamefont {T.~D.}\ \bibnamefont {Stanescu}}, \ and\
  \bibinfo {author} {\bibfnamefont {S.}~\bibnamefont {Das~Sarma}},\ }\href
  {\doibase 10.1103/PhysRevB.96.075161} {\bibfield  {journal} {\bibinfo
  {journal} {Phys. Rev. B}\ }\textbf {\bibinfo {volume} {96}},\ \bibinfo
  {pages} {075161} (\bibinfo {year} {2017}{\natexlab{b}})}\BibitemShut
  {NoStop}%
\bibitem [{\citenamefont {Liu}\ \emph {et~al.}(2017{\natexlab{c}})\citenamefont
  {Liu}, \citenamefont {Sau},\ and\ \citenamefont {Das~Sarma}}]{Liu2017Role}%
  \BibitemOpen
  \bibfield  {author} {\bibinfo {author} {\bibfnamefont {C.-X.}\ \bibnamefont
  {Liu}}, \bibinfo {author} {\bibfnamefont {J.~D.}\ \bibnamefont {Sau}}, \ and\
  \bibinfo {author} {\bibfnamefont {S.}~\bibnamefont {Das~Sarma}},\ }\href
  {\doibase 10.1103/PhysRevB.95.054502} {\bibfield  {journal} {\bibinfo
  {journal} {Phys. Rev. B}\ }\textbf {\bibinfo {volume} {95}},\ \bibinfo
  {pages} {054502} (\bibinfo {year} {2017}{\natexlab{c}})}\BibitemShut
  {NoStop}%
\bibitem [{\citenamefont {Chiu}\ \emph {et~al.}(2017)\citenamefont {Chiu},
  \citenamefont {Sau},\ and\ \citenamefont {Das~Sarma}}]{Chiu2017Conductance}%
  \BibitemOpen
  \bibfield  {author} {\bibinfo {author} {\bibfnamefont {C.-K.}\ \bibnamefont
  {Chiu}}, \bibinfo {author} {\bibfnamefont {J.~D.}\ \bibnamefont {Sau}}, \
  and\ \bibinfo {author} {\bibfnamefont {S.}~\bibnamefont {Das~Sarma}},\ }\href
  {\doibase 10.1103/PhysRevB.96.054504} {\bibfield  {journal} {\bibinfo
  {journal} {Phys. Rev. B}\ }\textbf {\bibinfo {volume} {96}},\ \bibinfo
  {pages} {054504} (\bibinfo {year} {2017})}\BibitemShut {NoStop}%
\bibitem [{\citenamefont {Prada}\ \emph {et~al.}(2017)\citenamefont {Prada},
  \citenamefont {Aguado},\ and\ \citenamefont {San-Jose}}]{Prada2017Measuring}%
  \BibitemOpen
  \bibfield  {author} {\bibinfo {author} {\bibfnamefont {E.}~\bibnamefont
  {Prada}}, \bibinfo {author} {\bibfnamefont {R.}~\bibnamefont {Aguado}}, \
  and\ \bibinfo {author} {\bibfnamefont {P.}~\bibnamefont {San-Jose}},\ }\href
  {\doibase 10.1103/PhysRevB.96.085418} {\bibfield  {journal} {\bibinfo
  {journal} {Phys. Rev. B}\ }\textbf {\bibinfo {volume} {96}},\ \bibinfo
  {pages} {085418} (\bibinfo {year} {2017})}\BibitemShut {NoStop}%
\bibitem [{\citenamefont {Clarke}(2017)}]{Clarke2017Experimentally}%
  \BibitemOpen
  \bibfield  {author} {\bibinfo {author} {\bibfnamefont {D.~J.}\ \bibnamefont
  {Clarke}},\ }\href {https://arxiv.org/abs/1702.01740} {\bibfield  {journal}
  {\bibinfo  {journal} {arXiv:1702.01740}\ } (\bibinfo {year}
  {2017})}\BibitemShut {NoStop}%
\bibitem [{\citenamefont {Lin}\ \emph {et~al.}(2012)\citenamefont {Lin},
  \citenamefont {Sau},\ and\ \citenamefont {Das~Sarma}}]{Lin2012Zero}%
  \BibitemOpen
  \bibfield  {author} {\bibinfo {author} {\bibfnamefont {C.-H.}\ \bibnamefont
  {Lin}}, \bibinfo {author} {\bibfnamefont {J.~D.}\ \bibnamefont {Sau}}, \ and\
  \bibinfo {author} {\bibfnamefont {S.}~\bibnamefont {Das~Sarma}},\ }\href
  {\doibase 10.1103/PhysRevB.86.224511} {\bibfield  {journal} {\bibinfo
  {journal} {Phys. Rev. B}\ }\textbf {\bibinfo {volume} {86}},\ \bibinfo
  {pages} {224511} (\bibinfo {year} {2012})}\BibitemShut {NoStop}%
\bibitem [{\citenamefont {Cole}\ \emph {et~al.}(2015)\citenamefont {Cole},
  \citenamefont {Das~Sarma},\ and\ \citenamefont {Stanescu}}]{Cole2015Effects}%
  \BibitemOpen
  \bibfield  {author} {\bibinfo {author} {\bibfnamefont {W.~S.}\ \bibnamefont
  {Cole}}, \bibinfo {author} {\bibfnamefont {S.}~\bibnamefont {Das~Sarma}}, \
  and\ \bibinfo {author} {\bibfnamefont {T.~D.}\ \bibnamefont {Stanescu}},\
  }\href {\doibase 10.1103/PhysRevB.92.174511} {\bibfield  {journal} {\bibinfo
  {journal} {Phys. Rev. B}\ }\textbf {\bibinfo {volume} {92}},\ \bibinfo
  {pages} {174511} (\bibinfo {year} {2015})}\BibitemShut {NoStop}%
\bibitem [{\citenamefont {Stanescu}\ and\ \citenamefont
  {Das~Sarma}(2017)}]{Stanescu2017Proximity}%
  \BibitemOpen
  \bibfield  {author} {\bibinfo {author} {\bibfnamefont {T.~D.}\ \bibnamefont
  {Stanescu}}\ and\ \bibinfo {author} {\bibfnamefont {S.}~\bibnamefont
  {Das~Sarma}},\ }\href {\doibase 10.1103/PhysRevB.96.014510} {\bibfield
  {journal} {\bibinfo  {journal} {Phys. Rev. B}\ }\textbf {\bibinfo {volume}
  {96}},\ \bibinfo {pages} {014510} (\bibinfo {year} {2017})}\BibitemShut
  {NoStop}%
\bibitem [{\citenamefont {Cole}\ \emph {et~al.}(2016)\citenamefont {Cole},
  \citenamefont {Sau},\ and\ \citenamefont {Das~Sarma}}]{Cole2016Proximity}%
  \BibitemOpen
  \bibfield  {author} {\bibinfo {author} {\bibfnamefont {W.~S.}\ \bibnamefont
  {Cole}}, \bibinfo {author} {\bibfnamefont {J.~D.}\ \bibnamefont {Sau}}, \
  and\ \bibinfo {author} {\bibfnamefont {S.}~\bibnamefont {Das~Sarma}},\ }\href
  {\doibase 10.1103/PhysRevB.94.140505} {\bibfield  {journal} {\bibinfo
  {journal} {Phys. Rev. B}\ }\textbf {\bibinfo {volume} {94}},\ \bibinfo
  {pages} {140505} (\bibinfo {year} {2016})}\BibitemShut {NoStop}%
\bibitem [{\citenamefont {Nichele}\ \emph {et~al.}(2017)\citenamefont
  {Nichele}, \citenamefont {Drachmann}, \citenamefont {Whiticar}, \citenamefont
  {O'Farrell}, \citenamefont {Suominen}, \citenamefont {Fornieri},
  \citenamefont {Wang}, \citenamefont {Gardner}, \citenamefont {Thomas},
  \citenamefont {Hatke}, \citenamefont {Krogstrup}, \citenamefont {Manfra},
  \citenamefont {Flensberg},\ and\ \citenamefont
  {Marcus}}]{nichele2017scaling}%
  \BibitemOpen
  \bibfield  {author} {\bibinfo {author} {\bibfnamefont {F.}~\bibnamefont
  {Nichele}}, \bibinfo {author} {\bibfnamefont {A.~C.~C.}\ \bibnamefont
  {Drachmann}}, \bibinfo {author} {\bibfnamefont {A.~M.}\ \bibnamefont
  {Whiticar}}, \bibinfo {author} {\bibfnamefont {E.~C.~T.}\ \bibnamefont
  {O'Farrell}}, \bibinfo {author} {\bibfnamefont {H.~J.}\ \bibnamefont
  {Suominen}}, \bibinfo {author} {\bibfnamefont {A.}~\bibnamefont {Fornieri}},
  \bibinfo {author} {\bibfnamefont {T.}~\bibnamefont {Wang}}, \bibinfo {author}
  {\bibfnamefont {G.~C.}\ \bibnamefont {Gardner}}, \bibinfo {author}
  {\bibfnamefont {C.}~\bibnamefont {Thomas}}, \bibinfo {author} {\bibfnamefont
  {A.~T.}\ \bibnamefont {Hatke}}, \bibinfo {author} {\bibfnamefont
  {P.}~\bibnamefont {Krogstrup}}, \bibinfo {author} {\bibfnamefont {M.~J.}\
  \bibnamefont {Manfra}}, \bibinfo {author} {\bibfnamefont {K.}~\bibnamefont
  {Flensberg}}, \ and\ \bibinfo {author} {\bibfnamefont {C.~M.}\ \bibnamefont
  {Marcus}},\ }\href {\doibase 10.1103/PhysRevLett.119.136803} {\bibfield
  {journal} {\bibinfo  {journal} {Phys. Rev. Lett.}\ }\textbf {\bibinfo
  {volume} {119}},\ \bibinfo {pages} {136803} (\bibinfo {year}
  {2017})}\BibitemShut {NoStop}%
\bibitem [{Hao()}]{HaoZhang2017PrivateCom}%
  \BibitemOpen
  \href@noop {} {}\bibinfo {note} {Hao Zhang, private
  communication}\BibitemShut {NoStop}%
\bibitem [{\citenamefont {Suominen}\ \emph {et~al.}(2017)\citenamefont
  {Suominen}, \citenamefont {Kjaergaard}, \citenamefont {Hamilton},
  \citenamefont {Shabani}, \citenamefont {Palmstr\o{}m}, \citenamefont
  {Marcus},\ and\ \citenamefont {Nichele}}]{suominen2017scalable}%
  \BibitemOpen
  \bibfield  {author} {\bibinfo {author} {\bibfnamefont {H.~J.}\ \bibnamefont
  {Suominen}}, \bibinfo {author} {\bibfnamefont {M.}~\bibnamefont
  {Kjaergaard}}, \bibinfo {author} {\bibfnamefont {A.~R.}\ \bibnamefont
  {Hamilton}}, \bibinfo {author} {\bibfnamefont {J.}~\bibnamefont {Shabani}},
  \bibinfo {author} {\bibfnamefont {C.~J.}\ \bibnamefont {Palmstr\o{}m}},
  \bibinfo {author} {\bibfnamefont {C.~M.}\ \bibnamefont {Marcus}}, \ and\
  \bibinfo {author} {\bibfnamefont {F.}~\bibnamefont {Nichele}},\ }\href
  {\doibase 10.1103/PhysRevLett.119.176805} {\bibfield  {journal} {\bibinfo
  {journal} {Phys. Rev. Lett.}\ }\textbf {\bibinfo {volume} {119}},\ \bibinfo
  {pages} {176805} (\bibinfo {year} {2017})}\BibitemShut {NoStop}%
\bibitem [{\citenamefont {Shabani}\ \emph {et~al.}(2016)\citenamefont
  {Shabani}, \citenamefont {Kjaergaard}, \citenamefont {Suominen},
  \citenamefont {Kim}, \citenamefont {Nichele}, \citenamefont {Pakrouski},
  \citenamefont {Stankevic}, \citenamefont {Lutchyn}, \citenamefont
  {Krogstrup}, \citenamefont {Feidenhans'l}, \citenamefont {Kraemer},
  \citenamefont {Nayak}, \citenamefont {Troyer}, \citenamefont {Marcus},\ and\
  \citenamefont {Palmstr\o{}m}}]{Shabani-2016-Two}%
  \BibitemOpen
  \bibfield  {author} {\bibinfo {author} {\bibfnamefont {J.}~\bibnamefont
  {Shabani}}, \bibinfo {author} {\bibfnamefont {M.}~\bibnamefont {Kjaergaard}},
  \bibinfo {author} {\bibfnamefont {H.~J.}\ \bibnamefont {Suominen}}, \bibinfo
  {author} {\bibfnamefont {Y.}~\bibnamefont {Kim}}, \bibinfo {author}
  {\bibfnamefont {F.}~\bibnamefont {Nichele}}, \bibinfo {author} {\bibfnamefont
  {K.}~\bibnamefont {Pakrouski}}, \bibinfo {author} {\bibfnamefont
  {T.}~\bibnamefont {Stankevic}}, \bibinfo {author} {\bibfnamefont {R.~M.}\
  \bibnamefont {Lutchyn}}, \bibinfo {author} {\bibfnamefont {P.}~\bibnamefont
  {Krogstrup}}, \bibinfo {author} {\bibfnamefont {R.}~\bibnamefont
  {Feidenhans'l}}, \bibinfo {author} {\bibfnamefont {S.}~\bibnamefont
  {Kraemer}}, \bibinfo {author} {\bibfnamefont {C.}~\bibnamefont {Nayak}},
  \bibinfo {author} {\bibfnamefont {M.}~\bibnamefont {Troyer}}, \bibinfo
  {author} {\bibfnamefont {C.~M.}\ \bibnamefont {Marcus}}, \ and\ \bibinfo
  {author} {\bibfnamefont {C.~J.}\ \bibnamefont {Palmstr\o{}m}},\ }\href
  {\doibase 10.1103/PhysRevB.93.155402} {\bibfield  {journal} {\bibinfo
  {journal} {Phys. Rev. B}\ }\textbf {\bibinfo {volume} {93}},\ \bibinfo
  {pages} {155402} (\bibinfo {year} {2016})}\BibitemShut {NoStop}%
\bibitem [{\citenamefont {Setiawan}\ \emph {et~al.}(2015)\citenamefont
  {Setiawan}, \citenamefont {Brydon}, \citenamefont {Sau},\ and\ \citenamefont
  {Das~Sarma}}]{Setiawan2015Conductance}%
  \BibitemOpen
  \bibfield  {author} {\bibinfo {author} {\bibfnamefont {F.}~\bibnamefont
  {Setiawan}}, \bibinfo {author} {\bibfnamefont {P.~M.~R.}\ \bibnamefont
  {Brydon}}, \bibinfo {author} {\bibfnamefont {J.~D.}\ \bibnamefont {Sau}}, \
  and\ \bibinfo {author} {\bibfnamefont {S.}~\bibnamefont {Das~Sarma}},\ }\href
  {\doibase 10.1103/PhysRevB.91.214513} {\bibfield  {journal} {\bibinfo
  {journal} {Phys. Rev. B}\ }\textbf {\bibinfo {volume} {91}},\ \bibinfo
  {pages} {214513} (\bibinfo {year} {2015})}\BibitemShut {NoStop}%
\bibitem [{\citenamefont {Wimmer}\ \emph {et~al.}(2011)\citenamefont {Wimmer},
  \citenamefont {Akhmerov}, \citenamefont {Dahlhaus},\ and\ \citenamefont
  {Beenakker}}]{Wimmer2011Quantum}%
  \BibitemOpen
  \bibfield  {author} {\bibinfo {author} {\bibfnamefont {M.}~\bibnamefont
  {Wimmer}}, \bibinfo {author} {\bibfnamefont {A.}~\bibnamefont {Akhmerov}},
  \bibinfo {author} {\bibfnamefont {J.}~\bibnamefont {Dahlhaus}}, \ and\
  \bibinfo {author} {\bibfnamefont {C.}~\bibnamefont {Beenakker}},\ }\href@noop
  {} {\bibfield  {journal} {\bibinfo  {journal} {New Journal of Physics}\
  }\textbf {\bibinfo {volume} {13}},\ \bibinfo {pages} {053016} (\bibinfo
  {year} {2011})}\BibitemShut {NoStop}%
\bibitem [{\citenamefont {Das~Sarma}\ \emph {et~al.}(2016)\citenamefont
  {Das~Sarma}, \citenamefont {Nag},\ and\ \citenamefont
  {Sau}}]{DasSarma2016How}%
  \BibitemOpen
  \bibfield  {author} {\bibinfo {author} {\bibfnamefont {S.}~\bibnamefont
  {Das~Sarma}}, \bibinfo {author} {\bibfnamefont {A.}~\bibnamefont {Nag}}, \
  and\ \bibinfo {author} {\bibfnamefont {J.~D.}\ \bibnamefont {Sau}},\ }\href
  {\doibase 10.1103/PhysRevB.94.035143} {\bibfield  {journal} {\bibinfo
  {journal} {Phys. Rev. B}\ }\textbf {\bibinfo {volume} {94}},\ \bibinfo
  {pages} {035143} (\bibinfo {year} {2016})}\BibitemShut {NoStop}%
\bibitem [{\citenamefont {Blonder}\ \emph {et~al.}(1982)\citenamefont
  {Blonder}, \citenamefont {Tinkham},\ and\ \citenamefont
  {Klapwijk}}]{Blonder1982Transition}%
  \BibitemOpen
  \bibfield  {author} {\bibinfo {author} {\bibfnamefont {G.~E.}\ \bibnamefont
  {Blonder}}, \bibinfo {author} {\bibfnamefont {M.}~\bibnamefont {Tinkham}}, \
  and\ \bibinfo {author} {\bibfnamefont {T.~M.}\ \bibnamefont {Klapwijk}},\
  }\href {\doibase 10.1103/PhysRevB.25.4515} {\bibfield  {journal} {\bibinfo
  {journal} {Phys. Rev. B}\ }\textbf {\bibinfo {volume} {25}},\ \bibinfo
  {pages} {4515} (\bibinfo {year} {1982})}\BibitemShut {NoStop}%
\bibitem [{\citenamefont {Beenakker}(1992)}]{Beenakker1992Transport}%
  \BibitemOpen
  \bibfield  {author} {\bibinfo {author} {\bibfnamefont {C.~W.~J.}\
  \bibnamefont {Beenakker}},\ }\href {\doibase 10.1103/PhysRevB.46.12841}
  {\bibfield  {journal} {\bibinfo  {journal} {Phys. Rev. B}\ }\textbf {\bibinfo
  {volume} {46}},\ \bibinfo {pages} {12841} (\bibinfo {year}
  {1992})}\BibitemShut {NoStop}%
\bibitem [{\citenamefont {Groth}\ \emph {et~al.}(2014)\citenamefont {Groth},
  \citenamefont {Wimmer}, \citenamefont {Akhmerov},\ and\ \citenamefont
  {Waintal}}]{Kwant}%
  \BibitemOpen
  \bibfield  {author} {\bibinfo {author} {\bibfnamefont {C.~W.}\ \bibnamefont
  {Groth}}, \bibinfo {author} {\bibfnamefont {M.}~\bibnamefont {Wimmer}},
  \bibinfo {author} {\bibfnamefont {A.~R.}\ \bibnamefont {Akhmerov}}, \ and\
  \bibinfo {author} {\bibfnamefont {X.}~\bibnamefont {Waintal}},\ }\href
  {http://stacks.iop.org/1367-2630/16/i=6/a=063065} {\bibfield  {journal}
  {\bibinfo  {journal} {New J. Phys.}\ }\textbf {\bibinfo {volume} {16}},\
  \bibinfo {pages} {063065} (\bibinfo {year} {2014})}\BibitemShut {NoStop}%
\bibitem [{\citenamefont {Cheng}\ \emph {et~al.}(2009)\citenamefont {Cheng},
  \citenamefont {Lutchyn}, \citenamefont {Galitski},\ and\ \citenamefont
  {Das~Sarma}}]{Cheng2009Splitting}%
  \BibitemOpen
  \bibfield  {author} {\bibinfo {author} {\bibfnamefont {M.}~\bibnamefont
  {Cheng}}, \bibinfo {author} {\bibfnamefont {R.~M.}\ \bibnamefont {Lutchyn}},
  \bibinfo {author} {\bibfnamefont {V.}~\bibnamefont {Galitski}}, \ and\
  \bibinfo {author} {\bibfnamefont {S.}~\bibnamefont {Das~Sarma}},\ }\href
  {\doibase 10.1103/PhysRevLett.103.107001} {\bibfield  {journal} {\bibinfo
  {journal} {Phys. Rev. Lett.}\ }\textbf {\bibinfo {volume} {103}},\ \bibinfo
  {pages} {107001} (\bibinfo {year} {2009})}\BibitemShut {NoStop}%
\bibitem [{\citenamefont {Das~Sarma}\ \emph {et~al.}(2012)\citenamefont
  {Das~Sarma}, \citenamefont {Sau},\ and\ \citenamefont
  {Stanescu}}]{DasSarma2012Splitting}%
  \BibitemOpen
  \bibfield  {author} {\bibinfo {author} {\bibfnamefont {S.}~\bibnamefont
  {Das~Sarma}}, \bibinfo {author} {\bibfnamefont {J.~D.}\ \bibnamefont {Sau}},
  \ and\ \bibinfo {author} {\bibfnamefont {T.~D.}\ \bibnamefont {Stanescu}},\
  }\href {\doibase 10.1103/PhysRevB.86.220506} {\bibfield  {journal} {\bibinfo
  {journal} {Phys. Rev. B}\ }\textbf {\bibinfo {volume} {86}},\ \bibinfo
  {pages} {220506} (\bibinfo {year} {2012})}\BibitemShut {NoStop}%
\bibitem [{\citenamefont {Law}\ \emph {et~al.}(2009)\citenamefont {Law},
  \citenamefont {Lee},\ and\ \citenamefont {Ng}}]{Law2009Majorana}%
  \BibitemOpen
  \bibfield  {author} {\bibinfo {author} {\bibfnamefont {K.~T.}\ \bibnamefont
  {Law}}, \bibinfo {author} {\bibfnamefont {P.~A.}\ \bibnamefont {Lee}}, \ and\
  \bibinfo {author} {\bibfnamefont {T.~K.}\ \bibnamefont {Ng}},\ }\href
  {\doibase 10.1103/PhysRevLett.103.237001} {\bibfield  {journal} {\bibinfo
  {journal} {Phys. Rev. Lett.}\ }\textbf {\bibinfo {volume} {103}},\ \bibinfo
  {pages} {237001} (\bibinfo {year} {2009})}\BibitemShut {NoStop}%
\bibitem [{\citenamefont {Flensberg}(2010)}]{Flensberg2010Tunneling}%
  \BibitemOpen
  \bibfield  {author} {\bibinfo {author} {\bibfnamefont {K.}~\bibnamefont
  {Flensberg}},\ }\href {\doibase 10.1103/PhysRevB.82.180516} {\bibfield
  {journal} {\bibinfo  {journal} {Phys. Rev. B}\ }\textbf {\bibinfo {volume}
  {82}},\ \bibinfo {pages} {180516} (\bibinfo {year} {2010})}\BibitemShut
  {NoStop}%
\bibitem [{\citenamefont {Kramer}(2012)}]{kramer2012quantum}%
  \BibitemOpen
  \bibfield  {author} {\bibinfo {author} {\bibfnamefont {B.}~\bibnamefont
  {Kramer}},\ }\href@noop {} {\emph {\bibinfo {title} {Quantum Transport in
  Semiconductor Submicron Structures}}},\ Vol.\ \bibinfo {volume} {326}\
  (\bibinfo  {publisher} {Springer},\ \bibinfo {year} {2012})\ p.\ \bibinfo
  {pages} {256}\BibitemShut {NoStop}%
\bibitem [{\citenamefont {Oda}\ and\ \citenamefont
  {Ferry}(2005)}]{oda2005silicon}%
  \BibitemOpen
  \bibfield  {author} {\bibinfo {author} {\bibfnamefont {S.}~\bibnamefont
  {Oda}}\ and\ \bibinfo {author} {\bibfnamefont {D.}~\bibnamefont {Ferry}},\
  }\href@noop {} {\emph {\bibinfo {title} {Silicon Nanoelectronics}}}\
  (\bibinfo  {publisher} {CRC press},\ \bibinfo {year} {2005})\ p.\ \bibinfo
  {pages} {163}\BibitemShut {NoStop}%
\bibitem [{\citenamefont {Datta}(1997)}]{datta1997electronic}%
  \BibitemOpen
  \bibfield  {author} {\bibinfo {author} {\bibfnamefont {S.}~\bibnamefont
  {Datta}},\ }\href@noop {} {\emph {\bibinfo {title} {Elecronic Transport in
  Mesoscopic Systems}}}\ (\bibinfo  {publisher} {Cambridge University Press},\
  \bibinfo {year} {1997})\BibitemShut {NoStop}%
\bibitem [{\citenamefont {Moore}\ \emph {et~al.}(2016)\citenamefont {Moore},
  \citenamefont {Stanescu},\ and\ \citenamefont {Tewari}}]{moore2016majorana}%
  \BibitemOpen
  \bibfield  {author} {\bibinfo {author} {\bibfnamefont {C.}~\bibnamefont
  {Moore}}, \bibinfo {author} {\bibfnamefont {T.~D.}\ \bibnamefont {Stanescu}},
  \ and\ \bibinfo {author} {\bibfnamefont {S.}~\bibnamefont {Tewari}},\ }\href
  {https://arxiv.org/abs/1611.07058} {\bibfield  {journal} {\bibinfo  {journal}
  {arXiv:1611.07058}\ } (\bibinfo {year} {2016})}\BibitemShut {NoStop}%
\bibitem [{\citenamefont {\ifmmode \check{Z}\else
  \v{Z}\fi{}uti\ifmmode~\acute{c}\else \'{c}\fi{}}\ and\ \citenamefont
  {Das~Sarma}(1999)}]{Zutic1999Spin}%
  \BibitemOpen
  \bibfield  {author} {\bibinfo {author} {\bibfnamefont {I.}~\bibnamefont
  {\ifmmode \check{Z}\else \v{Z}\fi{}uti\ifmmode~\acute{c}\else \'{c}\fi{}}}\
  and\ \bibinfo {author} {\bibfnamefont {S.}~\bibnamefont {Das~Sarma}},\ }\href
  {\doibase 10.1103/PhysRevB.60.R16322} {\bibfield  {journal} {\bibinfo
  {journal} {Phys. Rev. B}\ }\textbf {\bibinfo {volume} {60}},\ \bibinfo
  {pages} {R16322} (\bibinfo {year} {1999})}\BibitemShut {NoStop}%
\end{thebibliography}%


\end{document}